\documentclass[onecolumn]{IEEEtran}

\usepackage[utf8]{inputenc}
\usepackage[T1]{fontenc}
\usepackage{url}
\usepackage{cite}
\usepackage[cmex10]{amsmath}
\usepackage{algorithmic}
\usepackage{array}
\usepackage{textcomp}
\usepackage{stfloats}
\usepackage{verbatim}
\usepackage{graphicx}
\usepackage{color}
\usepackage{amssymb}
\usepackage{balance}
\usepackage{makecell}
\usepackage{subfigure}

\begin{document}

\title{An Information-Theoretic Framework for Receiver Quantization in Communication}

\author{Jing Zhou,~\IEEEmembership{Member,~IEEE}, Shuqin Pang, and Wenyi Zhang,~\IEEEmembership{Senior Member,~IEEE}

\thanks
{This work was supported in part by the National Natural Science Foundation of China through Grant 62231022 and in part by Henan Key Laboratory of Visible Light Communications through Grant HKLVLC2023-B03.
The material in this paper will be presented in part at the IEEE International Symposium on Information Theory (ISIT), Ann Arbor, MI, USA, June 2025 \cite{ISIT}. (\emph{Corresponding author: Wenyi Zhang.})

Jing Zhou is with the Department of Computer Science and Engineering, Shaoxing University, Shaoxing 312000, China (e-mail: jzhou@usx.edu.cn).

Shuqin Pang and Wenyi Zhang are with the Department of Electronic Engineering and Information Science, University of Science and Technology of China, Hefei 230027, China (e-mail: shuqinpa@mail.ustc.edu.cn, wenyizha@ustc.edu.cn).
} }

\markboth{Draft}{Draft}
\maketitle

\begin{abstract}
We investigate information-theoretic limits and design of communication under receiver quantization.
Unlike most existing studies that focus on low-resolution quantization, this work is more focused on the impact of weak nonlinear distortion due to resolution reduction from high to low.
We consider a standard transceiver architecture, which includes an independent and identically distributed (i.i.d.) complex Gaussian codebook at the transmitter, and a symmetric quantizer cascaded with a nearest neighbor decoder at the receiver.
Employing the generalized mutual information (GMI), an achievable rate under general quantization rules is obtained in an analytical form, which shows that the rate loss due to quantization is $\log\left(1+\gamma\mathsf{SNR}\right)$, where $\mathsf {SNR}$ is the signal-to-noise ratio at the receiver front-end, and $\gamma$ is determined by thresholds and levels of the quantizer.
Based on this result, the performance under uniform receiver quantization is analyzed comprehensively.
We show that the front-end gain control, which determines the loading factor (normalized one-sided quantization range) of quantization, has an increasing impact on performance as the resolution decreases.
In particular, we prove that the unique loading factor that minimizes the mean square error (MSE) of the uniform quantizer also maximizes the GMI, and the corresponding irreducible rate loss is given by $\log\left(1+\mathsf {mmse}\cdot\mathsf{SNR}\right)$, 
where $\mathsf {mmse}$ is the minimum MSE normalized by the variance of quantizer input, and it is equal to the minimum of $\gamma$.
A geometrical interpretation for the optimal uniform quantization at the receiver is further established.
Moreover, by asymptotic analysis, we characterize the impact of biased gain control, showing how small rate losses decay to zero and providing approximations for the achievable rate under large bias.
From asymptotic expressions of the optimal loading factor and $\mathsf {mmse}$, approximations and several ``per-bit rules'' for performance are also provided.
Finally we discuss more types of receiver quantization and show that the consistency between achievable rate maximization and MSE minimization does not hold in general.
\end{abstract}


\tableofcontents

\newpage
\section{Introduction}

\IEEEPARstart{T}{he} analog-to-digital conversion (ADC), including sampling and quantization, is essential for any digital receiver.
The power dissipation of state-of-the-art ADCs increases four times as the resolution increases by one bit, while every doubling of the sampling rate leads to a one-bit loss of resolution \cite{ADC1,ADC2}.
The impact of the ADC on the performance has received increasing attention along with the recent evolution of wireless communications, in which several challenges are faced, such as the increasing processing speed due to the utilization of a larger bandwidth in the mmWave and higher frequencies, the increasing scale of hardware due to the use of massive multiple-input-multiple-output (MIMO), and the critical need for low cost energy-efficient devices in emerging scenarios, e.g., massive machine-type communications (mMTC).

A majority of the studies on the ADC at communication receivers have focused on the performance and design under low-resolution output quantization, and one-bit quantization has been of particular interest due to its negligible power dissipation and simplicity of implementation, even without requiring automatic gain control (AGC).
In such studies the end-to-end channel is highly nonlinear, typically incurring a substantial performance loss, and necessitating a rethinking of the transceiver design.

On the other side, the transceiver architecture used in present wireless systems is built without considering the effect of output quantization.
It is thus necessary to ask, under such a conventional transceiver architecture, how much is the loss caused by output quantization with \emph{moderate to high} resolution?
In other words, if a small loss in achievable rate is acceptable, how fine need the quantization be?
Analytical results on these problems appear to be lacking.
Moreover, limited resolution of quantization leads to new problems, e.g., sensitivity of performance to the error of gain control, residual interference in multiuser systems, and so on.
These largely unexplored problems prompt us to revisit the topic of receiver quantization in communication in this work.

\subsection{Related Work on Quantization at Communication Receivers}

We begin from the impact of output quantization in the (discrete-time) additive white Gaussian noise (AWGN) channel, which is a benchmark model in communication theory.
The performance gain of using more output quantization levels in \emph{coded} transmission (i.e., soft-decision decoding) was observed very early in \cite{Null} via an information theoretic approach.
In the classic textbook of Wozencraft and Jacobs [\ref{WJ65}, Chap. 6.2], a cutoff rate analysis showed that, for an equiprobable uniformly spaced pulse amplitude modulation (PAM) input, the degradation due to output quantization is approximately $2$ dB when the alphabet size equals to the number of quantization levels, and the degradation vanishes when the quantization becomes increasingly fine.
Particularly, in the low-signal-to-noise-ratio (low-SNR) limit, hard-decision decoding (one-bit quantization that observes the sign of output) leads to a power loss of $\pi/2$ (approximately $2$ dB) \cite{Null,WJ65}; see also [\ref{VO79}, Chap. 2.11 and 3.4].\footnote{Interestingly, the loss can be fully recovered if we replace the hard-decision decoder (a.k.a. sign quantizer) by a carefully designed asymmetric one-bit quantizer (a.k.a. threshold quantizer) and employ asymmetric input \cite{KochLapidoth13}.}
For $K$-level output quantization, it was proved that a discrete input of at most $K+1$ mass points suffices to achieve the constrained capacity \cite{SDM09}.

In vector (MIMO) channels, the quantization loss in achievable rate can be very small at low-to-moderate SNR even if $1$\textasciitilde $3$-bit output quantization is used.
This fact was shown in information theoretic studies \cite{MR06,MC06,Nossek} by numerical examples for full-rank channels with multiplexing gains $2$\textasciitilde$4$.
Although the constrained capacity is still unknown even in the one-bit case, recent information theoretic studies have provided various results on MIMO systems with coarse output quantization \cite{MNS20,mo15,mo17,Rini17ITW,Khalili18,Laneman,Nam,BZEE22}, which show that proper transceiver design is critical in realizing efficient communication in such systems.
In particular, for one-bit output quantization, it was shown that the high-SNR capacity grows linearly with the rank of channel \cite{mo15}, while the low-SNR asymptotic power loss of $\pi/2$ still exists in the case of the vector channel \cite{MNS20}.

In light of these positive theoretical results, performance and design of wireless systems with low-resolution quantization have been extensively studied in recent years; see, e.g., \cite{Mezghani-Nossek12,Fan15,Roth-Nossek17,Jacobsson17,LL,Dutta,ARL,Chaaban}.
The most common approach therein, however, is not information-theoretic (since exact evaluation of mutual information can be difficult).
Instead, achievable rate estimation based on the additive quantization noise model (AQNM) has been widely used, which comes from Bussgang-like decomposition \cite{Bussgang,SPM-Bussgang,LozanoRangan}.
Results in \cite{Mezghani-Nossek12,mo15,mo17} suggested that such estimation approximates the mutual information well at low SNR, but becomes inaccurate at high SNR.\footnote{See the remark at the end of Sec. II for more discussions.}

Although mutual information is a fundamental performance measure, for communications under transceiver nonlinearity it has limited operational meaning, in the sense that the decoder that achieves the predicted rate can be too complex to implement, while that rate is not necessarily achievable by a standard transceiver architecture designed without considering nonlinearity because the decoder is typically \emph{mismatched} to the nonlinear channel.
In \cite{Zhang12,Liang-Zhang16}, a more meaningful performance measure that takes decoding rule into account, namely the generalized mutual information (GMI) \cite{ITFMD}, has been adopted, yielding analytical expressions of the achievable rate under output quantization and \emph{nearest neighbor decoding rule}.
Under a given (possibly mismatched) decoder, the GMI determines the highest rate below which the average probability of error, averaged over a given i.i.d. codebook ensemble, converges to zero as the block length $N$ grows without bound, and it is thus a lower bound on mismatch capacity \cite{ITFMD,GLT}.
The GMI has been applied in various scenarios for performance evaluation, including the bit-interleaved coded modulation (BICM) \cite{BICM}, fading channels \cite{Lapidoth-Shamai02,WSS04,Kramer}, and nonlinear fiber-optic channels \cite{Second}.
In fact, the rate estimation based on the AQNM (or Bussgang decomposition) is consistent with the GMI for scalar channel under Gaussian input and nearest neighbor decoding \cite{Zhang12}; see \cite{GNND} for more discussions.

\subsection{Related Work in Quantization Theory}

The rich theory of quantization was surveyed comprehensively in \cite{Gray-Neuhoff98}, in which two well-established asymptotic theories were emphasized.
The first is Shannon's information theoretic approach (rate distortion theory \cite{Berger}), which places quantization in the framework of lossy source coding and focuses on the high-dimension regime, thereby shedding light on vector quantization.
The second is the asymptotic quantization theory, which sheds light on quantizer design in the high-resolution regime.
The asymptotic quantization theory is more relevant to receiver quantization in communication which typically does not employ coding or vector quantization.
Some classical results in asymptotic quantization theory, especially those for uniform scalar quantization, are reviewed here.
Although not applicable directly to receiver quantization, they can be helpful in deriving new results (e.g., Corollary 10 in Sec. IV), closely related to our work (e.g., we will show the role of the mean square error (MSE) in the achievable rate analysis), and comparable to our findings (see, e.g., the remark in Sec. VII-A).

A basic result known in \cite{OPS48} and \cite{Bennett48} (rigorously proved in \cite{D12}) states that, for a high-resolution uniform quantizer with step size $\ell$, the MSE can be approximated by $\ell^2/12$.
This yields the ``$6$-dB-per-bit rule'' that each additional bit in resolution reduces the MSE by $6.02$ dB.
The rule reflects the impact of the step size that causes \emph{granular} distortion; but it ignores \emph{overload} distortion due to finite quantization range.
The quantization range is often represented by a parameter called \emph {loading factor} \cite{GG}, which is defined as the one-sided width of quantization range normalized by the standard deviation of the input of the quantizer.
The interplay between granular distortion and overload distortion dictates the optimal loading factor for a given resolution.
The asymptotic relationship between this optimal loading factor and the resolution has been characterized in the seminal work \cite{Hui-Neuhoff01} for several types of input densities.
Take the Gaussian source as an example.
In \cite{Hui-Neuhoff01} it has been shown that, for the optimal $2K$-level uniform quantization that minimizes the MSE,
1) the loading factor scales like $2\sqrt{\ln (2K)}$ (cf. the conventional ``four-sigma'' rule of thumb \cite{Bennett48,RoT,GG}), and
2) the granular distortion dominates and the overload distortion is asymptotically negligible.
Further properties of the uniform quantization have been analyzed in \cite{Na-Neuhoff12,Na-Neuhoff18,Na-Neuhoff19}.
For quantization at communication receivers, we need parallel results to characterize the optimal loading factor, which is essential for the design of the AGC.
We note that, although the importance of AGC design in the presence of output quantization has been recognized for a long time \cite{VO79,LozanoRangan}, it was  only investigated by numerical results in several works; see, e.g., \cite{MR06,MC06,AGCADC,Krone10,Roth-Nossek17}.

Similar to the AQNM, there is also an additive noise model for source quantization \cite{Bennett48,Gray-Neuhoff98,GG}, which approximates the quantization error as an independent white noise term added to the quantizer input (but does not include a scaling factor like that in the AQNM), though the ``noise'' is in fact a deterministic function of the input.
The quantization error can be white (uncorrelated between samples) when the source is i.i.d., and it is approximately uncorrelated with the input when the resolution is sufficiently high.
So the model may give a useful approximation under certain circumstances (see, e.g., \cite{Macro}).

\subsection{Our Work}

\subsubsection{Problem}

In this paper, we consider communications in the presence of output quantization, and focus on the effect of resolution reduction on the performance of a standard transceiver architecture designed without considering that effect.
Rather than considering only low-resolution output quantization ($1$\textasciitilde$3$ bits) as in most existing studies, we consider the entire region of resolution, especially the transition from high resolution (typically $8$\textasciitilde$12$ bits or more so that performance loss is negligible) to low resolution.
Since in the transition only weak to moderate nonlinearity is introduced, it is natural to keep the transceiver architecture unchanged rather than rebuild it.
The considered standard transceiver architecture includes an independent and identically distributed (i.i.d.) complex Gaussian codebook at the transmitter, and a uniform output quantizer cascaded with a (weighted) nearest neighbor decoder at the receiver, where the loading factor of the quantizer can be adjusted by gain control.
See Sec. II for details.
We choose such an architecture in view of the following facts.

\begin{itemize}
\item
When the impact of quantization is negligible, this architecture is capacity-achieving in several important channel models, such as the AWGN channel and the flat-fading channel with Gaussian noise and channel state information at the receiver (CSIR) \cite{TV05}.
It is also a very robust architecture in general noisy channels \cite{Lapidoth96}.
This architecture has been adopted in various performance evaluation problems, e.g., \cite{Lapidoth96,Lapidoth-Shamai02,WSS04} for linear channels with fading and \cite{Zhang12,Second,ZWSL19} for nonlinear channels.

\item
The nearest neighbor decoding (minimum Euclidean distance decoding) can be implemented efficiently and has been widely employed as a standard decoding rule in communication systems.
The uniform quantizer is also a standard component of practical receivers as well as a common assumption in performance analysis \cite{MR06,MC06,Nossek,C-UQ,Krone10,mo17,Rini17ITW,Dutta}.

\item
The achievable rate of a complex Gaussian input approximates that of regular high-order modulation schemes such as quadrature amplitude modulation (QAM).
In the AWGN channel, the high-SNR gap between their achievable rates is 1.53 dB, which can be further reduced by constellation shaping \cite{FU98}.
\end{itemize}

\subsubsection{Method}

The achievable rate results in this paper are derived based on the GMI, which, as discussed in Sec. I-A, is a convenient performance measure for the problem of information transmission under transceiver nonlinearity.
Moreover, as a performance measure under a \emph{mismatched} decoder (since the nearest neighbor decoding rule becomes suboptimal in the presence of receiver quantization), the GMI possesses optimality in the sense that it is the maximum achievable rate of the i.i.d. random code ensemble, thereby indicating the performance of a ``typical'' codebook \cite{ITFMD,Lapidoth-Shamai02}.
Specifically, for Gaussian codebook and nearest neighbor decoding, the GMI has a simple expression which can be evaluated by the correlation between the channel input and output.
Our asymptotic analyses also rely on methods and results in asymptotic (high-resolution) quantization theory.
A notable tool originating from numerical analysis is the Euler-Maclaurin summation formula \cite{NA}, which was initially introduced to source quantization theory in \cite{Na-Neuhoff12}.

\subsubsection{Summary of Contribution}
We provide information-theoretic results for the transceiver architecture including i.i.d. complex Gaussian codebook and nearest neighbor decoding, in the presence of complex Gaussian noise and symmetric receiver quantization, especially uniform quantization.
Our main contributions, including exact expressions, asymptotic formulas, and numerical results, are summarized as follows.

\begin{itemize}
\item
In Sec. III, for the considered transceiver architecture, we show that the GMI under a given SNR can be expressed by
\begin{align}\label{1}
I_\textrm{GMI} = C - \log( 1 + \gamma \mathsf{SNR} ),
\end{align}
where $C$ is the channel capacity when the resolution of quantization is unlimited, and the parameter $\gamma$, which does not depend on the SNR, is determined by thresholds and levels of the quantizer in an analytical form.

\item
In Sec. IV, for uniform quantization (with equispaced thresholds and mid-rise levels), we show that optimizing the loading factor by gain control before quantization (thereby minimizing $\gamma$ in (\ref{1})) is increasingly important as the resolution decreases, thus imposing a critical challenge to the AGC.
Interestingly, the problems of MSE minimization and achievable rate maximization are proven to be consistent, in the sense that there is a unique loading factor $L=L^*$ satisfying
\begin{align}
L^* = \arg\max\limits_L I_\textrm{GMI}(L) = \arg\min\limits_L {\mathsf{mse}}(L),
\end{align}
where $I_\textrm{GMI}(L)$ is the achievable rate as a function of the loading factor.
This fact, combined with existing result, implies that the optimal loading factor $L^*$ scales like $2\sqrt{b\ln2}$ as the resolution $b$ (in bits) increases.
We further prove that the minimum of $\gamma$ is exactly the minimum mean square error (MMSE) normalized by the variance of quantizer input (denoted by $\mathsf {mmse}$), so that the \emph{irreducible loss} in achievable rate due to uniform quantization is determined by
\begin{align}
C - I_\textrm{GMI}(L^*) = \log\left( 1 + \mathsf {mmse}\cdot\mathsf{SNR} \right).
\end{align}
For uniform quantization with the optimal loading factor, we establish a geometrical interpretation of how the MMSE and the additive Gaussian noise jointly determine the achievable rate.

\item
The uniform receiver quantization is further studied by asymptotic analysis.
In Sec. V, the impact of biased gain control is characterized by asymptotic behaviors and approximations of $I_\textrm{GMI}(L)$.
Approximations of $I_\textrm{GMI}(L)$ for overload region ($L<L^*$) and underload region ($L>L^*$) of the loading factor are proposed, respectively.
Specifically, in the high-resolution regime, we characterize the \emph{loading loss} in achievable rate, showing that i) the loading loss due to the overload distortion decays exponentially as the loading factor increases, and ii) the loading loss due to the granular distortion decays quadratically as the step size decreases.
In Sec. VI, focusing on the optimal uniform quantization, we provide a new approximation of $L^*$, an approximation of $I_\textrm{GMI}(L^*)$, and several per-bit rules for performance metrics such as saturation rate and irreducible rate loss;

\item
In Sec. VII, for general quantization rules, we illustrate that the consistency between achievable rate maximization and MSE minimization does not necessarily hold.
In fact, except for the uniform quantizer, in our setting only two specific quantizers that possess the consistency have been found; see Sec. VII-A.
We also discuss the possible gain of introducing more types of quantization for communication receiver.
\end{itemize}

\emph{Notation}:
We write $f(k)=o(g(k))$ to denote the asymptotic relationship $\lim_{k\to\infty}\frac{f(k)}{g(k)}=0$.
Specifically, we use $o_k(1)$ to denote a function $f(k)$ satisfying $\lim_{k\to\infty}f(k)=0$, and may omit the subscript if there is no danger of confusion.
We use $\phi(t)$ to denote the function $(2\pi)^{-1/2}\mathrm {exp}(-t^2/2)$, which is the probability density function (PDF) of the standard normal distribution, and use $Q(u)$ to denote the Q-function, i.e., $Q(u):=\int_u^\infty\phi(t)\mathrm d t$.
We further define $\phi(\infty)=0$ and $Q(\infty)=0$.
The complex conjugate of $A$ is denoted by $\overline A$.
The Euclidean norm of $\mathbf A$ is denoted by $\|\mathbf A\|$.
We use $X\perp Y$ to indicate the independence of two random variables $X$ and $Y$.

\section{Preliminaries}

\subsection{A Standard Transceiver Architecture}

\begin{figure}
\centering
\includegraphics[scale=0.28]{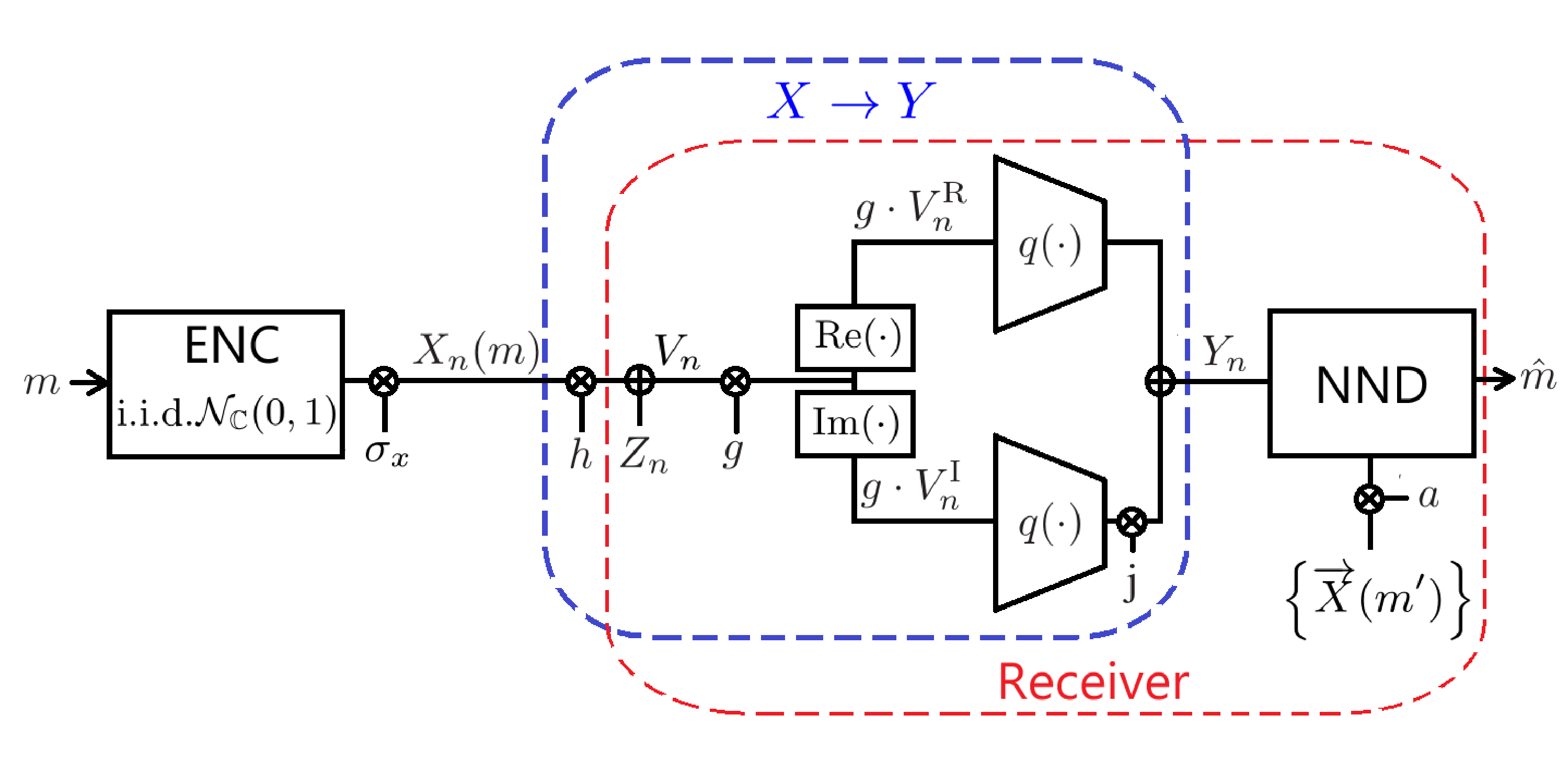}
\caption{Transceiver architecture.}
\label{Archi}
\end{figure}

The transceiver architecture we consider is shown in Fig. \ref{Archi}.
For a code rate $R$ bits/channel use (c.u.), a message is selected uniformly randomly from the index set $\mathcal M=\{1,2,...,\lceil 2^{NR}\rceil\}$.
If a message $m$ is selected, then the encoder maps it to a codeword $[X_1(m),...,X_N(m)]$ of block length $N$, which is generated according to a product complex Gaussian distribution $\mathcal {N}_\mathbb C(0,\sigma_x^2\mathbf I_N)$.
During transmission, each transmitted symbol is scaled by a channel gain $h\in\mathbb C$ which remains constant over the transmission duration of a codeword.
The scaled symbols are corrupted by i.i.d. complex Gaussian noise at the receiver front-end before quantization.
Then the channel output after quantization is given by
\begin{align}\label{YqXZ}
Y_n = q\left( g\cdot V_n^{\rm R}\right) + {\textrm j}\cdot q\left(g\cdot V_n^{\rm I} \right),
\end{align}
where $n=1,...,N$, $q(\cdot)$ denotes the quantizer which introduces nonlinear distortion, $g\in\mathbb R^+$ is a gain-control factor,
and
\begin{align}
V_n^{\textrm R} &= \mathrm {Re}( hX_n(m) + Z_n ), \\
V_n^{\textrm I} &= \mathrm {Im}( hX_n(m) + Z_n )
\end{align}
are real part and imaginary part of the received signal, respectively, where the noise $Z_n\sim \mathcal {N}_\mathbb C(0,\sigma^2)$ is independent of $X_n$.
In this model, we let the real and imaginary parts of the received signal be quantized by the same rule with the same gain-control factor.

\emph{Note (channel discretization)}:
In practical systems, the described transceiver architecture also includes Nyquist-type pulse shaping at the transmitter and matched filtering combined with symbol-rate sampling at the receiver.
It has a discrete-time memoryless model at symbol-level as (\ref{YqXZ}), if the channel does not introduce memory.
Note that oversampling (more accurately, sampling faster than the symbol rate) may improve the achievable rate with output quantization, especially in the case of one-bit quantization
(see \cite{Gilbert1993,Shamai1994} and \cite{Koch-Lapidoth10,Zhang12,landauJWCN,DZZ21}).
However, such performance improvement relies on several conditions, including significant nonlinearity (leading to frequency dispersion that can be utilized by oversampling), non-standard transceiver architecture (like new waveform design at the transmitter \cite{landauJWCN,DZZ21}), and sufficiently high SNR.
Receivers based on oversampling are beyond the scope of this paper since our focus is the effect of low-to-moderate nonlinearity on a standard transceiver architecture.

\subsubsection{Nearest Neighbor Decoding Rule}

The decoder selects a message according to the (scaled) nearest neighbor decoding rule \cite{Lapidoth96,Lapidoth-Shamai02} as
\begin{align}
\label{NND}
\hat{m} = \arg\min\limits_{m\in\mathcal M} \sum\limits_{n=1}^N \left|Y_n - aX_{n}(m)\right|^2.
\end{align}
That is, it selects a message corresponding to the codeword (scaled by a parameter $a$) with the minimum Euclidean distance to the received vector $[Y_1,...,Y_N]$.\footnote{Some more general forms of nearest neighbor decoding rule can be found in, e.g., \cite{Lapidoth96,Lapidoth-Shamai02,WSS04}. A more recent work including a detailed review and some generalizations is \cite{GNND}.}

\subsubsection{Symmetric Quantizer and Uniform Quantizer}\label{II2}

Let the quantizer in (\ref{YqXZ}) be symmetric with $2K$ representation points (levels) $\{\pm y_1,...,\pm y_K\}$ and normalized thresholds $\{0, \pm l_1,...,\pm l_{K-1}\}$, where $y_k\ge 0$ and $l_k> 0$ for $k=1,...,K$, and there are at least two nonzero levels.
Then its resolution (bit-width) is $b= \log_2{2K}$ bits, which typically satisfies $b\in\mathbb Z^+$.
Let $V\in\mathbb R$ be the input to be quantized with standard deviation $\sigma_v$.
The inputs of both quantizers in (\ref{YqXZ}) have the same standard deviation, namely $\sigma_v =\sqrt{(|h|^2\sigma_x^2+\sigma^2)/2}$.
Then for the input $V$, the output of the quantizer is
\begin{align}\label{qs}
q(g V) = y_k \cdot \mathrm{sgn}(V),\; \textrm{if} \; l_{k-1}\sigma_v \leq g|V| < l_k\sigma_v,
\end{align}
where the thresholds satisfy $l_0=0<l_1<...<l_{K-1}<l_K=\infty$.
Apparently, the quantizer output is a nonlinear function of its input, and the thus introduced nonlinearity degrades performance.
In the presence of gain control, we may turn our attention to an equivalent quantization rule for a normalized input $V/\sigma_v$ with adjustable thresholds $\{\ell_k=l_k/g,\; k=1,...,K-1\}$, where $g$ can be adjusted to optimize the performance.
A special case is the uniform quantizer, which has equispaced thresholds
\begin{align}
\ell_k = k\ell, \; k=1,...,K-1,
\end{align}
and mid-rise levels
\begin{align}
y_k = \left(k-\frac{1}{2}\right)\ell, \; k=1,...,K,
\end{align}
where $\ell$ is the step size.
Thus, we define its quantization range or support as $[-K\ell,K\ell]$.
Then the loading factor or support limit of the uniform quantizer is $L=K\ell$.

\subsection{An Achievable Rate Formula from Generalized Mutual Information}

Following the notation of \cite{GLT}, consider a memoryless channel $X\to Y$ with general alphabets $\mathcal X$ and $\mathcal Y$, input probability distribution $P_X(x)$, transition probability $P_{Y|X}(y|x)$, and decoding metric $d(X,Y)$.
The decoder selects a message according to
\begin{align}
\label{GD}
\hat{m} = \arg \min\limits_{m\in\mathcal M} \sum\limits_{n=1}^N d(X_{n}(m),Y_n).
\end{align}
As a lower bound on mismatch capacity, the GMI can be given by its dual expression as \cite{ITFMD,GLT}
\begin{align}\label{GMIg}
I_\text{GMI} = \sup\limits_{s\ge 0} \mathrm E\left[ \log \frac {e^{-s d(X,Y)}} {\mathrm E \left[ e^{-sd (X^\prime,Y) } | Y \right]} \right],
\end{align}
where $(X,Y,X^\prime) \sim P_X(x)P_{Y|X}(y|x)P_X(x^\prime)$.
The GMI gives the maximum rate below which the probability of decoding error, averaged over the i.i.d. random codebook emsemble, converges to zero as the coding block length grows without bound.
For communications with transceiver nonlinearity, if the input distribution $P_X(x)$ is complex Gaussian with variance $\sigma_x^2$ and the decoding rule (\ref{GD}) is specified by (\ref{NND}), then from (\ref{GMIg}) we obtain \cite{Zhang12}
\begin{align}\label{GMIGNND}
I_\text{GMI}=\sup\limits_{s\ge 0}\bigg(&\log(1+s|a|^2\sigma_x^2)-s\mathrm E\left[|Y-aX|^2\right]
+\frac{s\mathrm E\left[|Y|^2\right]}{1+s|a|^2\sigma_x^2}\bigg).
\end{align}
Maximizing the GMI by optimizing the scaling factor $a$ yields the following result [\ref{Zhang12}, Appendix C], which provides a general approach for the achievable rate analysis under transceiver nonlinearity with known transition probability.

{\textbf{Proposition 1}} \cite{Zhang12}:
\emph{For a memoryless scalar channel $X\to Y$ with transition probability $p_{Y|X}(y|x)$ and nearest neighbor decoding rule (\ref{NND}), where $X,Y\in\mathbb C$ and $\mathrm{Var}(X)=\sigma_x^2$, the maximum GMI under i.i.d. complex Gaussian codebook is given by}
\begin{align}\label{ZhangGMI}
I_\text{GMI} = \log\frac{1}{1-\Delta},
\end{align}
\emph{where}
\begin{align}
\Delta = \frac{ \left|\mathrm E\left[X \overline Y\right]\right|^2 } { \sigma_x^2 \mathrm E\left[|Y|^2\right] }.
\end{align}
\emph{To achieve the maximum GMI given in (\ref{ZhangGMI}), the scaling factor in (\ref{NND}) should be set as}
\begin{align}\label{scaling}
a = \alpha := \frac {\mathrm E\left[\overline XY\right]} {\sigma_x^2}.
\end{align}

Besides the proof in \cite{Zhang12} based on direct evaluation and optimization of the dual expression of the GMI, here we provide a sketch of an alternative proof.

\emph{Proof Sketch:}
It has been noted in \cite{Lapidoth-Shamai02} and \cite{Hassibi} that, for an additive \emph{uncorrelated} noise channel $Y=S+U$ (i.e., the noise $U$ satisfies $\mathrm E\left[S\overline U\right]=0$, but is not necessarily independent of the input $S$), if $S\sim\mathcal N_\mathbb C(0,\mathrm E\left[|S|^2\right])$, then
\begin{align}
I(X;Y)\ge \log\left( 1 + \frac {\mathrm E\left[|S|^2\right]} {\mathrm E\left[|U|^2\right]} \right). \label{LapidothGMI}
\end{align}
In \cite{Lapidoth96}, under Gaussian codebook and nearest neighbor decoding rule, the achievability of the RHS of (\ref{LapidothGMI}) and a random coding converse for it are established by a geometric argument,\footnote{The proof in \cite{Lapidoth96} was intended for independent noise. However, as noted in \cite{Lapidoth-Shamai02}, it goes through verbatim for uncorrelated noise.}
where the Gaussian codebook can either be an i.i.d. Gaussian codebook or an equienergy one.
In the former case the rate (\ref{LapidothGMI}) is the GMI.
For the scalar channel $X\to Y$, $X\sim\mathcal N_\mathbb C(0,\sigma_x^2)$, by a Bussgang decomposition, we can always write $Y=\alpha X+D$, where $\alpha$ is given in (\ref{scaling}), and $D=Y-\alpha X$ satisfies $\mathrm E\left[X\overline D\right]=0$.
That is, we eliminate the correlation between the scaled input $\alpha X$ and the corresponding distortion $D$ by a carefully chosen scaling factor.
It is straightforward to show that (\ref{ZhangGMI}) can be obtained by (\ref{LapidothGMI}) when $S=\alpha X$ and $U=D$. $\hfill\blacksquare$

We note that, in Proposition 1, if $\mathrm E[Y]\neq0$, then a pre-processing of the channel output as $Y\to \tilde Y=Y-\mathrm E[Y]$ improves the GMI (\ref{ZhangGMI}).
When $\mathrm E[Y]=0$ has been satisfied (e.g., when $q(\cdot)$ is a symmetric quantizer, the output $Y_n$ in (\ref{YqXZ}) satisfies $\mathrm E[Y_n]=0$), we have $\Delta=|\rho_{XY}|^2$, where
\begin{align}\label{Pearson}
\rho_{XY} = \frac {\mathrm{cov}(X,Y)} {\sigma_x\sigma_y}
\end{align}
is the \emph{Pearson correlation coefficient} between $X$ and $Y$, where $\sigma_y$ is the standard deviation of $Y$.
Also note that the equality in (\ref{LapidothGMI}) holds if and only if $X\perp D$ and $D\sim\mathcal{N}_\mathbb C(0,\sigma^2)$ (e.g., when there is no quantization in (\ref{YqXZ})), so that the channel $X\to Y$ reduces to the AWGN channel, the nearest neighbor decoding rule is optimal, and the GMI equals to the mutual information (also the channel capacity) $\log\big(1+\frac{\sigma_x^2}{\sigma^2}\big)$.

\emph{Remark (Bussgang decomposition and AQNM)}:
Proposition 1 and its proof provide an information-theoretic interpretation to the achievable rate estimation based on Bussgang decomposition $Y=\alpha X+D$.
The estimated rate is typically given in terms of an effective SNR (or signal to noise-and-distortion ratio) as
\begin{align}\label{Bussgang}
R = \log \left( 1 + \mathsf{SNR}_\mathrm e \right),
\end{align}
where
\begin{align}\label{eSNR}
\mathsf{SNR}_\mathrm e =\sup\limits_{a\in\mathbb R}\frac{\mathrm E[|a X|^2]}{\mathrm E[|Y-a X|^2]}=\frac{\mathrm E[|\alpha X|^2]}{\mathrm E[|D|^2]},
\end{align}
and $\alpha$ is called the Bussgang gain.
It is straightforward to verify that $\mathsf{SNR}_\mathrm e = \Delta/(1-\Delta)$, so that the rate (\ref{Bussgang}) is equal to the GMI expression (\ref{ZhangGMI}) in Proposition 1.
In \cite{GNND} this interpretation is extended to a more general setting, which includes a memoryless channel state, imperfect CSIR, and a vector (rather than scalar) channel output (but still assume i.i.d. Gaussian input).
In brief, when the receiver \emph{linearly} combines elements of the channel output into a scalar for nearest neighbor decoding,
the linear combiner that maximizes the GMI is exactly the linear minimum MSE (LMMSE) estimator of the channel input upon observing the channel output, and the maximum GMI is consistent with the achievable rate estimation based on Bussgang decomposition; see [\ref{GNND}, Sec. IV-C] for details.\footnote{In the description in [\ref{GNND}, Sec. IV-C], it was incorrectly stated that the distortion is zero-mean. This may not hold for general nonlinear channels. But this does not affect the conclusion since in the effective SNR (cf. (\ref{eSNR})), the distortion power is defined as the second-order moment, instead of the variance, of distortion. (Discussion with Yuhao Liu of Tsinghua University is acknowledged)}
Since the interpretation relies on i.i.d. Gaussian codebook, the nearest neighbor decoding rule, and a certain receiver processing, the following issues should be noted.
\begin{itemize}
  \item
  To achieve accurate rate estimation, the AQNM should be derived rigorously according to the Bussgang decomposition. In fact, this is not always satisfied in the literature.
  For example, in multiantenna systems, treating the residual noise in the AQNM as i.i.d. Gaussian vector (omitting the correlation among its elements) leads to inaccurate results, as pointed out in, e.g., \cite{Mezghani-Nossek12}, \cite{LLZ} (which includes an example of significant overestimate of the achievable rate), and \cite{SPM-Bussgang}.
  \item
  The GMI does not necessarily approach the mutual information with the same input distribution in general channels.
  Thus, the AQNM should not be expected to provide an accurate estimate of the mutual information, especially at high SNR, as observed in \cite{Mezghani-Nossek12,mo15,mo17}.
  \item
  No information-theoretic foundation has been established for the generalized Bussgang decomposition \cite{SPM-Bussgang} for non-Gaussian input signals.
\end{itemize}

\section{Achievable Rate Under Receiver Quantization: Exact and Asymptotic Results}

Based on Proposition 1, we establish the following result which provides an analytical expression for the achievable rate of the transceiver architecture considered in this paper.

\textbf{Theorem 2}:
\emph{For the channel (\ref{YqXZ}) where $q(\cdot)$ is the symmetric quantizer described in Sec. \ref{II2}, the achievable rate under i.i.d. complex Gaussian codebook and nearest neighbor decoding rule (\ref{NND}) is}
\begin{align}\label{main}
I_\text{GMI} = \log(1 + \mathsf{SNR}) - \log(1 + \gamma\mathsf{SNR}),
\end{align}
\emph{where $\mathsf{SNR}=|h|^2\sigma_x^2/\sigma^2$ is the SNR at the receiver front-end, and $\gamma$ is a parameter determined by the quantizer as
\begin{align}\label{gammaAB}
\gamma = 1 - \frac {\mathcal A^2} {\mathcal B},
\end{align}
in which
\begin{align}\label{As}
\mathcal A = \sqrt{2\pi} \sum\limits_{k=1}^{K} y_k \left( \phi(\ell_{k-1})-\phi(\ell_k) \right),
\end{align}
and}
\begin{align}\label{Bs}
\mathcal B=\pi \sum\limits_{k=1}^{K} y_k^2 \left( Q(\ell_{k-1})-Q(\ell_k) \right).
\end{align}

\begin{IEEEproof}
Applying Proposition 1, it is sufficient to show that
\begin{align}\label{Delta}
\Delta = \frac {\mathsf {SNR}} {1+\mathsf {SNR}} \frac {\mathcal A^2} {\mathcal B},
\end{align}
which can be obtained by showing that
\begin{subequations}
\begin{align}
\mspace{-12mu}\mathrm E\left[X\overline Y\right] &= 2 \mathrm E \left[ \mathrm {Re}(X)\cdot\mathrm {Re}(Y) + \mathrm j \cdot\mathrm {Im}(X)\cdot\mathrm {Re}(Y) \right]\\
&= \frac {2\mathrm{Re}(h)\sigma_x^2} {\sqrt{\pi(|h|^2\sigma_x^2 + \sigma^2)}} \mathcal A
-\frac {2\mathrm j \cdot\mathrm{Im}(h)\sigma_x^2} {\sqrt{\pi(|h|^2\sigma_x^2 + \sigma^2)}} \mathcal A\\
&= \frac{2}{\sqrt{\pi}}\frac {\overline h\sigma_x^2} {\sqrt{|h|^2\sigma_x^2 + \sigma^2}} \mathcal A\label{XYA}
\end{align}
\end{subequations}
and
\begin{align}\label{YB}
\mathrm E\left[|Y|^2\right] = 2 \mathrm E \left[\mathrm {Re}(Y)^2\right] = \frac{4}{\pi}\mathcal B.
\end{align}
The identities (\ref{XYA}) and (\ref{YB}) can be obtained by lengthy but straightforward evaluations of expectations (cf. [\ref{Zhang12}, Appendix D]).
\end{IEEEproof}

\emph{Remark}: In [\ref{Zhang12}, Sec. V], a parallel result of Theorem 2 for \emph{real-valued} channel with symmetric output quantization was given by deriving its effective SNR, while Theorem 2 shows that the GMI in the complex-valued case has exactly the same expression except that the pre-log factor is doubled.

If the impact of quantization is omitted, then the model (\ref{YqXZ}) reduces to a linear Gaussian channel $Y_n=hX_n(m)+Z_n$, and the achievable rate of the considered transceiver architecture is $\log(1+\mathsf{SNR})$, which is also the channel capacity $C$ when the power of the input is $\sigma_x^2$.
The expression of $I_\text{GMI}$ in Theorem 2 shows explicitly that the rate loss due to quantization is
\begin{align}\label{loss}
C-I_\text{GMI}=\log\left(1+\gamma\mathsf{SNR}\right),
\end{align}
which has the same expression as $C$ except for a pre-SNR factor $\gamma$.
For a given SNR, the rate loss is determined by $\gamma$, which reflects the impact of nonlinearity and does not depend on the SNR.
For the GMI (\ref{main}) given in Theorem 2, the effective SNR is
\begin{align}\label{SNRe}
\mathsf {SNR}_\mathrm e = \frac { (1-\gamma) |h|^2\sigma_x^2} {\gamma |h|^2\sigma_x^2 + \sigma^2}
= \frac {(1-\gamma)} {\gamma \mathsf {SNR}+1} \mathsf{SNR}.
\end{align}
From this expression, we see that the effect of quantization is to transfer a fraction $\gamma$ of the power to the denominator of the effective SNR.

From Theorem 2, we immediately obtain the following corollary, which shows that the parameter $\gamma$ dominates the asymptotic behavior of performance, especially the low-SNR slope of the achievable rate and the saturation rate.

\textbf{Corollary 3}: \emph{The achievable rate} $I_\text{GMI}$ \emph{given in (\ref{main}) has the following properties. }
\begin{itemize}
\item \emph{The parameter $\gamma$ in (\ref{main}) satisfies $0<\gamma<1$.
As the quantization becomes increasingly fine, we have $\gamma\to 0$ (from above), $\mathsf {SNR}_\mathrm e \to \mathsf{SNR}$ (from below), and}
\begin{align}\label{lossa}
C - I_\text{GMI} = \mathsf{SNR} \cdot \gamma - \frac {\mathsf{SNR}^2} {2} \gamma^2 + o(\gamma^2)\; \;\textrm {nats/c.u.}
\end{align}
\item
\emph{High- and low-SNR asymptotics: As $\mathsf {SNR}\to \infty$, we have $\mathsf{SNR}_\mathrm e \to \frac{1-\gamma}{\gamma}$ and}
\begin{align}\label{saturate}
I_\text{GMI} = \log \frac{1}{\gamma} - \frac{1-\gamma}{\gamma} \frac{1}{\mathsf{SNR}} + o\left(\frac{1}{\mathsf{SNR}}\right).
\end{align}
\emph{So the saturation rate (i.e. the high-SNR limit of the achievable rate) is given by}
\begin{align}
\bar{I}_\text{GMI} = \log\frac{1}{\gamma}.
\end{align}
\emph{As $\mathsf {SNR}\to 0$, we have}
\begin{align}\label{lowsnr}
I_\textrm{GMI} = (1-\gamma) \mathsf {SNR} &- \frac{1-\gamma^2}{2} \mathsf{SNR}^2
+ o\left(\mathsf{SNR}^2\right) \;\;\textrm {nats/c.u.}
\end{align}
\end{itemize}

\begin{IEEEproof}
The upper bound $\gamma<1$ follows from the fact that $\mathcal A$ and $\mathcal B$ are always positive (guaranteed by the setting of the quantizer in Sec. II-A-2).
The lower bound $\gamma>0$ follows from the fact $\mathcal A^2<\mathcal B$, which can be proved by combining (\ref{XYA}) and (\ref{YB}) and then applying the Cauchy-Schwartz inequality.
The upper bound is consistent with the nonnegativity of the GMI (\ref{main}), while the lower bound can also be seen directly from the fact that the GMI (\ref{main}) cannot exceed the capacity $C$.
For increasingly fine quantization, a formal proof of $\gamma\to 0$ can be found in [\ref{Zhang12}, Sec. V] (although it considered a real-valued channel, the proof goes through verbatim in our setting except for some different notations).
Since $\gamma$ is always nonnegative, we can infer that $\gamma$ tends to zero from above, and thus $\mathsf {SNR}_\mathrm e$ tends to $\mathsf{SNR}$ from below.
The properties (\ref{lossa}), (\ref{saturate}), and (\ref{lowsnr}) can be obtained from the expression (\ref{main}) by standard asymptotic analysis.
\end{IEEEproof}

The following result considers extreme cases of GMI with gain control.

\textbf{Proposition 4}:
\emph{The achievable rate} $I_\textrm{GMI}$ \emph{given in (\ref{main}), as a function of the gain control factor $g$, satisfies}
\begin{align}
\lim\limits_{g\to0} I_\text{GMI}(g) = \lim\limits_{g\to\infty} I_\text{GMI}(g) = I_\textrm{GMI}^{\textrm{1-bit}},
\end{align}
\emph{where}
\begin{align}\label{GMI1bit}
I_\textrm{GMI}^\textrm{1-bit} = \log\frac {1 + \mathsf{SNR}} {1 + \frac {\pi-2} {\pi} \mathsf{SNR}}
\end{align}
\emph{is the GMI under one-bit quantization. }
\begin{IEEEproof}
See Appendix A.
\end{IEEEproof}

This result can be interpreted intuitively as follows.
Note that as $g\to\infty$ we have
\begin{align}
\Pr\left( \min \left\{|V_n^{\rm R}|, |V_n^{\rm I}|\right\} \ge \ell_{K-1}\sigma_v \right) \to 1,
\end{align}
and as $g\to 0$ we have
\begin{align}
\Pr\left( \max \left\{|V_n^{\rm R}|, |V_n^{\rm I}|\right\} < \ell_1\sigma_v \right)\to 1.
\end{align}
Thus, in both limits, the effective resolution of the quantizer reduces to one, so that the achievable rate converges to  $I_\textrm{GMI}^{\textrm{1-bit}}$.

\emph{Remark}:
At high SNR the GMI $I_\textrm{GMI}^{\textrm{1-bit}}$ converges to $\log_2\frac{\pi}{\pi-2}=1.4604$ bits/c.u., while in the low-SNR limit it has a slope of $2/\pi$.\footnote{The channel capacity under one-bit receiver quantization, given by $1-H_2(Q(\mathsf{SNR}))$ \cite{VO79,SDM09}, has the same low-SNR asymptotic behavior, i.e., the well-known ``2-dB loss'' result (see Sec. I-A).
Note that the GMI is derived under Gaussian input and nearest neighbor decoder, and the capacity is achieved by antipodal signaling (the decoding rule is unrestricted).
The difference becomes evident at high-SNR, where the capacity converges to $2$ bits/c.u., while the GMI converges to $1.4604$ bits/c.u.}
We note that $I_\textrm{GMI}^{\textrm{1-bit}}$ is not the infimum of $I_\textrm{GMI}(g)$ in general;
see an example given in Sec. IV-D, Fig. \ref{gmimmsenus}.

Intuitively, a smaller MSE is preferred in quantization.
However, for communications the main performance metric is the achievable rate.
Thus the relationship of MSE minimization and achievable rate maximization is of interest.
Due to symmetry, we may define the normalized MSE of the quantizer $q(\cdot)$ in the channel (\ref{YqXZ}) in terms of the in-phase component of $Y_n$ as
\begin{align}\label{MSE}
\mathsf{mse} := \mathrm E \left[ \left( q(g\cdot V^\mathrm R) - \frac{ V^\mathrm R}{\sigma_v} \right)^2 \right],
\end{align}
where we omit the index of $V^\mathrm R$ and $Y$ since we assume i.i.d. input, and we normalize the input since we use normalized levels of quantization.
The following result shows that the normalized MSE can be expressed by $\mathcal A$ and $\mathcal B$, and is lower bounded by $\gamma$.

\textbf{Proposition 5}:
\begin{align}\label{MSENU}
\mathsf{mse} = 1 - \frac{2}{\pi} \left( \sqrt{2\pi}\mathcal A - \mathcal B \right) \ge \gamma,
\end{align}
\emph{where equality holds if and only if}
\begin{align}
\frac {\mathcal A} {\mathcal B} = \sqrt\frac{2}{\pi}.
\end{align}.
\begin{IEEEproof}
See Appendix B.
\end{IEEEproof}

We define the normalized MMSE as
\begin{align}\label{normmmse}
\mathsf{mmse} := \min\limits_g\mathsf{mse}.
\end{align}
From (\ref{MSENU}), the normalized MSE is minimized only if
\begin{align}
\sqrt{2\pi} \frac {\mathrm d\mathcal A} {\mathrm d g} = \frac {\mathrm d\mathcal B} {\mathrm d g}.
\end{align}
The different expressions of $\mathsf{mse}$ in (\ref{MSENU}) and $\gamma$ in (\ref{gammaAB}) show that, in general, the gain control factor that minimizes the MSE does not necessarily maximize the achievable rate.
However, for uniform quantization, the next section will show that the two optimization problems are consistent; i.e., there is a unique gain control factor (and correspondingly, a unique loading factor) that solves both problems simultaneously.

\section{Optimal Uniform Quantization at the Receiver}

\subsection{Achievable Rate Evaluation under Uniform Quantization}

\textbf{Corollary 6}:
\emph{If $q(\cdot)$ is the uniform quantizer described in Sec. \ref{II2}, then
\begin{align}\label{Au}
\mathcal A = \sqrt{2\pi} \sum\limits_{k=0}^{K-1} \ell \cdot \phi(k\ell) - \frac{\ell}{2},
\end{align}
and}
\begin{align}\label{Bu}
\mathcal B = \pi \sum\limits_{k=0}^{K-1} 2k\ell^2 Q(k\ell) + \frac{1}{8}\pi\ell^2.
\end{align}

\begin{IEEEproof}
See Appendix C.
\end{IEEEproof}

For a given resolution, (\ref{Au}) and (\ref{Bu}) show that $\gamma$ is determined solely by the step size $\ell$ or equivalently by the loading factor $L$, which can be optimized by adjusting the gain-control factor $g$ in (\ref{YqXZ}) according to the channel gain $h$.

\emph{Remark}:
The quantities $\mathcal A$ and $\mathcal B$ are important in our analysis.
For uniform quantization, we note that the summation $\sum_{k=0}^{K-1}\ell\cdot\exp\frac{-k^2\ell^2}{2}$ in $\mathcal A$ is exactly the left Riemann sum of $\exp\frac{-t^2}{2}$ over $[0, K\ell]$ with a regular partition,
and similarly, the summation $\pi\sum_{k=0}^{K-1}2k\ell^2Q(k\ell)$ in $\mathcal B$ is exactly the left Riemann sum of $2\pi tQ(t)$ over $[0, K\ell]$ with a regular partition.
Therefore, for a fixed loading factor $L$ we have the following high-resolution limits:
\begin{subequations}\label{Aa}
\begin{align}
\lim\limits_{K\to\infty} \mathcal A &= \sqrt{2\pi}
\lim\limits_ {K\to\infty} \left( \sum\limits_{k=0}^{K-1} \frac{L}{K} \phi\left( \frac{kL}{K} \right)-\frac{L}{2K} \right)\\
&=\int_{0}^L \exp\frac{-t^2}{2}\mathrm d t\\
&=\sqrt{2\pi} \left( \frac{1}{2}-Q(L) \right),\label{Aa1}
\end{align}
\end{subequations}
\begin{subequations}\label{Ba}
\begin{align}
\mspace{-8mu}\lim\limits_{K\to\infty} \mathcal B
&= 2\pi \lim\limits_{K\to\infty} \left( \sum\limits_{k=0}^{K-1} \frac{L}{K} \frac{kL}{K} Q\left( \frac{kL}{K} \right) + \frac{L^2}{16K^2} \right)\\
&= 2\pi \int_0^L tQ(t)\mathrm d t\\
&= \frac{\pi}{2} - 2\pi\int_L^\infty tQ(t)\mathrm d t.\label{Ba1}
\end{align}
\end{subequations}

\begin{figure}
\centering
\includegraphics[scale=0.7]{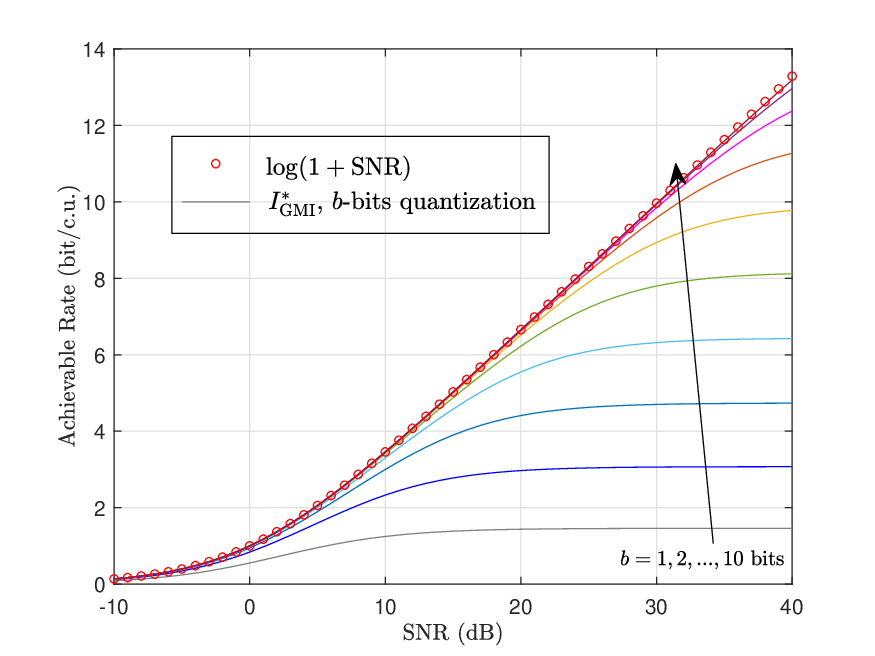}
\caption{Achievable rate with uniform output quantization and optimal loading factor.}\label{Rate}
\end{figure}

\begin{figure}
\centering
\includegraphics[scale=0.7]{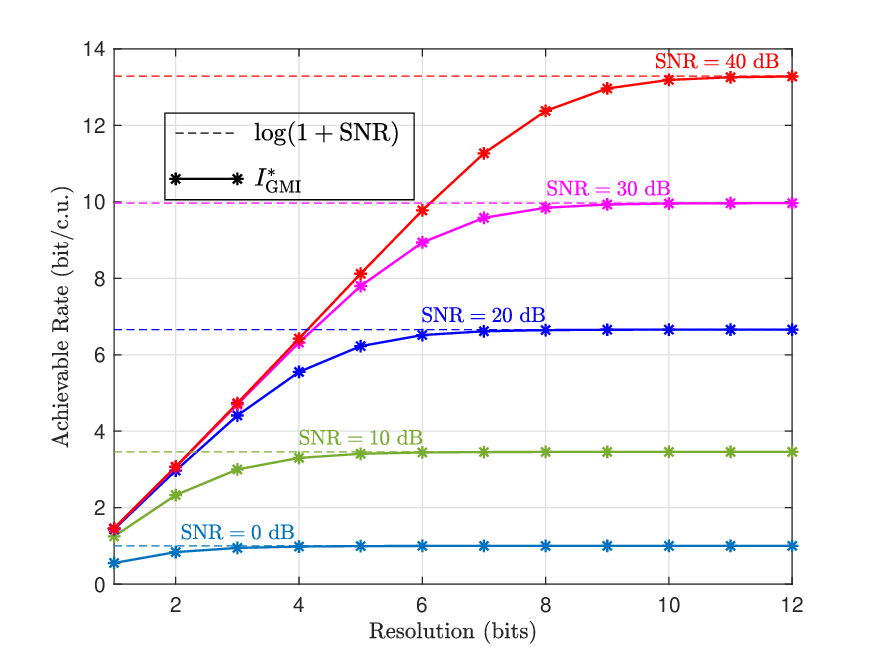}
\caption{Convergence of achievable rate to channel capacity.}\label{Rate1}
\end{figure}

\begin{figure*}
\centering
\subfigure[$b=12$ bits. Simple gain control strategy $\mspace{32mu}$ is always enough.]{\includegraphics[scale=0.4]{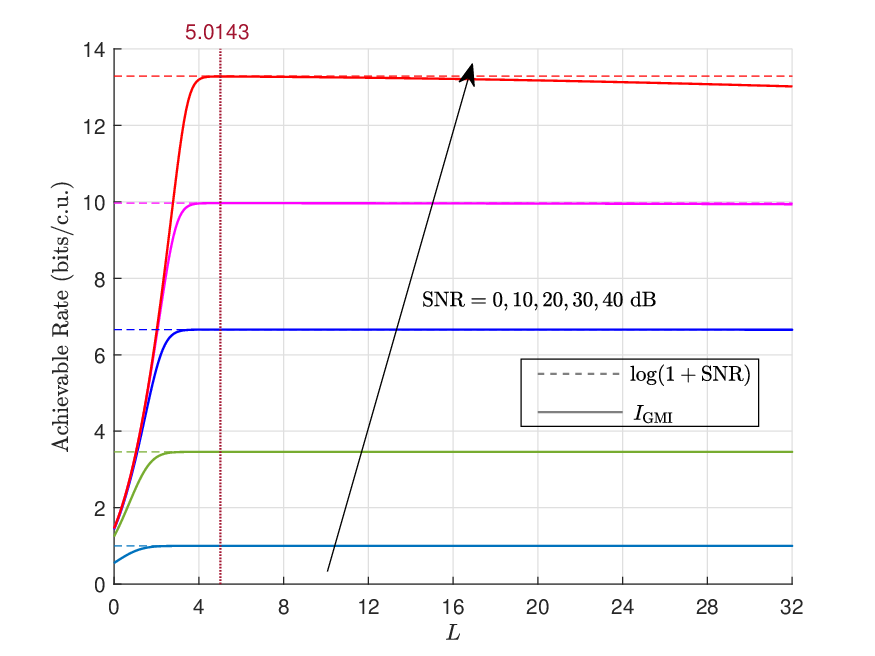}\label{b12}}
\subfigure[$b=10$ bits. Simple gain control strategy $\mspace{32mu}$ is usually enough.]{\includegraphics[scale=0.4]{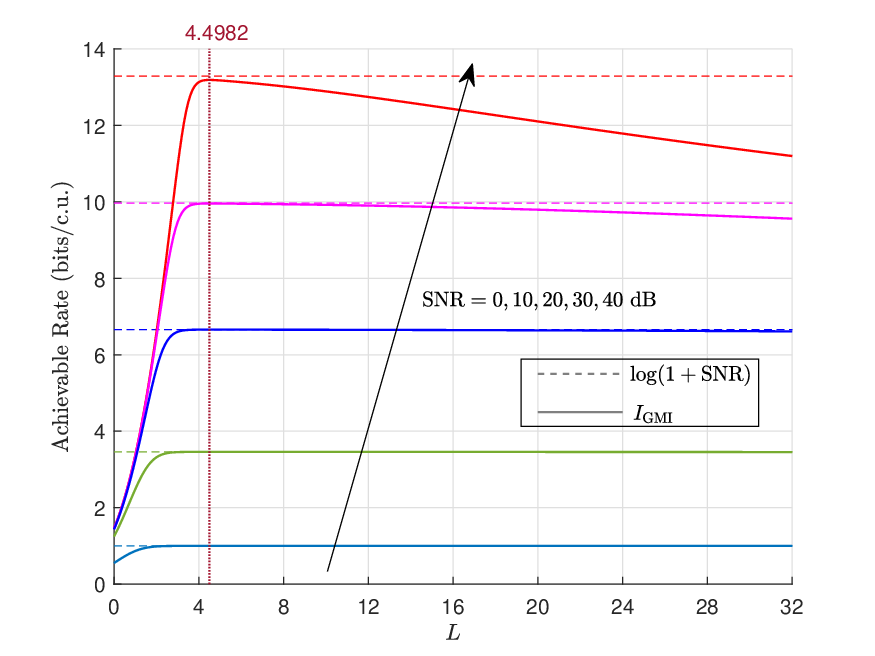}\label{b10}}
\subfigure[$b=8$ bits. Four-sigma rule of thumb is convenient.]{\includegraphics[scale=0.4]{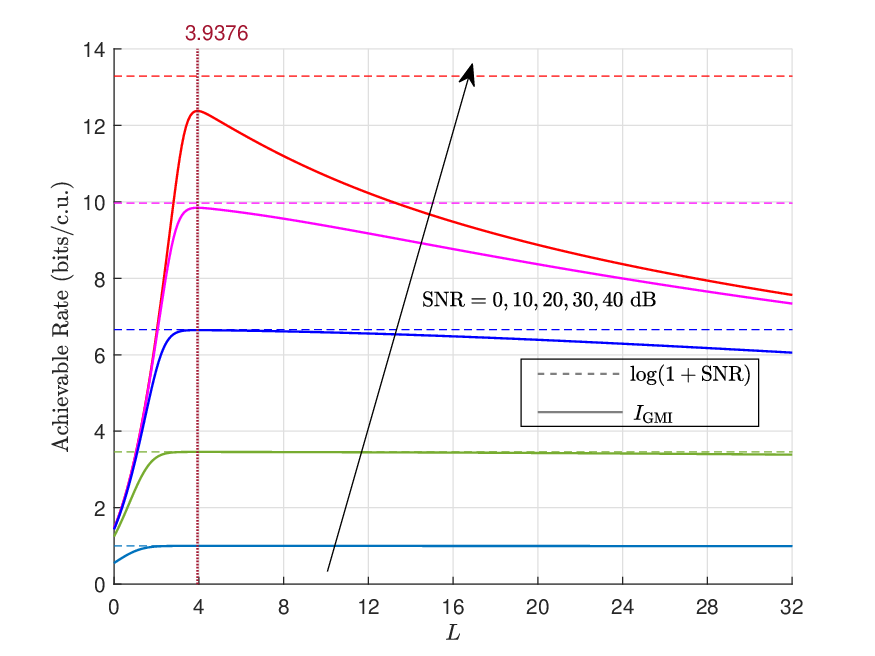}\label{b8}}
\newline
\subfigure[$b=6$ bits. Four-sigma rule of thumb $\mspace{32mu}$ causes small loss at high SNR.]{\includegraphics[scale=0.4]{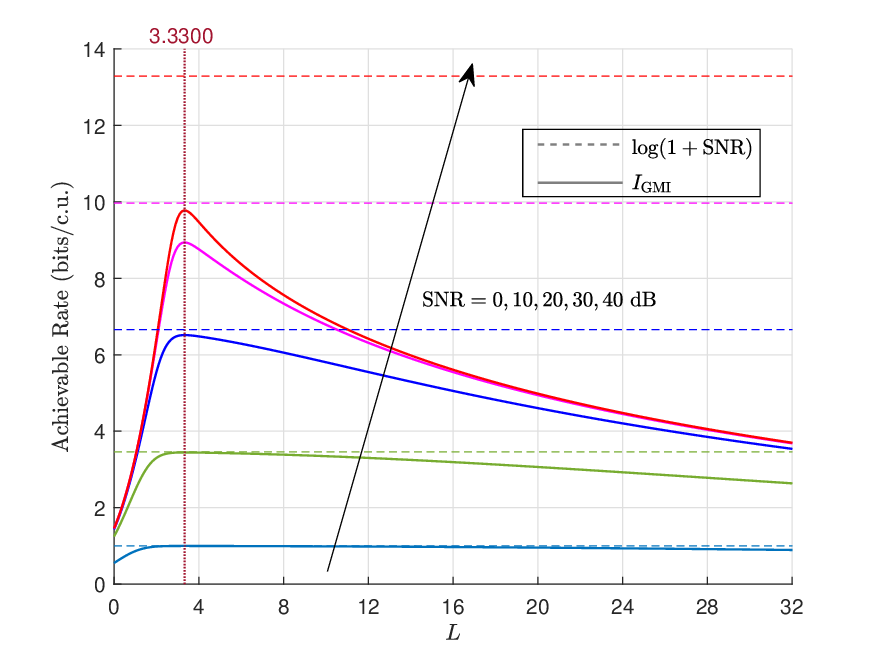}\label{b6}}
\subfigure[$b=4$ bits. Four-sigma rule of thumb $\mspace{32mu}$ causes considerable loss.]{\includegraphics[scale=0.4]{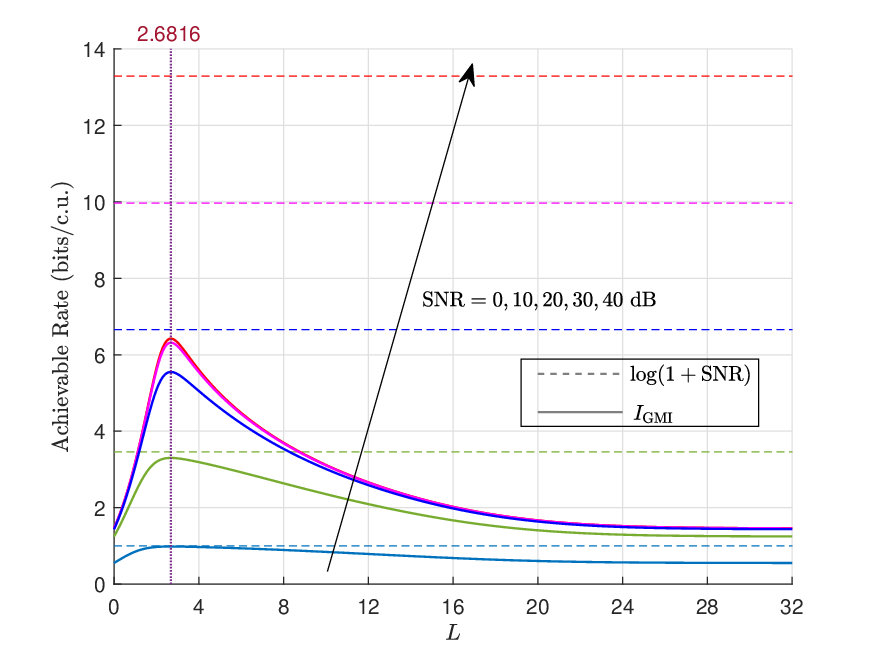}\label{b4}}
\subfigure[$b=2$ bits. Four-sigma rule of thumb causes significant loss.]{\includegraphics[scale=0.4]{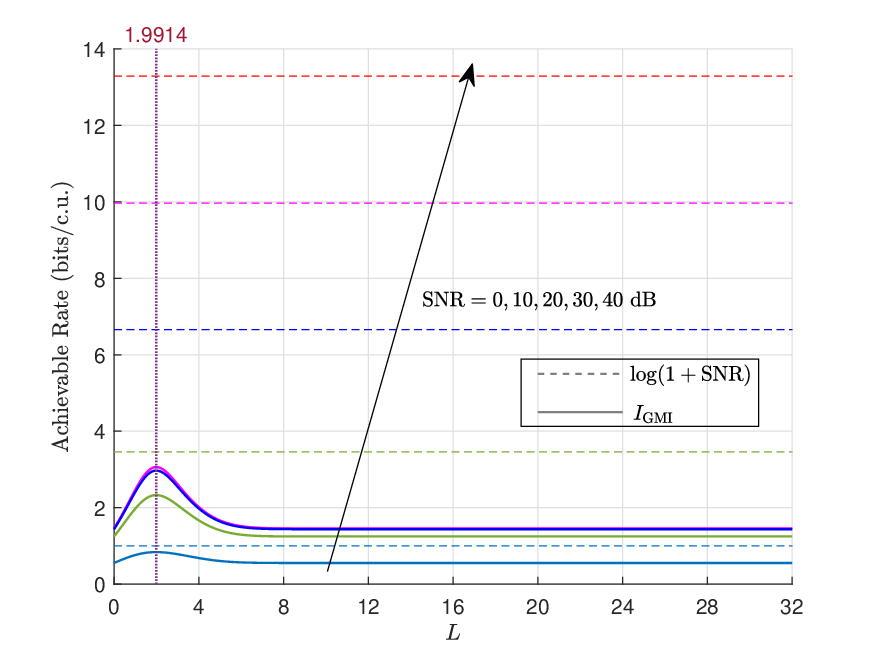}\label{b2}}
\caption{Impact of loading factor on achievable rate: Fixed resolution, varying SNR. Vertical lines and corresponding values show the optimal loading factors.}\label{bits}
\end{figure*}

\begin{figure*}
\centering
\subfigure[$\mathsf{SNR}=-10 \textrm {dB}$.]{\includegraphics[scale=0.4]{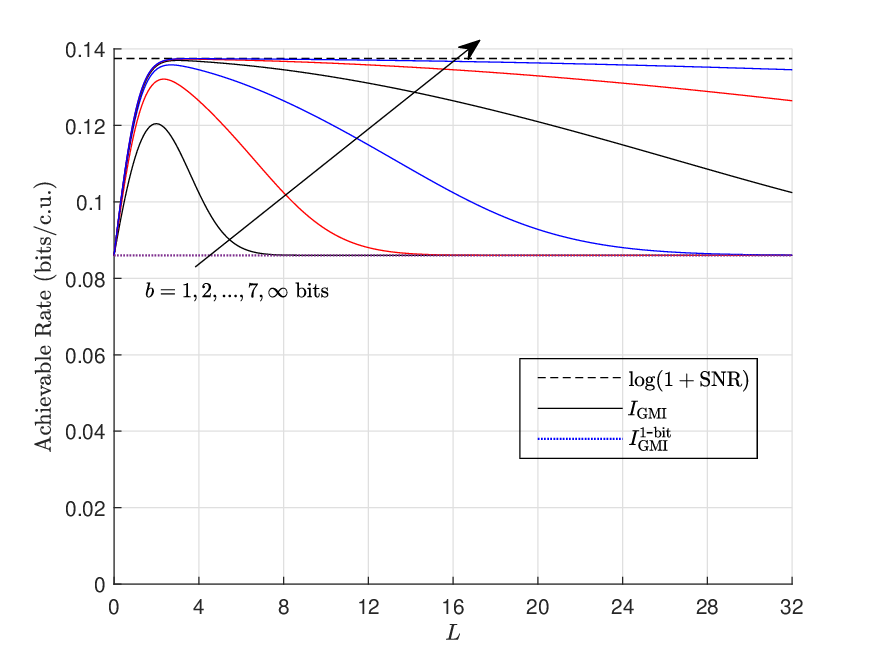}\label{-10dB}}
\subfigure[$\mathsf{SNR}=0 \textrm {dB}$.]{\includegraphics[scale=0.4]{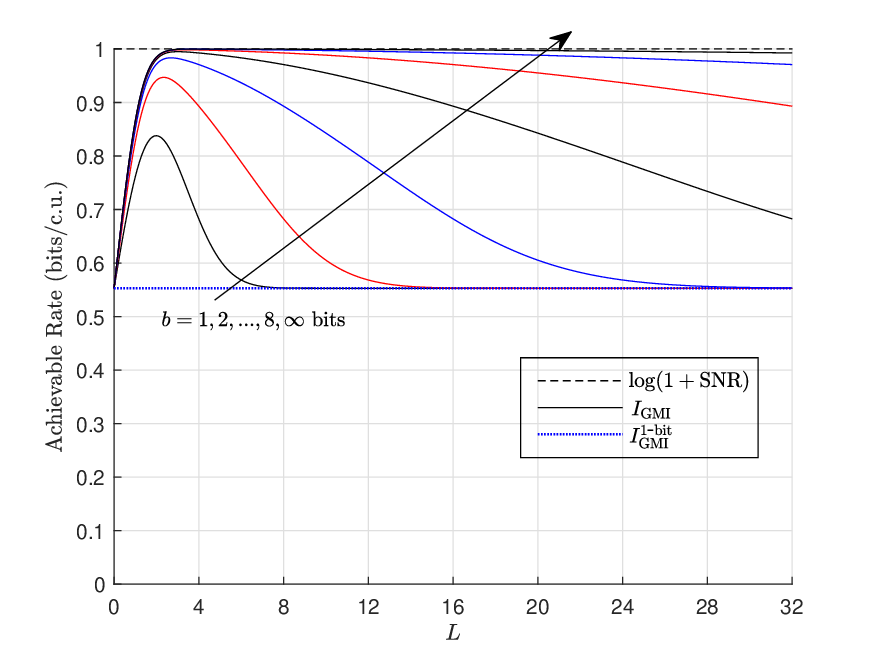}\label{0dB}}
\subfigure[$\mathsf{SNR}=10 \textrm {dB}\approx \mathsf{SNR}_q(2)$.]{\includegraphics[scale=0.4]{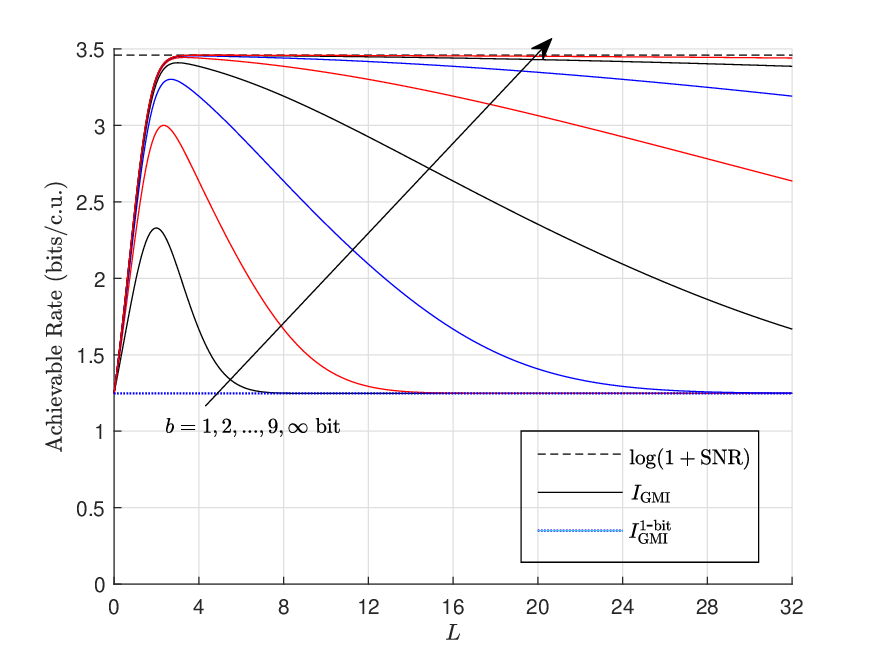}\label{10dB}}
\newline
\subfigure[$\mathsf{SNR}=20 \textrm {dB}\approx \mathsf{SNR}_q(4)$.]{\includegraphics[scale=0.4]{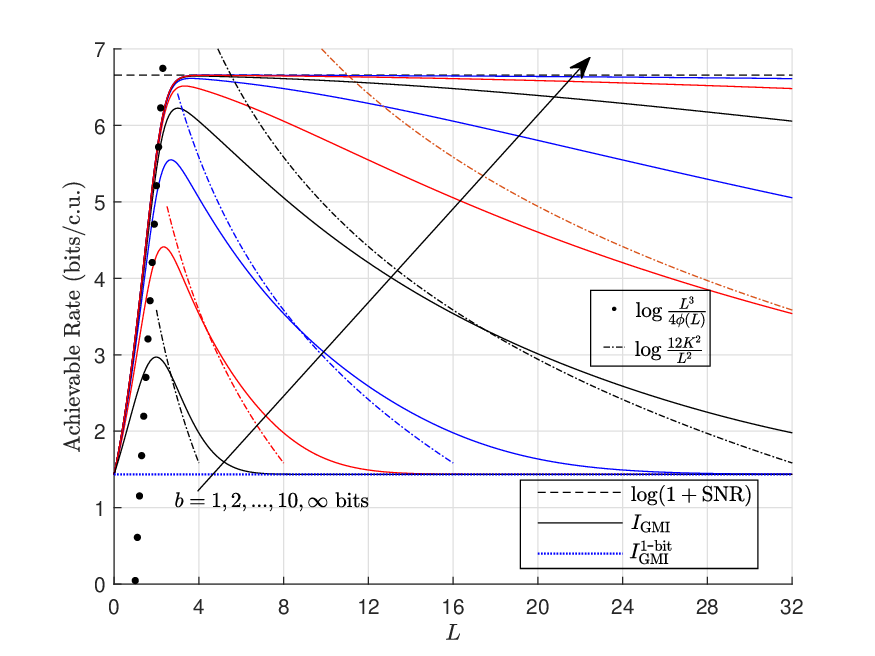}\label{20dB}}
\subfigure[$\mathsf{SNR}=30 \textrm {dB}\approx \mathsf{SNR}_q(6)$.]{\includegraphics[scale=0.4]{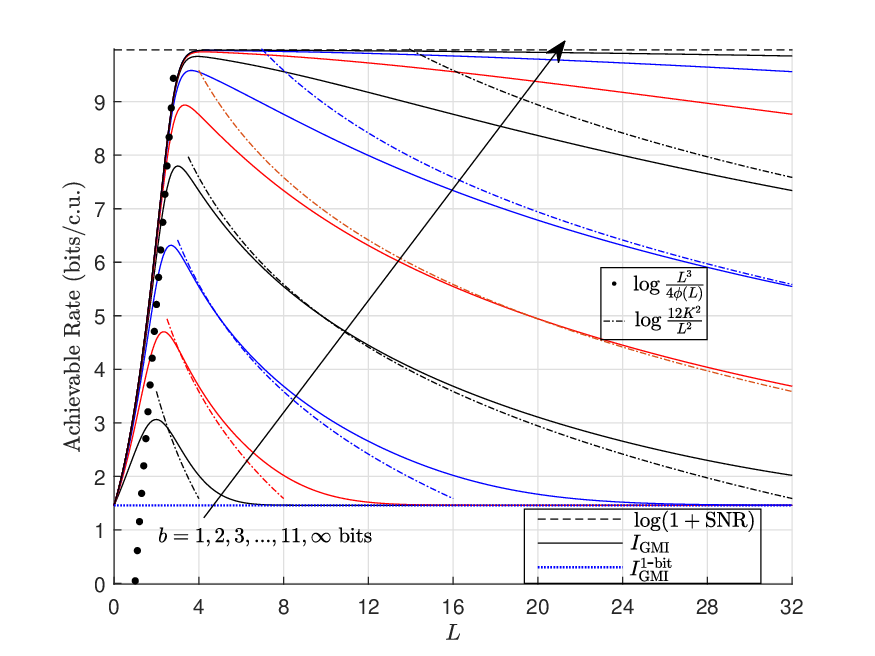}\label{30dB}}
\subfigure[$\mathsf{SNR}=40 \textrm {dB}\approx \mathsf{SNR}_q(8)$.]{\includegraphics[scale=0.4]{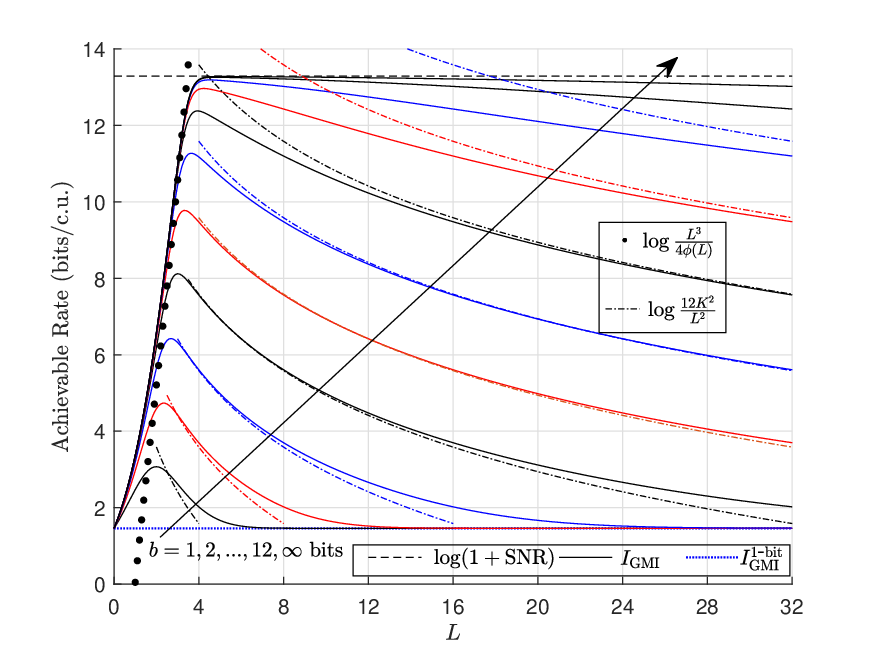}\label{40dB}}
\caption{Impact of loading factor on achievable rate: Fixed SNR, varying resolution (the notation $\mathsf{SNR}_q(b)$ stands for the `SNR' of a uniform quantizer with resolution $b$; see the remark after Corollary 10 for its definition, and see Table I for numerical results.)}\label{snrs}
\end{figure*}

In Fig. \ref{Rate} and Fig. \ref{Rate1} we show numerical evaluations of $I_\textrm{GMI}$ in Theorem 2, which has been maximized over $L>0$ for each resolution and is then denoted by $I_\textrm{GMI}^*$.
The corresponding unique loading factor is denoted by $L^*$.
Its uniqueness will be proved later.
In Fig. \ref{bits} and Fig. \ref{snrs}, we show how the achievable rate $I_\textrm{GMI}$ varies with the loading factor $L$ (some more numerical results and details therein will be interpreted in subsequent sections).
As the resolution increases, the increasing of $L^*$ is clear.
A ``waterfall'' near $L=0$ can be observed in all figures, implying that an underestimate of the optimal loading factor always causes serious rate loss.
To interpret this phenomenon, we write the normalized MSE of uniform quantization as the sum of overload distortion and granular distortion as
\begin{align}\label{mseog}
\mathsf{mse} = \mathsf{mse}_\textrm o + \mathsf{mse}_\textrm g,
\end{align}
where
\begin{subequations}
\begin{align}\label{mseo}
\mathsf{mse}_\textrm o &:= \int_{L}^\infty \left(t-L+\frac{\ell}{2}\right)^2\phi(t)\mathrm d t\\
&=\int_{K\ell}^\infty \left(t-\left(K-\frac{1}{2}\right)\ell\right)^2\phi(t)\mathrm d t,
\end{align}
\end{subequations}
and
\begin{subequations}
\begin{align}\label{mseg}
\mathsf{mse}_\textrm g &:= \sum\limits_{k=1}^{K} \int_{(k-1)\ell}^{k\ell} \left( t - \left( k-\frac{1}{2} \right)\ell \right)^2 \phi(t)\mathrm d t\\
&= \sum\limits_{k=1}^{K} \int_{(k-1)L/K}^{kL/K} \left( t - \frac{k-\frac{1}{2}}{K} L \right)^2 \phi(t)\mathrm d t.
\end{align}
\end{subequations}
Clearly, decreasing $L$ reduces the granular distortion but increases the overload distortion.
When $L<L^*$ the overload distortion increases quickly, thereby causing the waterfall of the achievable rate.


The numerical results in Fig. \ref{bits} and Fig. \ref{snrs} reveal the increasing importance of gain control (realized by an AGC module in practical systems) as the resolution decreases:
1) Under high-resolution output quantization, we only require a rough estimate of the channel gain to guarantee that the loading factor remains \emph{no less than} a predefined threshold, say $4$ (from the four-sigma rule of thumb \cite{Bennett48,GG}), so that we can stay away from the waterfall;
2) Under low-resolution output quantization, such a simple strategy may increase rate loss considerably, but perfect gain control needs accurate channel estimation which is also challenging in this case.

\subsection{Consistency Between Rate Maximization and MSE Minimization and a GMI-MMSE Formula}

From source quantization theory \cite{Hui-Neuhoff01,Na-Neuhoff18} we know that, for Gaussian input, the MSE of the uniform quantizer is a strictly convex function of the step size (or loading factor) and the minimum is located at the \emph{unique} solution of $\frac{\mathrm d \mathsf {MSE}}{\mathrm d\ell}=0$, denoted by $\ell^*$.
In fact, that solution is also the unique step size that maximizes the GMI; that is, for uniform quantization at the receiver there is a consistency between achievable rate maximization and MSE minimization.
This consistency further leads to a GMI-MMSE formula which connects the maximum achievable rate and the MMSE of quantization in a simple and closed form.
We first give the following lemma, and then use it to establish the aforementioned findings in Theorem 8.
Both results exclude the special case $b=1$ because
for symmetric one-bit quantization we have $\mathcal A=\ell/2$, $\mathcal B=\pi\ell^2/8$, and $\gamma \equiv 1-2/\pi$ for $0<\ell<\infty$; i.e., the gain control is unnecessary since it has no impact on the achievable rate.

\textbf{Lemma 7}:
\emph{For uniform quantization at the receiver, for $b > 1$ (i.e., except for the one-bit quantization), we have}
\begin{align}\label{condition}
\frac {\mathrm d I_\text{GMI}} {\mathrm d \ell} = 0 \Leftrightarrow \frac {\mathrm d \mathsf{mse}} {\mathrm d \ell} = 0,
\end{align}
\emph{and both of them are equivalent to}
\begin{align}\label{AB}
\frac{\mathcal A}{\mathcal B} = \sqrt {\frac{2}{\pi}}.
\end{align}

\begin{IEEEproof}
See Appendix D.
\end{IEEEproof}

\textbf{Theorem 8}:
\emph{For the channel (\ref{YqXZ}) where $q(\cdot)$ is the uniform quantizer described in Sec. \ref{II2} with resolution $b>1$, the input employs i.i.d. complex Gaussian codebook, and the receiver employs the nearest neighbor decoding rule (\ref{NND}), we have the following properties.}
\begin{itemize}
\item
\emph{The loading factor $L^*$ that maximizes the GMI (\ref{main}) is unique, and it is also the unique loading factor that minimizes the MSE (\ref{MSE}), i.e.,}
\begin{align}\label{equiv}
\arg \max \limits_L I_\textrm{GMI}(L) =\arg \min \limits_L {\mathsf{mse}}(L).
\end{align}

\item
\emph{The minimum of $\gamma$ as a function of $L$, namely $\gamma(L^*)$,  is exactly the normalized MMSE (\ref{normmmse}) and satisfies}
\begin{align}\label{gammammse}
0 < \mathsf{mmse} = \gamma(L^*) \leq \gamma,
\end{align}
\emph{and the maximum GMI can be written as}
\begin{equation}\label{GMIMMSE}
I_\textrm{GMI}^* = \log (1 + \mathsf{SNR}) - \log \left( 1 + \mathsf{mmse}\cdot\mathsf{SNR} \right).
\end{equation}
\end{itemize}

\begin{IEEEproof}
First, in \cite{Hui-Neuhoff01}, it has been proved that the MSE is minimized if and only if the loading factor is set to be the unique solution of $\frac{\mathrm d \mathsf{mse}}{\mathrm d L}=0$, denoted by $L^*$.
Specifically, the uniqueness of $L^*$ has been confirmed in [\ref{Hui-Neuhoff01}, Sec. V-A] by showing that the second derivative of the MSE is strictly positive.
Lemma 7 implies that the loading factor that satisfies $\frac{\mathrm d I_\text{GMI}}{\mathrm d L}=\frac{\mathrm d I_\text{GMI}}{K\mathrm d \ell}=0$ is also unique.
Noting that the achievable rate $I_\text{GMI}$ is a continuous and differentiable function of $L$, from Proposition 4 we can infer that $L^*$ also maximizes $I_\text{GMI}(L)$ (it is not difficult to exclude the other possible case that $L^*$ minimizes $I_\text{GMI}(L)$).
Then the first part of Theorem 8 is proved.

According to Lemma 7, if $L=L^*$ then (\ref{AB}) holds.
Combining this fact with (\ref{MSENU}) and (\ref{gammaAB}), we have
\begin{align}
\gamma(L^*) = 1 - \frac{\pi}{2} \mathcal B(\ell^*) = \mathsf{mmse},
\end{align}
thereby completing the proof of the second part of Theorem 8.
\end{IEEEproof}

\textbf{Corollary 9}:
\emph{For uniform receiver quantization with resolution $b\ge 1$, the achievable rate given in (\ref{main}), as a function of the loading factor $L$, satisfies}
\begin{align}\label{infell}
\inf\limits_L I_\text{GMI}(L)=
\lim\limits_{L\to0} I_\text{GMI}(L) = \lim\limits_{L\to\infty} I_\text{GMI}(L) = I_\textrm{GMI}^{\textrm{1-bit}},
\end{align}
\emph{where} $I_\textrm{GMI}^{\textrm{1-bit}}$ \emph{is given in (\ref{GMI1bit}).}
\emph{Correspondingly, the parameter $\gamma$ satisfies}
\begin{align}\label{gammarange}
0 < \gamma \leq 1 - \frac{2}{\pi},
\end{align}
\emph{where the equality holds when $b=1$.}

\begin{IEEEproof}
The second and third equality in (\ref{infell}) follows directly from Proposition 4.
Combining these two extreme cases with the uniqueness of the solution of $\frac{\mathrm d I_\text{GMI}}{\mathrm d L}=0$ when $b>1$ (see the proof of Theorem 8), we can infer that, for $0<L<\infty$, the infimum of $I_\text{GMI}(L)$ is $I_\textrm{GMI}^{\textrm{1-bit}}$.
Correspondingly, $\gamma$ is upper bounded by its one-bit case, namely $1 - \frac{2}{\pi}$.
\end{IEEEproof}

For the case $b=1$, from Proposition 5 it can be shown that, when $\ell = 4/\sqrt{2\pi} = 1.5958$, the normalized MSE achieves its minimum as $\mathsf{mmse}=1-2/\pi=\gamma$, implying that the GMI in this case can also be written as (\ref{GMIMMSE}).

\emph{Remark}:
From Theorem 2, the rate loss due to uniform quantization with a loading factor $L$ is given by $\log(1+\gamma(L)\mathsf{SNR})$.
For uniform quantization, Theorem 8 characterizes the minimum of rate loss when $L=L^*$.
Thus we should distinguish two parts of the total rate loss as follows.
\begin{itemize}
\item
Irreducible loss:\footnote{We note that the loss (\ref{IRL}) is irreducible in the sense of GMI, and it is not necessarily irreducible in general since the GMI is only a lower bound on the mismatch capacity of the channel (\ref{YqXZ}).}
the unavoidable part for given resolution and $\mathsf {SNR}$, given by
\begin{align}\label{IRL}
C - I_\textrm{GMI}^* = \log( 1 + \mathsf{mmse}\cdot\mathsf{SNR} ).
\end{align}
Numerical results in Fig. \ref{Rate} and Fig. \ref{Rate1} consider only irreducible loss.

\item
Loading loss: the remaining part, given by
\begin{align}\label{LL}
I_\textrm{GMI}^* - I_\textrm{GMI}(L) = \log \frac {1 + \gamma(L)\mathsf{SNR}} {1 + \mathsf{mmse}\cdot\mathsf{SNR}},
\end{align}
which is due to a suboptimal loading factor and can be reduced by improving the accuracy of gain control.
Numerical results in Fig. \ref{bits} and Fig. \ref{snrs} show the importance of reducing loading loss.
\end{itemize}

The equivalence (\ref{equiv}) established in Theorem 8 enables us to utilize existing results on MSE-optimal uniform quantization.
In particular, combining (\ref{equiv}) and results in \cite{BG80,Hui-Neuhoff01}, we have the following corollary with respect to GMI, which implies that the optimal loading factor $L^*$ grows with the resolution $b$ like $2\sqrt{b\ln2}$.

\textbf{Corollary 10}: \emph{In the channel (\ref{YqXZ}) under i.i.d. complex Gaussian codebook and nearest neighbor decoding rule (\ref{NND}), the optimal step size $\ell^*$ that maximizes the GMI (\ref{main}) satisfies}
\begin{align}\label{llimit}
\lim\limits_{K\to\infty} K\ell^* = \infty, \;\lim \limits_ {K\to\infty} \ell^* = 0.
\end{align}
\emph{The optimal loading factor $L^*=K\ell^*$ grows monotonically with $K$ and satisfies}
\begin{align}
\label{asy}
\lim \limits_{K\to\infty} \frac {L^\ast} { 2\sqrt{\ln (2K)} } = 1.
\end{align}
\emph{The corresponding MMSE consisting of the granular distortion} $\mathsf{mmse}_\textrm g$ \emph{and the overload distortion} $\mathsf{mmse}_\textrm o$
\emph{satisfies }
\begin{align}\label{mmseell}
\lim\limits_{K\to\infty} \frac {\mathsf{mmse}} {{\ell^*}^2/12} = \lim\limits_{K\to\infty} \frac {\mathsf{mmse}_\textrm g} {{\ell^*}^2/12} = 1;
\end{align}
\emph{that is, the granular distortion dominates the MMSE as the resolution increases: }
\begin{align}
\lim\limits_{K\to\infty} \frac {\mathsf{mmse}_\textrm o} {\mathsf{mmse}_\textrm g} = 0.
\end{align}

For (\ref{llimit}) and (\ref{asy}), we also provide new proofs, respectively, by asymptotic results on the achievable rate; see Sec. VI-A.

\emph{Remark}:
In quantization theory, the ``SNR'' for a quantizer is often defined as the ratio between the variance of the quantizer input and the MSE\cite{Gray-Neuhoff98,GG}.
For the optimal uniform quantizer we denote
\begin{align}
\mathsf{SNR}_q = \frac {1} {\mathsf{mmse}}.
\end{align}
Theorem 8 implies that the maximum of the saturation rate given in (\ref{saturate}) is
\begin{align}\label{satrate}
\bar I_\textrm{GMI}^* = \log \frac {1} {\mathsf{mmse}} = \log \mathsf{SNR}_q,
\end{align}
and the corresponding effective SNR is given by
\begin{align}
\sup_{\mathsf{SNR}>0} {\mathsf{SNR}}_\textrm{e} = \lim\limits_{\mathsf{SNR}\to\infty} {\mathsf{SNR}}_\textrm{e} = \mathsf{SNR}_q - 1.
\end{align}
For finite SNR we have
\begin{align}
\mathsf{SNR}_\textrm{e} = \frac {1-\mathsf{mmse}} {\mathsf{mmse}\cdot\mathsf {SNR}+1} \mathsf{SNR}
=\frac {(1-\mathsf{mmse}) |h|^2\sigma_x^2} {\mathsf{mmse}\cdot|h|^2\sigma_x^2 + \sigma^2}.
\end{align}

\begin{table*}
\caption{Numerical Results of Optimal Parameters and Performance Metrics for Uniform Quantization}
\centering
\setlength{\extrarowheight}{6pt}
\begin{tabular}{c|c|c|c|c|c|c|c|c}
\hline $b$ &$2K$ &$L^*$ &$\ell^*$ &$\gamma(L^*)=\mathsf{mmse}$ &$\ln(2K)/3K^2$&$\mathsf{SNR}_q$ (dB) &$\bar I^*_\textrm{GMI}$(bits/c.u.)&$\hat {\bar I}_\textrm{GMI}$(bits/c.u.)\\
\hline $1$ &2&   1.5958&1.5958&0.3634&0.2310&4.40&1.4604&2.11\\
\hline $2$ &4&   1.9914&0.9957&0.1188&0.1155&9.25&3.0728&3.11\\
\hline $3$ &8&   2.3441&0.5860&3.7440$\times10^{-2}$&4.3322$\times10^{-2}$&14.27&4.7393&4.53\\
\hline $4$ &16&  2.6816&0.3352&1.1543$\times10^{-2}$&1.4441$\times10^{-2}$&19.38&6.4369&6.11\\
\hline $5$ &32&  3.0102&0.1881&3.4952$\times10^{-3}$&4.5127$\times10^{-3}$&24.57&8.1604&7.79\\
\hline $6$ &64&  3.3300&0.1041&1.0400$\times10^{-3}$&1.3538$\times10^{-3}$&29.83&9.9091&9.53\\
\hline $7$ &128& 3.6395&5.6868$\times10^{-2}$&3.0433$\times10^{-4}$&3.9486$\times10^{-4}$&35.17&11.6821&11.31\\
\hline $8$ &256& 3.9376&3.0762$\times10^{-2}$&8.7686$\times10^{-5}$&1.1282$\times10^{-4}$&40.57&13.4773&13.11\\
\hline $9$ &512& 4.2237&1.6499$\times10^{-2}$&2.4919$\times10^{-5}$&3.1730$\times10^{-5}$&46.03&15.2924&14.94\\
\hline $10$&1024&4.4982&8.7855$\times10^{-3}$&6.9970$\times10^{-6}$&8.8138$\times10^{-6}$&51.55&17.1248&16.79\\
\hline $11$&2048&4.7614&4.6498$\times10^{-3}$&1.9444$\times10^{-6}$&2.4238$\times10^{-6}$&57.11&18.9722&18.65\\
\hline $12$&4096&5.0143&2.4484$\times10^{-3}$&5.3554$\times10^{-7}$&6.6104$\times10^{-7}$&62.71&20.8325&20.53\\
\hline
\end{tabular}
\end{table*}

In Table I,\footnote{Some numerical results therein have been given in [\ref{Hui-Neuhoff01}, Table II], in which an error occurred in actual SNR computation for Gaussian source, $b=8$.} we show numerical results of the optimal uniform quantization under our transceiver architecture (an approximation of $\mathsf{mmse}$ given by $\ln(2K)/3K^2$, and an approximation of $\bar I_\textrm{GMI}$, denoted by $\hat I_\textrm{GMI}$, will be introduced in Sec. VI-B).

As the SNR decreases, numerical results in Fig. \ref{snrs} show that the supremum and infimum of $I_\textrm{GMI}$ have a decreasing ratio which converges to $\pi/2$, coinciding with the ``2 dB loss'' result for hard-decision decoding.
Thus, a major part of the low-SNR capacity can always be utilized.
At moderate-to-high SNR (Figs. (\ref{10dB}-\ref{40dB})), we may roughly separate different scenarios of receiver quantization as follows, which show that channel estimation and AGC design become more challenging as the resolution decreases.

\begin{itemize}
\item
SNR-limited (high resolution) scenario:
$b> 2\log_{10}\mathsf{SNR}$ $+b_1$ bits and $\mathsf {SNR}_q\gg \mathsf {SNR}$, where $b_1$ can be $2$\textasciitilde$3$.
The irreducible loss (\ref{IRL}) is negligible, and significant rate improvement can be obtained by increasing the SNR.
In fact, we have
\begin{align}
\frac {\textrm d(C-I^*_\textrm{GMI})} {\textrm d\mathsf{SNR}} = \mathsf{mmse} + o(\mathsf{mmse}^2),
\end{align}
which suggests that the irreducible loss increases slowly with the SNR.
A large overestimate of the optimal gain control factor does not cause significant loading loss, thereby allowing the simple gain control strategy described at the end of Sec. IV-A.
In this scenario the impact of receiver quantization is not important.
There is still room to reduce the resolution if accurate gain control is possible.

\item
Resolution-limited (low resolution) scenario:
$b\leq 2\log_{10}\mathsf{SNR}$ bits and $\mathsf {SNR}_q<\mathsf {SNR}$.
The achievable rate is seriously limited due to large irreducible loss, while accurate gain control is required to avoid large loading loss.
Significant rate improvement can be obtained by increasing the resolution.
\item Moderate resolution scenario (the remaining cases):
The resolution is enough to maintain a small irreducible loss, but an overestimate of the optimal gain control factor may cause considerable loading loss.
\end{itemize}

We note that the consistency between rate maximization and MSE minimization does not hold in general; see discussions on nonuniform quantization in Sec. VII.

\subsection{Geometry of Optimal Uniform Quantization at the Receiver}
\begin{figure}
\centering
{\includegraphics[scale=0.32]{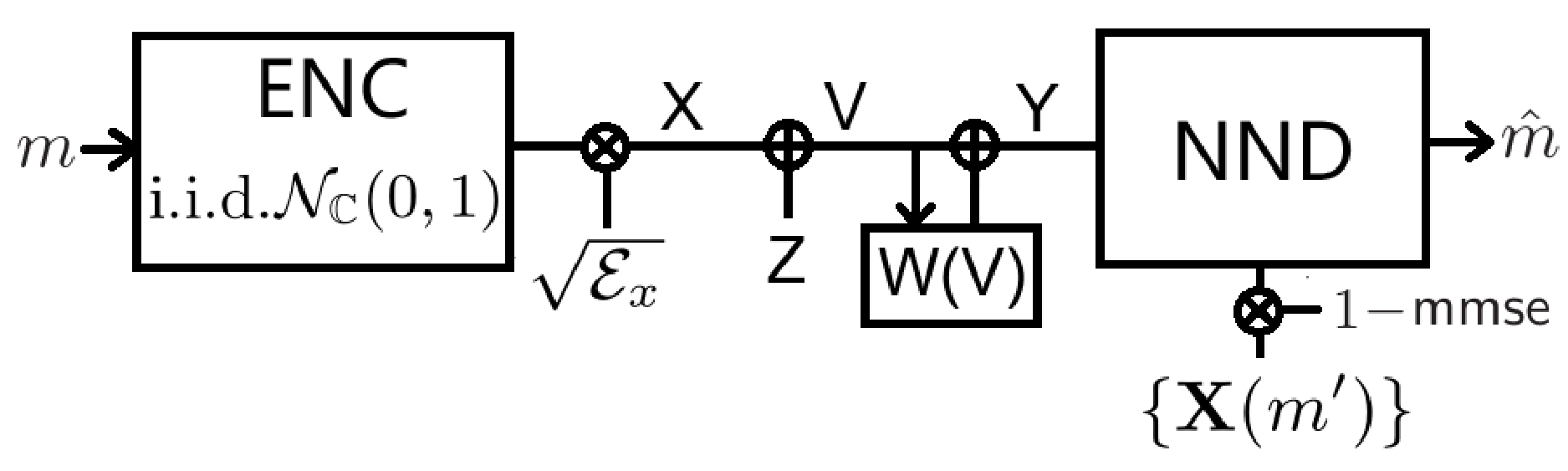}}
\caption{A simplified equivalent model of the transceiver architecture when $L=L^*$.}\label{Archie}
\centering
\includegraphics[width=3.52in,height=3.96in]{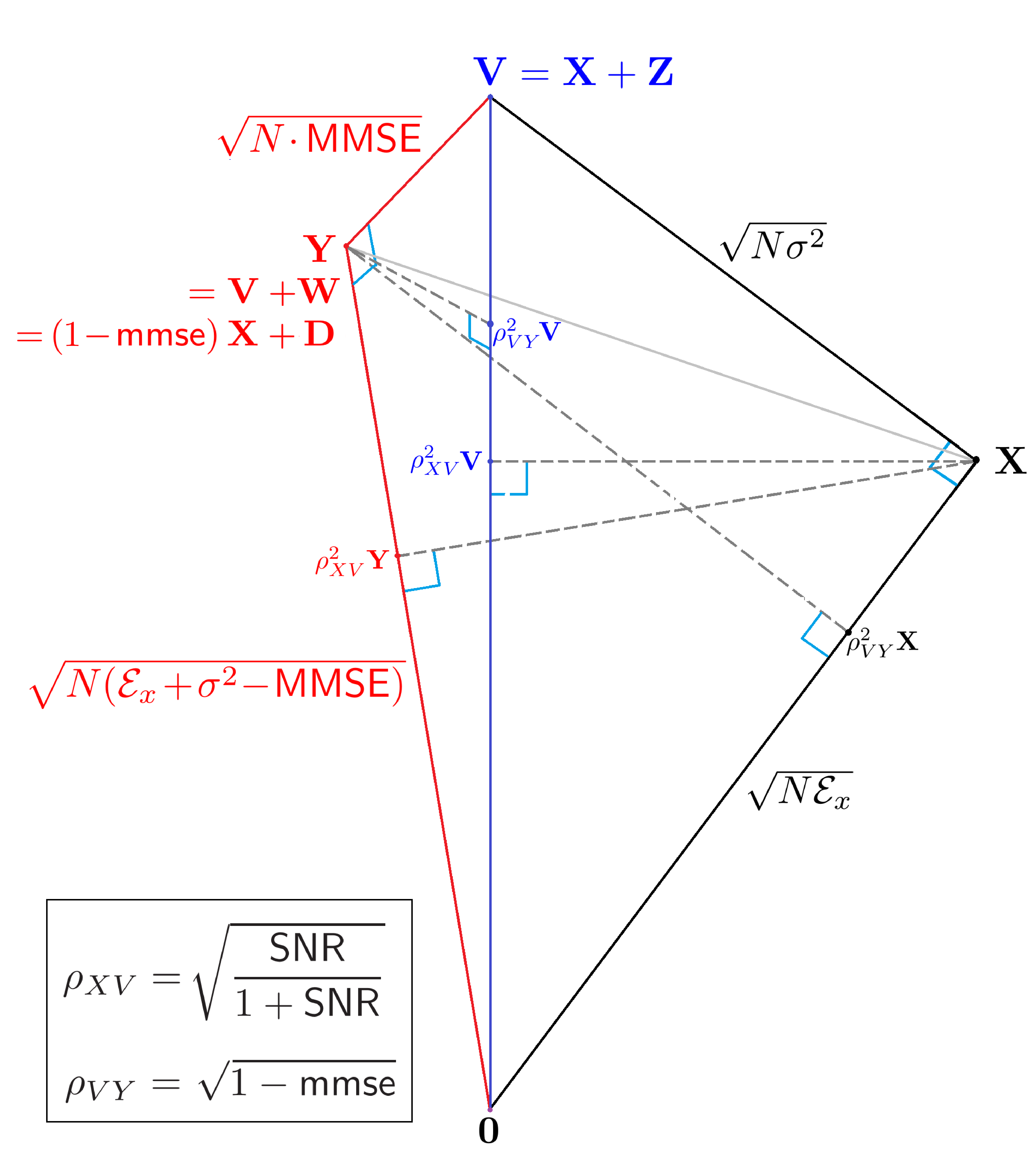}
\caption{Geometry of optimal uniform quantization at the receiver.}\label{Geometry}
\end{figure}

We have shown that, in the standard transceiver architecture shown in Fig. \ref{Archi}, if the gain control factor $g$ is set appropriately so that the loading factor of the quantizer $q(\cdot)$ is equal to the optimal value $L^*$, then the achievable rate attains its maximum given by the GMI-MMSE formula (\ref{GMIMMSE}).
In fact, a geometrical interpretation for such optimal uniform quantization can be established.
For simplification we first introduce an equivalent model shown in Fig. \ref{Archie}, where we let $\mathsf X=hX$, $\mathcal E_x=|h|^2\sigma_x^2$, $\mathsf{Y}=\sigma_v Y=Y\cdot\sqrt{(\mathcal E_x+\sigma^2)/2}$, $\mathsf V=V^{\mathrm R}+\mathrm j V^\mathrm I$, $\mathsf Z=Z$, and $\mathsf W=\mathsf Y-\mathsf V$ (we write $\mathsf W(\mathsf V)$ in  Fig. \ref{Archie} to emphasize that $\mathsf W$ is a function of $\mathsf V$), so that
\begin{align}\label{YXZW}
\mathsf Y = \mathsf X + \mathsf Z + \mathsf W = \mathsf V + \mathsf W,
\end{align}
where the quantization error $\mathsf W$ satisfies
\begin{align}
\mspace{-8mu}\mathrm E\left[|\mathsf W|^2\right] = \mathsf{MMSE} = \mathsf{mmse}\cdot\mathrm E\left[|\mathsf V|^2\right] = 2\cdot \mathsf{mmse} \cdot \sigma_v^2.
\end{align}
Thus, the high-resolution limit of $\mathsf Y$ is $\mathsf V$, and the high-SNR limit of $\mathsf V$ is $\mathsf X$.

Now we are ready to illustrate the geometry of the optimal uniform quantization at the receiver in the $N$-dimensional Euclidean space, as shown in Fig. \ref{Geometry}.
We use boldface letters to denote codewords or signal vectors in the equivalent model (\ref{YXZW}), e.g., $\mathbf X =[\mathsf X_1,...,\mathsf X_N]$.
Since i.i.d. codebook is considered, the vector $\mathbf X$ and other vectors in Fig. \ref{Geometry} are all i.i.d. random vectors.
Thus, we have $\mathrm E\left[\|\mathbf X\|^2\right]=N\mathcal E_x$, and the empirical average power of the input codeword converges in probability to $\mathcal E_x$, i.e.,
\begin{align}
\lim_{N\to\infty} \left( \left| \frac{1}{N} \|\mathbf X\|^2 - \mathcal E_x \right| > \epsilon \right)=0, \;\forall \epsilon>0.
\end{align}
We thus briefly say that the length of $\mathbf X$ (in asymptotic sense) is $\sqrt{\mathrm E\left[\|\mathbf X\|^2\right]}=\sqrt{N\mathcal E_x}$.
Similarly, the length of $\mathbf Z$ is $\sqrt{N\mathcal \sigma^2}$.
The geometry in Fig. \ref{Geometry} includes two Pythagorean relations as follows.

\begin{itemize}
\item
For the additive noise channel $\mathsf V=\mathsf X+\mathsf Z$, we have $\mathrm E\left[\mathsf X\overline{\mathsf Z}\right]=0$, implying
\begin{align}\label{VXZ}
\mathrm E \left[\|\mathbf V\|^2\right] = \mathrm E \left[\|\mathbf X\|^2\right] + \mathrm E \left[\|\mathbf Z\|^2\right].
\end{align}

\item
For the quantization channel $\mathsf Y=\mathsf V+\mathsf W$, it has been shown in \cite{BG80} that the MMSE uniform quantization satisfies $\mathrm E\left[\mathsf Y\overline{\mathsf W}\right]=0$; i.e.,
the error $\mathsf W$ is uncorrelated with the \emph{output} of the quantizer, yielding
\begin{align}\label{VYe}
\mathrm E \left[\|\mathbf V\|^2\right] = \mathrm E \left[\|\mathbf Y\|^2\right] + \mathrm E \left[\|\mathbf W\|^2\right].
\end{align}
\end{itemize}

Thus, the lengths of $\mathbf V$, $\mathbf W$, and $\mathbf Y$ are $\sqrt{N(\mathcal E_x + \mathcal \sigma^2)}$, $\sqrt{N\cdot\mathsf{MMSE}}$, and $\sqrt{N(\mathcal E_x + \mathcal \sigma^2-\mathsf{MMSE})}$, respectively.
We then have two right triangles in Fig. \ref{Geometry}:
the \emph{quantization triangle} $\mathbf{0VY}$\footnote{Note that the angle between $\mathbf Y$ and $\mathbf V-\mathbf Y$ is a right angle. Since Fig. \ref{Geometry} is a one-dimensional illustration of the geometry in the $N$-dimensional space, we cannot make every right angle therein look like a right angle.}, which determines the saturation rate $\bar I^*_\textrm{GMI}$ in (\ref{satrate}), and the \emph{noise triangle} $\mathbf{0XV}$, which determines the channel capacity $C$, i.e., the limit of the achievable rate under fine quantization.
These two triangles jointly determine the triangle $\mathbf{0XY}$, and consequently determine the achievable rate $I^*_\textrm{GMI}$.

We note that $\mathbf{YVX}$ and $\mathbf{0XY}$ are not right triangles.
In fact, for finite $\mathsf {SNR}$ and finite resolution, $\mathrm E\left[|\mathsf Y-\mathsf X|^2\right]$ is strictly smaller than $\mathrm E\left[|\mathsf Z|^2\right]+\mathrm E\left[|\mathsf W|^2\right]$, because the error $\mathsf W$ is correlated with the noise $\mathsf Z$.
Treating the error as an independent noise always reduces the achievable rate.

In Fig. \ref{Geometry} there are also some perpendicular lines (dashed) which indicate some projections.
For example, the projection of $\mathbf X$ onto $\mathbf V$ is the LMMSE estimator of $\mathbf X$ from $\mathbf V$ as
\begin{align}
\hat{\mathbf X}_{\textrm{LMMSE}}(\mathbf V)=\frac{\mathrm E\left[\mathsf X\overline{\mathsf V}\right]}{\mathrm E\left[|\mathsf V|^2\right]}\mathbf V,
\end{align}
i.e., a scaling of $\mathbf V$ with a scalar (called Wiener coefficient), which is equal to $\rho_{XV}^2=\mathsf{SNR}/(1+\mathsf{SNR})$ since $\mathrm E\left[\mathsf X\overline{\mathsf V}\right]=\mathrm E\left[|\mathsf X|^2\right]=\mathsf {SNR}\cdot \sigma^2$.
Similarly, the Wiener coefficient of the LMMSE estimator of $\mathbf Y$ from $\mathbf V$ is $\rho_{VY}^2=1-\mathsf{mmse}$.
The following result provides some equivalent expressions of the condition (\ref{AB}) for GMI maximization, which also interprets the Wiener coefficients of the other two projections shown in Fig. \ref{Geometry}.

\textbf{Proposition 11}:
\emph{The following four conditions are equivalent to each other:}
\begin{align}
\frac {\mathcal A} {\mathcal B} = \sqrt{\frac{2}{\pi}},
\end{align}
\begin{align}\label{ww}
\frac {\mathrm E\left[\mathsf{X}\overline{\mathsf{Y}}\right]} {\mathrm E\left[|\mathsf{Y}|^2\right]}
= \frac {\mathrm E\left[\mathsf{X}\overline{\mathsf V}\right]} {\mathrm E\left[|\mathsf V|^2\right]},
\end{align}
\begin{align}\label{betabeta}
\frac {\mathrm E\left[\mathsf{Y}\overline{\mathsf{X}}\right]} {\mathrm E\left[|\mathsf{X}|^2\right]}
= \frac{\mathrm E\left[\mathsf{V}\overline{\mathsf Y}\right]} {\mathrm E\left[|\mathsf V|^2\right]},
\end{align}
\emph{and a relationship between Pearson correlation coefficients as}
\begin{align}\label{rhorhorho}
\rho_{XY} = \rho_{XV} \cdot \rho_{VY},
\end{align}
\emph{where}
\begin{align}\label{rhoxv}
\frac {\mathrm E\left[\mathsf{X}\overline{\mathsf V}\right]} {\mathrm E\left[|\mathsf V|^2\right]} = \frac {\mathsf{SNR}} {1+\mathsf{SNR}} = \rho^2_{XV},
\end{align}
\emph{and}
\begin{align}\label{rhovy}
\frac {\mathrm E\left[\mathsf{V}\overline{\mathsf Y}\right]} {\mathrm E\left[|\mathsf V|^2\right]} = 1-\mathsf{mmse}=\rho^2_{VY},
\end{align}
\emph{respectively.}

\begin{IEEEproof}
See Appendix E.
\end{IEEEproof}

\emph{Remark}:
Proposition 11 shows that the optimal uniform quantization rule should let
\begin{align}\label{XY}
\hat{\mathbf X}_\textrm{LMMSE} (\mathbf Y) = \frac {\mathsf{SNR}} {1+\mathsf{SNR}} \mathbf Y
\end{align}
hold; i.e., it should let the Wiener coefficient be equal to the one in (\ref{rhoxv}).
Similarly, the optimal uniform quantization rule should let
\begin{align}\label{YX}
\hat{\mathbf Y}_\textrm{LMMSE} (\mathbf X) = (1-\mathsf{mmse}) \mathbf X
\end{align}
hold, where the Wiener coefficient is equal to the one in (\ref{rhovy}).
Moreover, from (\ref{YX}) we obtain an additive uncorrelated noise model as
\begin{align}\label{YmmseXd}
\mathsf Y = (1-\mathsf{mmse}) \mathsf X + \mathsf D,
\end{align}
where $\mathsf D $ satisfies $\mathrm E\left[\mathsf X\overline{\mathsf D}\right]=0$ and
\begin{align}
\mathrm E\left[|\mathsf D|^2\right] = (1-\mathsf{mmse}) (\mathsf{mmse} \cdot \mathcal E_x+\sigma^2).
\end{align}
The channel (\ref{YmmseXd}) is a scaling of the channel $Y=\alpha X+D$ in the proof of Proposition 1.
We can apply Proposition 1 directly to (\ref{YmmseXd}) and obtain exactly the same GMI expression as (\ref{GMIMMSE}) in Theorem 8.
Correspondingly, the scaling factor for the nearest neighbor decoder in Fig. \ref{Geometry} should be set as $1-\mathsf{mmse}$  to achieve (\ref{GMIMMSE}), while in the original model (Fig. \ref{Archi}) we should set it according to (\ref{scaling}).

We next discuss how the triangles in Fig. \ref{Geometry} determine the performance via three angles as follows.

1) $\theta_{XV}\in(0,\pi/2)$: it is determined by the SNR and satisfies
\begin{align}
\mathsf{SNR} = \cot^2\theta_{XV}
\end{align}
and
\begin{align}
\rho_{XV} = \cos\theta_{XV} = \sqrt { \frac {\mathsf {SNR}} {1+\mathsf{SNR}} }.
\end{align}
The high-resolution limit of the achievable rate (i.e., the capacity) can be expressed as
\begin{align}
C = \log(1+\mathsf{SNR}) = \log\left(\csc^2\theta_{XV}\right).
\end{align}

2) $\theta_{VY}\in\left(0,\arccos\sqrt{\frac{2}{\pi}}\right)$: it is determined by the resolution, or equivalently by $\mathsf{mmse}$, and satisfies
\begin{align}
\mathsf{mmse} = \sin^2\theta_{VY}
\end{align}
and
\begin{align}
\rho_{VY} = \cos\theta_{VY} = \sqrt{1-\mathsf{mmse}}.
\end{align}
The high-SNR limit of the achievable rate (i.e., the saturation rate) can be expressed as
\begin{align}
\bar I_\textrm{GMI}^* = \log \frac{1}{\mathsf{mmse}} = \log\left(\csc^2\theta_{VY}\right).
\end{align}

3) $\theta_{XY}\in(\max(\theta_{XV},\theta_{VY}),\pi/2)$: it is determined by $\theta_{XV}$ and $\theta_{VY}$ as (see Proposition 11)
\begin{align}\label{coscoscos}
\cos\theta_{XY} = \cos\theta_{XV} \cos\theta_{VY},
\end{align}
so that
\begin{align}\label{rhoXY}
\rho_{XY} = \cos\theta_{XY} = \sqrt { \frac {\mathsf {SNR}} {1+\mathsf{SNR}} } \sqrt{1-\mathsf{mmse}}.
\end{align}
The maximum achievable rate can be expressed as
\begin{align}
I^*_\textrm{GMI} = \log\left(\csc^2\theta_{XY}\right),
\end{align}
corresponding to an effective SNR as $\mathsf {SNR}^*_\textrm e=\cot^2\theta_{XY}$.

The preceding relations show that, under varying resolution and fixed SNR ($\theta_{VY}$ varies while the triangle $\mathbf{0XV}$ is fixed), according to the relationship (\ref{coscoscos}), the optimal loading factor lets $\cos\theta_{XY}$ be directly proportional to $\cos\theta_{VY}$.
For increasingly fine quantization, the limit of the triangle $\mathbf{0XY}$ is the triangle $\mathbf{0XV}$, the limit of $\rho_{XV}^2\mathbf Y$ is $\rho_{XV}^2\mathbf V$, and the limit of achievable rate is the channel capacity $C$.
Similarly, under varying SNR and fixed resolution ($\theta_{XV}$ varies while the triangle $\mathbf{0VY}$ is fixed), the optimal loading factor lets $\cos\theta_{XY}$ be directly proportional to $\cos\theta_{XV}$.
For increasing SNR, the limit of the triangle $\mathbf{0XY}$ is the triangle $\mathbf{0VY}$, the limit of $\rho_{VY}^2 \mathbf X$ is $\rho_{VY}^2 \mathbf V$, and the achievable rate tends to the saturation rate.
The low-SNR slope of the achievable rate is $\cos^2\theta_{VY}=1-\mathsf{mmse}$.

Finally we note that a suboptimal loading factor always breaks the geometry in Fig. \ref{Geometry}.
For example, if we fix the loading factor and let $b$ increase without bound, then the granular distortion vanishes but the overload distortion keeps unchanged, so that $\mathbf Y$ cannot converge to $\mathbf V$.
For a given resolution, as the SNR increases, a suboptimal loading factor breaks the Pythagorean relation (\ref{VYe}).
In this case the length of $\mathbf Y$ can be larger or smaller than $\sqrt{N(\mathcal E_x+\sigma^2-\mathsf{MMSE})}$,
but the distance $\mathrm E\left[\|\mathbf Y-\mathbf V\|^2\right]$ and the angle $\theta_{VY}$ always become larger, thereby reducing the achievable rate.

\section{Asymptotic Analysis: Characterization of $I_\textrm{GMI}(L)$}

From previous numerical results, we have noted that adjusting the loading factor by gain control is essential in approaching the maximum achievable rate.
To understand the impact of biased gain control, this section investigates the achievable rate $I_\textrm{GMI}(L)$ as a function of the loading factor $L$ when the resolution and the SNR are given.
Our basic approach is asymptotic analysis, by which we establish analytical results and interpret some phenomena observed in numerical results.
In source quantization, asymptotic analysis yields high-resolution quantization theory, which provides fairly accurate results for resolutions equal to or greater than $3$ bits \cite{Gray-Neuhoff98}.
Our results for receiver quantization exhibit similar accuracy.

In summary, the behavior of $I_\textrm{GMI}(L)$ can be characterized by two regions as follows.

\emph{1) Overload Region ($L<L^*$)}:
In this region the overload distortion dominates the performance and the achievable rate behaves like a waterfall.
In particular, for high-resolution quantization at high SNR, the GMI is approximated by $\log(L^3/4\phi(L))$ (except when $L$ is very close to $L^*$), which appears almost linear; see (119a) and Fig. 5, (d)-(f).
When $L$ approaches $L^*$, the impact of granular distortion becomes increasingly important, so that the GMI grows slower and approaches its maximum.
As $L\to 0$, the effective resolution of the quantizer reduces to one bit and $I_\textrm{GMI}(L)$ converges to $I_\textrm{GMI}^\textrm{1-bit}$ given in (\ref{GMI1bit}).

\emph{2) Underload Region ($L>L^*$)}:
In this region the granular distortion dominates the performance.
As $L$ increases, the rate loss increases like $\log(1+L^2K^{-2}\mathsf{SNR}/12)$ (see (\ref{app4})) over a wide range.
When $L\gg L^*$, the effective resolution reduces to one bit and $I_\textrm{GMI}(L)$ converges to $I_\textrm{GMI}^\textrm{1-bit}$ again.

Specifically, when the resolution is sufficiently high so that the irreducible loss is negligible, the loading loss (\ref{LL}) can be expressed as $\log(1+\gamma(L)\mathsf{SNR})$, which is characterized as follows.

\begin{itemize}
\item
When $L\to L^*$ from below (in the overload region), the loading loss decays like $O(4\phi(L)/L^3)\mathsf{SNR}$; i.e., it decays exponentially with $L$; see Theorem 12.

\item
When $L\to L^*$ from above (in the underload region), the loading loss decays like $L^2K^{-2}\mathsf{SNR}/12$; i.e., it decays quadratically with $L$; see Theorem 14.
\end{itemize}

We first establish the preceding high-resolution asymptotic results for loading loss in Sec. V-A, and then utilize them to characterize the behavior of $I_\textrm{GMI}(L)$ in Sec. V-B.

\subsection{Decay of Rate Loss under High-Resolution Receiver Quantization}
Now consider the impact of bias in gain control when the irreducible loss is negligible.
In other words, we consider how $I_\textrm{GMI}(L)$ converges to $I_\textrm{GMI}^*\approx C$ as $L\to L^*$.
Our method is to characterize asymptotic behaviors of rate loss under finite loading factor and finite step size, respectively, in the high-resolution limit.
Then the accuracy of the obtained asymptotic formulas is evaluated numerically for finite-resolution receiver quantization.

The following result shows that, the rate loss due to only the overload distortion decays exponentially as the loading factor increases.
Asymptotically, the loading loss is directly proportional to the overload distortion, and is also directly proportional to the SNR.

\textbf{Theorem 12}:
\emph{In the channel (\ref{YqXZ}) under i.i.d. complex Gaussian codebook and nearest neighbor decoding rule (\ref{NND}), the rate loss due to uniform quantization with a loading factor $L$ satisfies}
\begin{subequations}
\begin{align}\label{Rlossasy}
\mspace{-15mu} C - I_\text{GMI} &= \left( \underline{\mathsf{mse}}_\textrm{o} + (1+o_L(1))\frac{4\phi^2(L)}{L^2} \right) \mathsf{SNR} \;\textrm {nats/c.u.}\\
&= \left( (1+o_L(1))\frac{4\phi(L)}{L^3} \right) \mathsf{SNR} \;\;\textrm {nats/c.u.}
\end{align}
\end{subequations}
\emph{in the high-resolution limit, where}
\begin{align}\label{inod}
\underline{\mathsf{mse}}_\textrm{o} := 2\int_L^\infty(t-L)^2\phi(t) \mathrm d t
\end{align}
\emph{is the infimum of the normalized overload distortion} $\mathsf{mse}_\textrm{o}$ \emph{given in (\ref{mseo}).}

\begin{IEEEproof}
Combining (\ref{Aa1}) and (\ref{Ba1}), we obtain the high-resolution limit of $\gamma$ as a function of $L$ as
\begin{subequations}
\begin{align}\label{barL}
\bar{\gamma}(L) &:= \lim\limits_{K\to\infty}\gamma\\
&= \frac { \frac{1}{4} - \int_L^\infty tQ(t)\mathrm d t - \left(\frac{1}{2}-Q(L)\right)^2 }
{ \frac{1}{4} - \int_L^\infty tQ(t)\mathrm d t } \\
&= \frac{ 4\int_L^\infty\left(\phi(t)-tQ(t)\right) \mathrm d t - 4Q^2(L) } { 1 - 4\int_L^\infty tQ(t) \mathrm d t }.
\end{align}
\end{subequations}
In the high-resolution limit, the granular distortion vanishes and the normalized MSE includes only the overload distortion, namely
$\underline{\mathsf{mse}}_\textrm{o}$.
From Proposition 5, (\ref{Aa1}), and (\ref{Ba1}), we obtain another expression of the overload distortion as
\begin{align}\label{intphitq}
\underline{\mathsf{mse}}_\textrm{o} = 4 \int_L^\infty \left(\phi(t)-tQ(t)\right) \mathrm d t.
\end{align}
From [\ref{Hui-Neuhoff01}, Lemma 7 and Eqn. A9], we can infer that\footnote{The derivation in \cite{Hui-Neuhoff01} begins from the original form (\ref{inod}). However, we can also begin from (\ref{intphitq}) and confirm (\ref{1o1}) directly by bounds of Q-function \cite{Borjesson-Sundberg79}, e.g.,
$\frac{1}{t+1/t}\phi(t)<Q(t)\leq \frac{1}{3t/4+\sqrt{t^2+8}/4}\phi(t)$.}
\begin{align}\label{1o1}
\underline{\mathsf{mse}}_\textrm{o} = (1+o_L(1)) \frac{4\phi(L)}{L^3}.
\end{align}
We thus obtain
\begin{subequations}\label{gammamse}
\begin{align}
\bar\gamma(L) &= \frac {\underline{\mathsf{mse}}_\textrm{o} - 4Q^2(L)} {1 - 4\int_L^\infty tQ(t)\mathrm d t} \\
&=\underline{\mathsf{mse}}_\textrm{o} + (1+o_L(1)) \frac{4\phi^2(L)}{L^2}\\
&=(1+o_L(1)) \frac{4\phi(L)}{L^3},
\end{align}
\end{subequations}
where we utilize the fact
\begin{align}\label{Qasy}
Q(t) = (1+o_t(1)) \frac{\phi(t)}{t},
\end{align}
which follows from bounds for the Q function as \cite{Borjesson-Sundberg79}
\begin{align}\label{ULBQ}
\frac{\phi(t)}{t} > Q(t) > \frac{t}{1+t^2}\phi(t).
\end{align}
The proof is completed by combining (\ref{gammamse}) and (\ref{lossa}).
\end{IEEEproof}

\emph{Remark}:
By the lower bound in (\ref{ULBQ}), one can show that $\phi(t)-tQ(t)<\phi(t)/t^2$.
Thus the integral in (\ref{intphitq}) can be upper bounded by
\begin{align}
\int_L^\infty \frac{\phi(t)}{t^2} \mathrm d t = L^{-1} \phi(L)-Q(L).
\end{align}
Using the lower bound in (\ref{ULBQ}) again, we obtain
\begin{align}\label{UBmseo}
\underline{\mathsf{mse}}_\textrm o < \frac{4\phi(L)}{L^3},
\end{align}
which implies that, as $L$ increases, $\underline{\mathsf{mse}}_\textrm o$ converges to $4\phi(L)L^{-3}$ from below (cf. (\ref{1o1})).

On the other hand, the rate loss due to only the granular distortion decays quadratically as the step size vanishes.
To prove this we need the following lemma which is one of the various forms of the Euler-Maclaurin summation formula.

\textbf{Lemma 13}\cite{NA}:
\emph{For a real-valued continuously differentiable function} $f(t)$ \emph{defined on} $[a,b]$\emph{, we have}
\begin{align}
\int_a^b f(t) \mathrm d t
=\;& \ell \left( \frac{f(a)}{2} + \sum\limits_{k=1}^{K-1} f(a+k\ell) + \frac{f(b)}{2} \right)
- \frac{\ell^2}{12} \left(f'(b)-f'(a)\right) + o(\ell^2),
\end{align}
\emph{where} $\ell=\frac{b-a}{K}$.

This lemma characterizes the error of numerical integration using the composite trapezoidal rule with an evenly spaced (uniform) grid.
Based on Lemma 13, the following result can be obtained.

\textbf{Theorem 14}:
\emph{In the channel (\ref{YqXZ}) under i.i.d. complex Gaussian codebook and nearest neighbor decoding rule (\ref{NND}), the rate loss due to uniform quantization with a step size $\ell$ satisfies}
\begin{subequations}\label{Lossl}
\begin{align}
C - I_\text{GMI} &= \left( \overline{\mathsf{mse}}_\textrm g + o(\ell^2) \right) \mathsf{SNR}\;\;\textrm {nats/c.u.}\label{msegloss}\\
&= \left( \frac{\ell^2}{12} + o(\ell^2) \right)\mathsf{SNR} \;\;\textrm {nats/c.u.}\label{quadloss}
\end{align}
\end{subequations}
\emph{in the high-resolution limit, where}
\begin{align}
\overline{\mathsf{mse}}_\textrm g := \lim\limits_{K\to\infty} \mathsf{mse}_\textrm g
\end{align}
\emph{is the supremum of the normalized granular distortion} $\mathsf{mse}_\textrm g$ \emph{given in (\ref{mseg}).}

\begin{IEEEproof}
Applying the Euler-Maclaurin summation formula in Lemma 13, for $\mathcal A$ we have
\begin{subequations}
\label{A}
\begin{align}
 \int_0^{K\ell} \exp \frac{-t^2}{2} \mathrm d t
=\;& \ell\left( \frac{1}{2} + \sum\limits_{k=1}^{K-1} \exp\frac{-k^2\ell^2}{2} + \frac{1}{2}\exp\frac{-K^2\ell^2}{2} \right)
 + \frac{\ell^2}{12} K\ell\exp\frac{-K^2\ell^2}{2} + o(\ell^2)\\
=\;& \sum\limits_{k=0}^{K-1} \ell \cdot \exp\frac{-k^2\ell^2}{2} - \frac{\ell}{2}
 +\frac{\ell}{2} \exp\frac{-K^2\ell^2}{2} + \frac{K\ell^3}{12} \exp\frac{-K^2\ell^2}{2} + o(\ell^2)\\
=\;& \mathcal A + o(\ell^2),
\end{align}
\end{subequations}
and for $\mathcal B$ we have
\begin{subequations}
\label{B}
\begin{align}
\int_0^L 2tQ(t) \mathrm d t
=\;& \ell \left( \sum\limits_{k=1}^{K-1} 2k\ell Q(k\ell) + K\ell Q(K\ell) \right)
 -\frac{\ell^2}{6} \left( Q(K\ell) - \frac{K\ell}{\sqrt{2\pi}}\exp\frac{-K^2\ell^2}{2} - \frac{1}{2} \right)+ o(\ell^2)\\
=\;& \sum\limits_{k=0}^{K-1} 2k\ell^2 Q(k\ell) + \frac{\ell^2}{8} - \frac{\ell^2}{24} + o(\ell^2)\\
=\;& \frac{\mathcal B}{\pi} - \frac{\ell^2}{24} + o(\ell^2).
\end{align}
\end{subequations}
As $K\to\infty$, we have $L=K\ell=\infty$. Then in the high-resolution limit we have $\mathcal A=\sqrt{\pi/2}+o(\ell^2)$ and $\mathcal B/\pi=1/2+\ell^2/24+o(\ell^2)$, which yield
\begin{align}\label{mseasy}
\mathsf{mse} = \overline{\mathsf{mse}}_g = \frac{\ell^2}{12}+o(\ell^2)
\end{align}
and
\begin{align}\label{BAB}
\bar\gamma = 1 - \frac{1}{\pi} \frac{\frac{\pi}{2} + o(\ell^2)} {\frac{1}{2} + \frac{\ell^2}{24} + o(\ell^2)}
=\frac{\ell^2}{12} + o(\ell^2).
\end{align}
The proof of (\ref{msegloss}) is completed by combining (\ref{mseasy}) and (\ref{lossa}), while the proof of (\ref{quadloss}) is completed by combining (\ref{BAB}) and (\ref{lossa}).
\end{IEEEproof}

\begin{figure}
\centering
\includegraphics[scale=0.7]{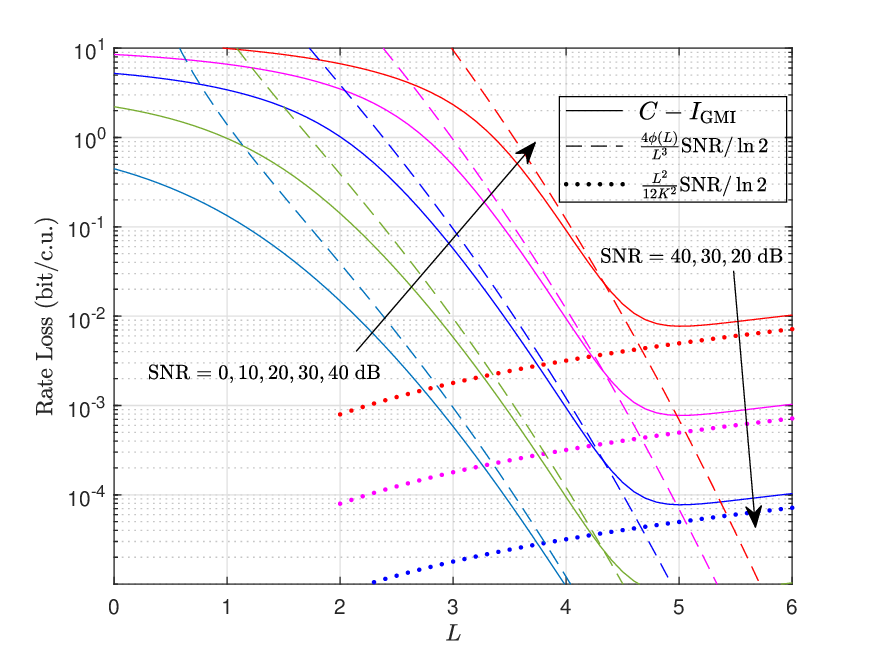}
\caption{Approximations of small irreducible rate loss: $b=12$ bits. }\label{RlossL}
\centering
\includegraphics[scale=0.7]{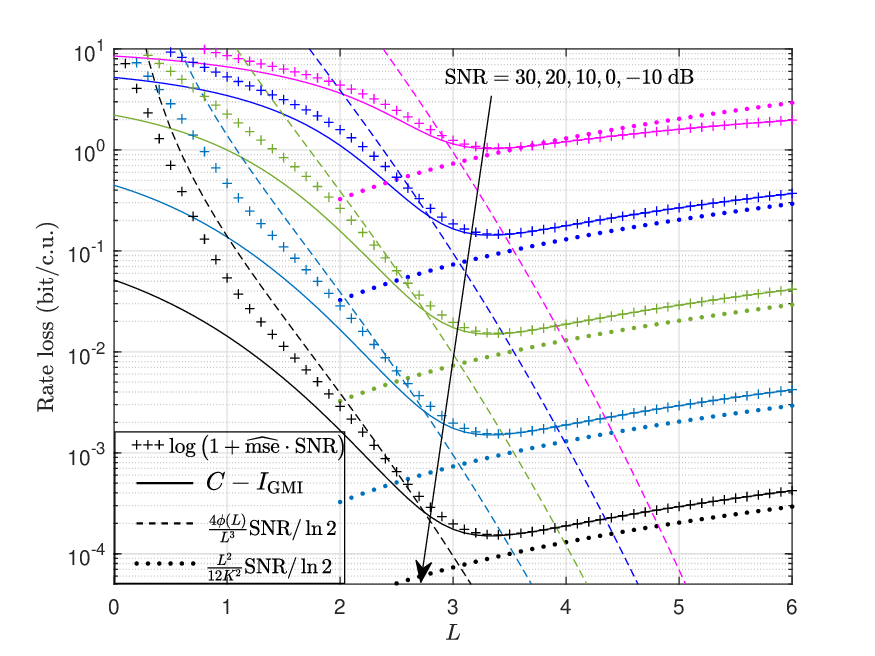}
\caption{Approximations of small irreducible rate loss: $b=6$ bits. }\label{RlossL6}
\end{figure}

In Theorem 12 and Theorem 14 we assume output quantization with \emph{unlimited} resolution.
We now check whether they provide useful approximations of rate loss under finite resolutions.
In Fig. \ref{RlossL}, we show the rate loss $C-I_\textrm{GMI}$ due to 12 bits output quantization.
According to Theorem 12, a small irreducible loss in overload region can be approximated by
\begin{align}\label{appmmseo}
C - I_\textrm{GMI}(L) \approx 4\phi(L)L^{-3} \mathsf{SNR} \;\;\textrm {nats/c.u.},
\end{align}
and according to Theorem 14, a small irreducible loss in underload region can be approximated by
\begin{align}\label{appmmseg}
C - I_\textrm{GMI}(L) \approx L^2\mathsf{SNR}/12K^2 \;\;\textrm {nats/c.u.}
\end{align}
Clearly, these approximations successfully capture the decay of rate loss when it is dominated by the overload distortion.
When the loading factor $L$ exceeds $4$, the increasing granular distortion kicks in and dominates the performance quickly and the rate loss is well approximated by (\ref{appmmseg}).
When the resolution decreases, e.g., in Fig. \ref{RlossL6} where the resolution is 6 bits, the approximation (\ref{appmmseo}) becomes less useful, especially at high SNR.
But the approximation (\ref{appmmseg}) is still satisfactory.
For higher accuracy, by considering both the overload distortion and the granular distortion, we propose an approximation given by
\begin{align}
C - I_\textrm{GMI}(L) \approx \log \left( 1 + \widehat{\mathsf{mse}} \cdot \mathsf{SNR} \right),
\end{align}
where
\begin{align}\label{hatmse}
\widehat{\mathsf{mse}} = \frac{4\phi(L)}{L^{3}} + \frac{4\phi^2(L)}{L^2} + \frac{L^2}{12K^2},
\end{align}
in which the first two terms of the RHS come from (\ref{Rlossasy}), and the last term comes from (\ref{Lossl}).
As shown in Fig. \ref{RlossL6}, the proposed formula well approximates the transition from the overload region to the underload region.

\subsection{Properties and Approximations of $I_\textrm{GMI}(L)$}

Now discuss some general properties of $I_\textrm{GMI}(L)$.
The derivative of the achievable rate
\begin{align}
\frac {\mathrm d I_\text{GMI}} {\mathrm d L}
=\frac {-\mathsf{SNR}} {\gamma\mathsf{SNR}+1} \frac {\mathrm d \gamma} {\mathrm d L}
\end{align}
shows that, for the SNR-limited scenario (defined in Sec. IV-B), where $\gamma\mathsf{SNR}\ll 1$ over a wide range of $L$, we have
\begin{align}
\frac {\mathrm d I_\text{GMI}} {\mathrm d L} \approx -\mathsf{SNR} \frac {\mathrm d \gamma} {\mathrm d L},
\end{align}
which implies that the penalty of a small bias of gain control increases approximately linearly with the SNR; see Fig. \ref{RlossL} and Fig. \ref{RlossL6}.
This is consistent with Theorem 12 and Theorem 14.
For the resolution-limited scenario (also defined in Sec. IV-B), where $\gamma\mathsf{SNR}\gg 1$, we have
\begin{align}
\frac {\mathrm d I_\text{GMI}} {\mathrm d L} \approx -\frac{1}{\gamma} \frac{\mathrm d \gamma} {\mathrm d L},
\end{align}
which does not vary with the SNR, implying that the achievable rate saturates; see high-SNR curves in Fig. \ref{b4} and Fig. \ref{b2}.

Apart from the aforementioned cases, in general it is not easy to establish a global property for $I_\textrm{GMI}(L)$ from the derivative.
We next propose approximations for $I_\textrm{GMI}(L)$ by asymptotic analysis of $\gamma(L)$.

\begin{figure}
\centering
\includegraphics[scale=0.7]{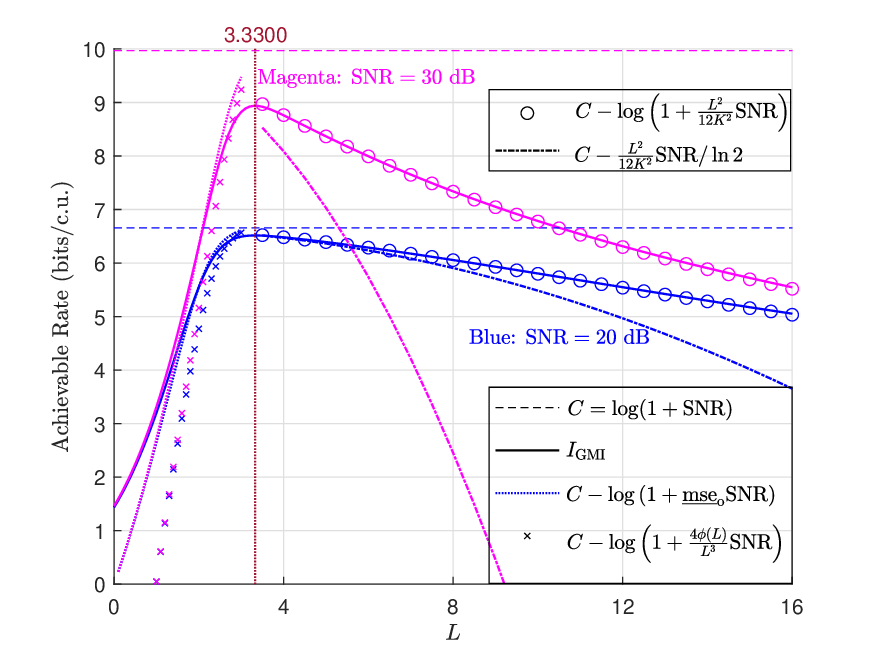}
\caption{Approximations of $I_\textrm{GMI}(L)$: $b=6$ bits. }\label{ILapp}
\centering
\includegraphics[scale=0.7]{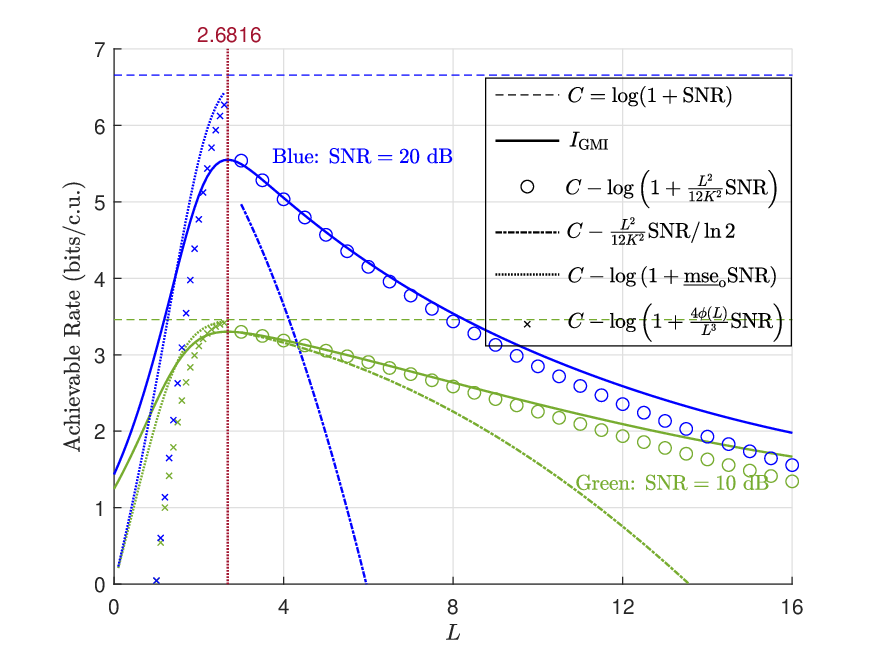}
\caption{Approximations of $I_\textrm{GMI}(L)$: $b=4$ bits. }\label{ILapp4}
\end{figure}

\subsubsection{Overload Region}
In this region it is not easy to find a simple and accurate approximation for $I_\textrm{GMI}(L)$.
In the range of resolution of practical interest (e.g., $b\leq 12$ bits), the loading factor in the overload region is small, so that the asymptotics (\ref{Qasy}) can be inaccurate and the asymptotic expression of $\bar\gamma(L)$ in (\ref{gammamse}) is less useful.
For example, Fig. \ref{ILapp} and Fig. \ref{ILapp4} show that $C-\log\left(1+4\phi(L)L^{-3}\mathsf{SNR}\right)$ is not a satisfactory approximation.
Instead, using
\begin{align}\label{msewaterfall}
C - \log\left( 1 + \underline{\mathsf{mse}}_\textrm o\mathsf{SNR} \right).
\end{align}
may approximate the waterfall better, but it cannot be expressed in a closed form.
A special case is high-resolution quantization at high SNR, where the waterfall in the overload region can be approximated by the high-resolution limit of the saturation rate $\log(1/\gamma)$ as
\begin{subequations}
\begin{align}\label{waterfall}
\log \frac{1}{\bar\gamma(L)}
&= \log\left( (1+o_L(1))\frac{L^3}{4\phi(L)} \right)\\
&\approx \log\left( \frac{L^3}{4\phi(L)} \right)\\
&= L^2/2 + 3\ln L + \ln\frac{\sqrt{2\pi}}{4} \;\textrm{nats/c.u.},
\end{align}
\end{subequations}
which does not depend on resolution or SNR; see Figs. (\ref{20dB}-\ref{40dB}).

\subsubsection{Underload Region}
It is easier to approximate $I_\textrm{GMI}(L)$ in this region.
We utilize the asymptotic expression (\ref{BAB}) of $\bar\gamma$ and let the resolution be finite with $2K$ levels, yielding
\begin{align}\label{app4}
I_\textrm{GMI}(L) &\approx C - \log \left( 1+\frac{L^2}{12K^2} \mathsf{SNR} \right),
\end{align}
which is highly accurate even when the resolution is low; see Fig. \ref{ILapp} and Fig. \ref{ILapp4}.
We have also considered an even simpler approximation $C-L^2\mathsf{SNR}/12K^2 \;\textrm{nats/c.u.}$ in the high-resolution scenario; see Fig. \ref{RlossL} and Fig. \ref{RlossL6}.
However, it becomes less useful as the loading loss increases, as shown in Fig. \ref{ILapp} and Fig. \ref{ILapp4}.
The approximation (\ref{app4}) also implies that the saturation rate can be approximated by
\begin{align}\label{app4sa}
\bar I_\textrm{GMI}(L) = \log (1/\gamma) \approx\log\frac{12K^2}{L^2}.
\end{align}
In Figs. \ref{20dB}-\ref{40dB}, we can observe that (\ref{app4sa}) well approximates $I_\text{GMI}$ in the resolution-limited scenario.

\section{Asymptotic Analysis: Characterization of $L^*$ and $I_\textrm{GMI}^*$}

This section focuses on $I_\textrm{GMI}^*$, the maximum achievable rate of the transceiver architecture described in Sec. II-A when the receiver quantization is uniform with the optimal loading factor $L^*$.
In Sec. VI-A, we consider approximations of the optimal loading factor $L^*$, which is essential for achieving $I_\textrm{GMI}^*$ by accurate gain control.
In Sec. VI-B, we characterize $I_\textrm{GMI}^*$ as a function of the resolution $b$, and further provide some per-bit rules for different performance metrics such as saturation rates and irreducible rate loss.

\subsection{Approximations of Optimal Loading Factor}

We first provide a new proof of the property (\ref{llimit}) for the step size and loading factor of the optimal uniform quantizer.
In \cite{BG80}, it was shown that the uniform quantizer that minimizes the MSE must satisfy (\ref{llimit}) if the input density has an infinite support.
Here we show that (\ref{llimit}) is a natural corollary of the achievable rate results in Theorem 12 and Theorem 14.

\emph{New Proof of (\ref{llimit})}:
We consider two possible cases other than (\ref{llimit}) and exclude them respectively.
First, if the high-resolution limit of the optimal step size, namely $\lim_{K\to\infty}\ell^*$, is strictly larger than zero, then $\lim_{K\to\infty}L^*=\infty$.
However, we can infer from Theorem 14 that, for a step size bounded away from zero, if the resolution is high enough, then there must exist a smaller step size that reduces the loss in achievable rate.
Thus $\lim_{K\to\infty}\ell^*$ must be equal to zero.
Second, on the other side, if $\lim_{K\to\infty}L^*<\infty$, then $\lim_{K\to\infty}\ell^*=0$.
However, we can infer from Theorem 12 that, given a finite $L$, if the resolution is high enough, then there must exist a larger $L$ that reduces the loss in achievable rate.
Thus $\lim_{K\to\infty}L^*$ cannot be finite.
Therefore, we conclude that $\lim_{K\to\infty}\ell^*=0$ and $\lim_{K\to\infty}L^*=\infty$ hold simultaneously.
$\hfill\blacksquare$

The scaling law (\ref{asy}) was obtained in \cite{Hui-Neuhoff01} (as a special case of a more general result) by analyzing the derivative of the MSE.
Here we give a simpler proof of (\ref{asy}) from the condition (\ref{AB}) for achievable rate maximization.
The proof utilizes the Euler-Maclaurin formula.

\emph{New Proof of (\ref{asy})}:
In the high-resolution regime, by combining (\ref{A}), (\ref{B}), and
 (\ref{AB}) we obtain
\begin{align}
\label{fraclimit}
\frac{\int_{K\ell^\ast}^\infty \exp \frac{-t^2}{2} \mathrm d t + o({\ell^\ast}^2)}
{\int_{K\ell^\ast}^\infty 2tQ(t)\mathrm d t - \frac{1}{24}{\ell^\ast}^2 + o({\ell^\ast}^2)}
= \sqrt{2\pi},
\end{align}
yielding
\begin{align}
\label{conditionK}
\lim\limits_{K\to\infty} \frac{ Q(L^\ast) } { \int_{L^\ast}^\infty 2tQ(t)\mathrm d t -\frac{1}{24K^2}{L^\ast}^2 } = 1.
\end{align}
Noting that
\begin{align}
\lim\limits_{K\to\infty} \frac {Q(L^\ast)} {\int_{L^\ast}^\infty tQ(t) \mathrm d t}
= \lim\limits_{K\to\infty} \frac{-\phi(L^\ast)} {-L^\ast Q(L^\ast)}
= 1,
\end{align}
where the second equality follows from (\ref{ULBQ}), we obtain
\begin{align}
\label{elementary}
\lim\limits_{K\to\infty} \frac{ 24K^2 \exp\frac{{-L^\ast}^2}{2} } { \sqrt{2\pi}{L^\ast}^3 } = 1,
\end{align}
which is equivalent to
\begin{align}
\label{elementary1}
\lim\limits_{K\to\infty} \frac{ 2\ln(2K)+\ln6 } { {L^\ast}^2/2+3\ln{L^\ast}+\ln\sqrt{2\pi} } = 1.
\end{align}
The proof is completed by noting that (\ref{elementary1}) implies (\ref{asy}).
$\hfill\blacksquare$

Although (\ref{asy}) provides a simple approximation of $L^*$ as $\hat L_1=2\sqrt{\ln(2K)}$, it can be refined by more elaborate techniques.
In [\ref{Hui-Neuhoff01}, Sec. V] three approximations have been proposed.
For Gaussian input, the first is just $\hat L_1$, and
the other two satisfy a stronger condition
\begin{align}
\lim\limits_{K\to\infty} \left\{L^* - \hat L_i \right\}=0, \; i=2,3,
\end{align}
where
\begin{align}\label{appMSE}
{ \hat L_2 } = \sqrt{ 4\ln (2K) -3\ln\ln (2K) -\ln \frac{32\pi}{9} }
\end{align}
and
\begin{align}\label{L3}
{ \hat L_3 }= \sqrt{\hat L_2^2+\epsilon}, \; K>1,
\end{align}
with an error term
\begin{align}
\epsilon &= 2\ln \Bigg[ \left( 1+\frac{4\ln (2K)}{2K} \right) \left( 1-\frac{3}{4\ln (2K)} \right)
\cdot \left( 1 + \frac{1}{2\ln (2K)} \left( \frac{3}{2}\ln\ln (2K) + \ln\frac{4\sqrt{2\pi}}{3} \right) \right)^{3/2} \Bigg]
\end{align}
included, which satisfies $\lim_{K\to\infty}\epsilon=0$.

In the following result, we adopt a new way to approximate $L^*$, which begins from (\ref{conditionK}) in the preceding new proof of (\ref{asy}).

{\textbf{Proposition 15}}:
\emph{In the channel (\ref{YqXZ}) under i.i.d. complex Gaussian codebook and nearest neighbor decoding rule (\ref{NND}), the optimal loading factor} $L^\ast$ \emph{that maximizes the GMI (\ref{main}) satisfies}
\begin{align}\label{approx}
\lim\limits_{K\to\infty} \left\{ L^\ast- \hat{L}_0 \right\} = 0,
\end{align}
\emph{where} $\hat{L}_0$ \emph{is the unique real-valued solution of the transcendental equation}\footnote{A transcendental equation $x+\ln x=\ln a$ can be solved by the Lambert W function as $x=W(a)$. Similarly, the solution $\hat L$ in Proposition 15 can be given by a special case of the generalized Lambert W function, although less is known about this function \cite{LW}.}
\begin{align}\label{app}
{L}^2 + 6\ln L - \ln\frac{18}{\pi} = 4\ln(2K).
\end{align}

\begin{IEEEproof}
According to (\ref{llimit}) and (\ref{fraclimit}), the difference between the numerator and the denominator of (\ref{conditionK}) vanishes as $K$ tends to infinity.
Therefore, $L^\ast$ can be approximated with vanishing error by the solution of
\begin{align}\label{app1}
\int_{L}^\infty 2tQ(t) \mathrm d t - Q(L) = \frac{L^2}{24K^2},
\end{align}
since its LHS and RHS are continuous functions of $L$.
According to (\ref{ULBQ}) and its variation
\begin{align}
\left(1-\frac{1}{1+t^2}\right)\phi(t) < 2tQ(t)-\phi(t) < \phi(t),
\end{align}
we note that, instead of (\ref{app1}), we can turn to the solution of equation
\begin{align}\label{Lasy}
24K^2\phi(L)=L^3,
\end{align}
which is equivalent to (\ref{app}).
The uniqueness of $\hat L$ is clear since the LHS of (\ref{app}) is a smooth and monotonically increasing function of $L$.
\end{IEEEproof}

\emph{Remark}: The equation (\ref{app}) coincides with (\ref{elementary1}) in the aforementioned new proof of (\ref{asy}).
In fact, (\ref{app}) or (\ref{Lasy}) can also be heuristically argued via an approximate analysis of the MSE: Just let the derivative of the MSE approximation (\ref{hatmse}) be zero, and omit the exponential terms therein except for the dominated one.

\begin{figure*}
\centering
\includegraphics[scale=1]{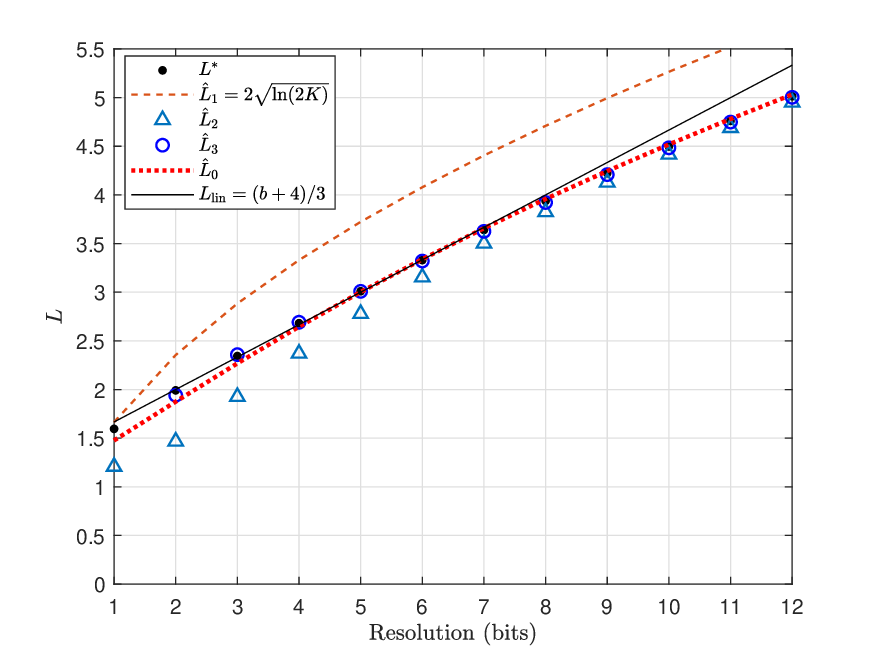}
\caption{Numerical results, asymptotics, and approximations of $L^\ast$.}\label{L}
\end{figure*}

In Fig. \ref{L}, we show $L^*$, its approximation $\hat L_0$ from (\ref{approx}) in Proposition 15, 
the simple approximation $2\sqrt{\ln (2K)}$, and the approximations $\hat L_2$ and $\hat L_3$ from \cite{Hui-Neuhoff01}.
Note that the case $b=1$ does not require gain control.
We see that the new approximation $\hat L_0$ is accurate for all resolutions $b\ge 2$.
It is much better than $\hat L_2$ and close to the more complicated approximation $\hat L_3$.
Interestingly, for moderate resolutions it can be observed that $L^*$ increases approximately linearly.
Therefore, to approach $I^*_\textrm{GMI}$, it is sufficient to use a simple linear approximation of $L^*$ as
\begin{align}\label{Llin}
\hat L_\textrm{lin} = \frac{b+4}{3}.
\end{align}
Fig. \ref{L} shows that it is accurate for $2\leq b\leq7$.
When $b\ge 8$, its accuracy degrades, but the loading loss caused is negligible since it grows quadratically with $L$; see numerical results in Fig. \ref{bits}.

\subsection{Maximum Achievable Rate Approximations and Per-Bit Rules}
We have found a simple relationship between $I_\textrm{GMI}^*$ and $\mathsf {mmse}$ in Theorem 8, and we have also given numerical results of $\mathsf {mmse}$ for $b=1,2,...,12$ in Table I.
However, a direct connection between $I_\textrm{GMI}^*$ and the resolution is still very useful.
The following result provides such a connection via approximating $\mathsf {mmse}$ by $b$.

{\textbf{Proposition 16}}:
\emph{The achievable rate given in (\ref{GMIMMSE}) satisfies}
\begin{align}\label{GMIbs}
I_\textrm{GMI}^* = \hat I_\textrm{GMI}(b) + o_b(1),
\end{align}
\emph{where}
\begin{align}\label{hatgmi}
\hat I_\textrm{GMI}(b) = C - \log\left( 1 + \frac{4b\ln2}{3\cdot 4^b}\mathsf{SNR} \right).
\end{align}

\begin{IEEEproof}
From (\ref{asy}) and (\ref{mmseell}) we obtain
\begin{align}\label{mmseK}
\lim\limits_{K\to\infty} \frac{3K^2}{\ln(2K)} \mathsf{mmse} = 1.
\end{align}
Replacing $K$ by $2^{b-1}$ yields a high-resolution approximation of MMSE given by
\begin{align}\label{appmmse}
\mathsf{mmse} = (1+o_b(1)) \frac{4b\ln2}{3\cdot 4^b} \approx \frac{4b\ln2}{3\cdot 4^b}.
\end{align}
If we replace $\mathsf{mmse}$ in (\ref{GMIMMSE}) by $\frac{4b\ln2}{3\cdot 4^b}$, then we obtain $\hat I_\textrm{GMI}(b)$.
According to (\ref{appmmse}), it is straightforward to show that the gap between $I_\textrm{GMI}^*$ and $\hat I_\textrm{GMI}(b)$ is $o_b(1)$ for an arbitrary finite $\mathsf{SNR}$, thereby completing the proof.
\end{IEEEproof}

\begin{figure}
\centering
\includegraphics[scale=0.7]{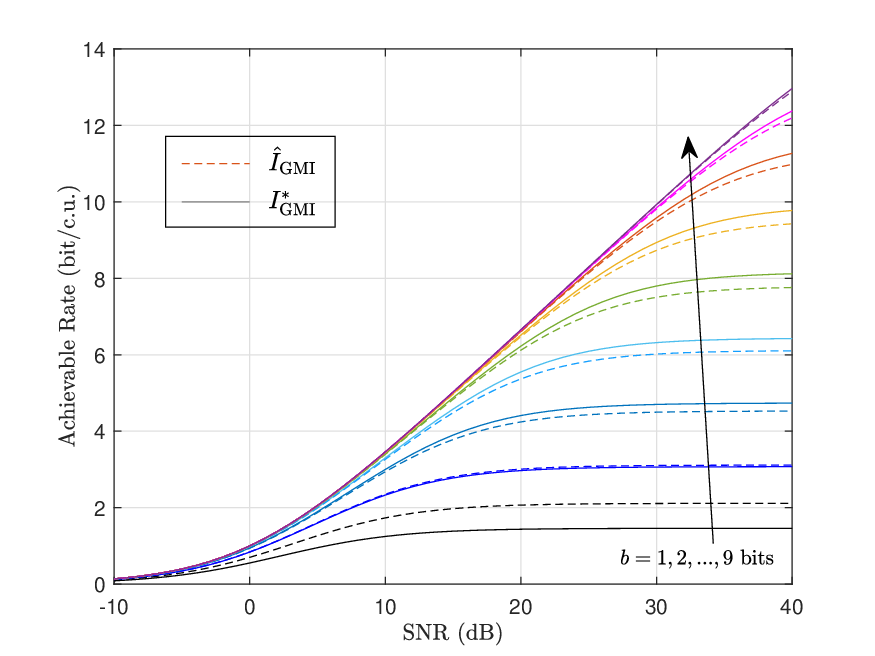}
\caption{Approximation of $I^*_\textrm{GMI}$.}
\label{GMIapp}
\end{figure}

The approximation (\ref{appmmse}) implies a $6$-dB-per-bit-rule as
\begin{align}\label{6db}
10\log_{10}\mathsf {SNR}_q \approx 6.02 b - 10\log_{10}b + 0.34 \; (\textrm{dB}),
\end{align}
which has been known since \cite{Hui-Neuhoff01}.
Therefore, in the high-resolution regime, each additional bit in resolution reduces the MSE by four times:
\begin{align}\label{mmse6}
\lim\limits_{b\to\infty} \frac {\mathsf {mmse}(b)} {\mathsf {mmse}(b-1)} = 4.
\end{align}
The approximation (\ref{appmmse}) is still not accurate enough for moderate to low resolutions; see Table I.
More approximations for MMSE or $\mathsf{SNR}_q$ can be found in [\ref{Hui-Neuhoff01}, Sec. IV] based on refined asymptotic formulas of $L^*$ and its asymptotic relationship with MMSE.
However, using (\ref{appmmse}) is enough to get the simple and useful approximation of $I_\textrm{GMI}^*$ in Proposition 16.
In Fig. \ref{GMIapp} we compare $I^*_\textrm{GMI}$ with its approximation $\hat I_\textrm{GMI}(b)$.
It is shown that the accuracy is acceptable when $b\ge 2$.

We next introduce some per-bit rules for other performance metrics.
The following one is obtained from Proposition 16 immediately.

\begin{itemize}
\item
\emph{A 2-bpcu-per-bit rule for saturation rate}:
The saturation rate $\bar I_\textrm{GMI}^*(b)$ given in (\ref{satrate}) satisfies
\begin{align}\label{SRapp}
\bar I_\textrm{GMI}^*(b) = \hat{\bar I}_\textrm{GMI}(b) + o(1),
\end{align}
where
\begin{align}
\hat{\bar I}_\textrm{GMI}(b) = 2b - \log_2 b + 0.11 \;\textrm{bits/c.u.},
\end{align}
which implies that
\begin{align}
\lim\limits_{b\to\infty} \left(\bar I_\textrm{GMI}^*(b) - \bar I_\textrm{GMI}^*(b-1)\right) = 2\;\textrm{bits/c.u.}
\end{align}
\end{itemize}

\begin{figure}
\centering
\includegraphics[scale=0.7]{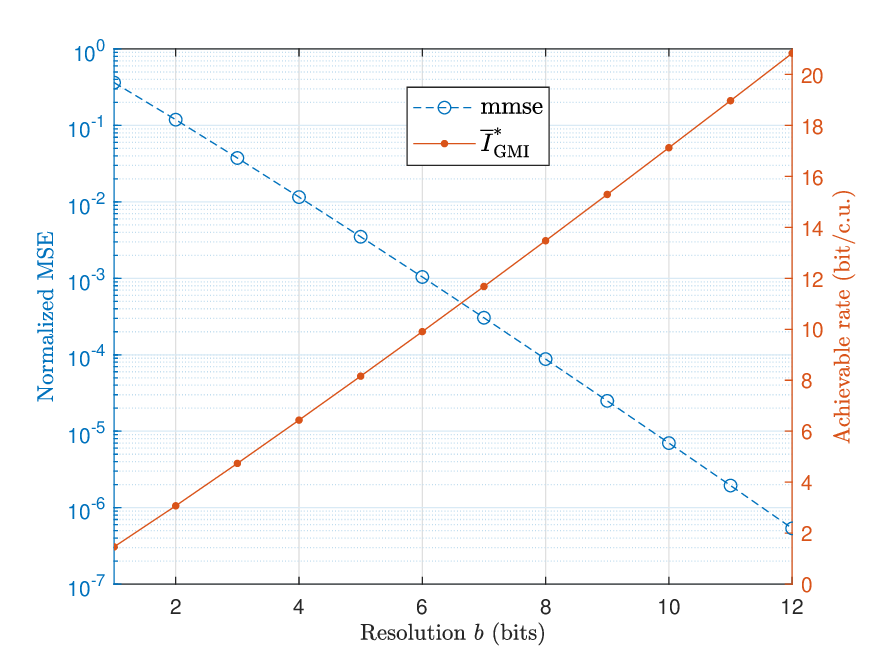}
\caption{Normalized MMSE and saturation rate under different resolutions.}
\label{gammaGMI}
\end{figure}

Numerical evaluations of $\hat{\bar I}_\textrm{GMI}(b)$ for $1\leq b\leq 12$ are shown in Table I.
In Fig. \ref{gammaGMI}, it is shown that $\mathsf{mmse}$ decreases exponentially as the resolution increases,
and correspondingly, the saturation rate $\bar {I}^*_\text{GMI}$ grows approximately linearly with the resolution.
The improvement per bit is $1.6$-$1.9$ bits/c.u. in our range of interest, although the high-resolution limit is $2$ bits/c.u.

\begin{figure}
\centering
\includegraphics[scale=0.7]{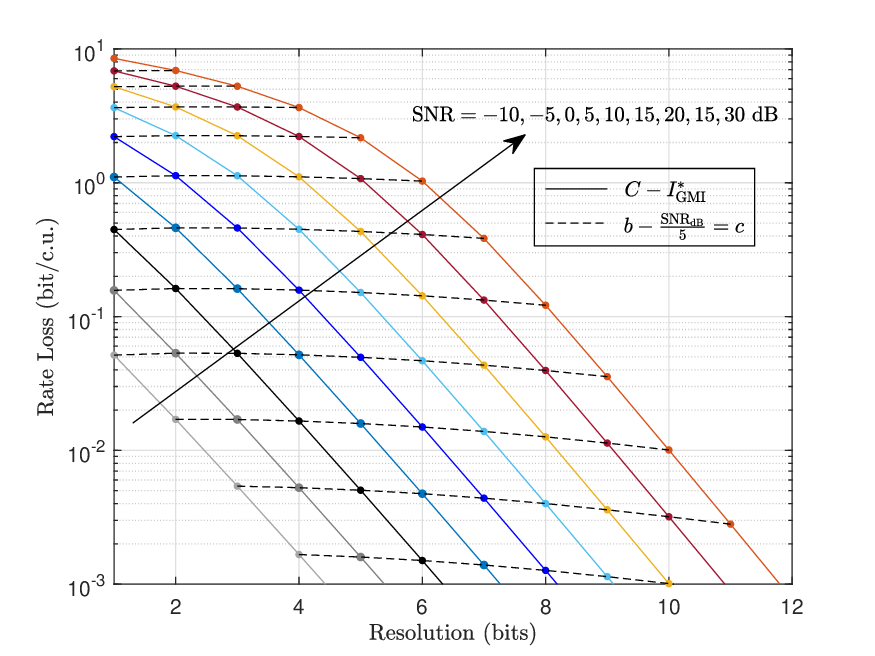}
\caption{A 5-dB-per-bit rule for irreducible rate loss.}\label{5dB}
\centering
\includegraphics[scale=0.7]{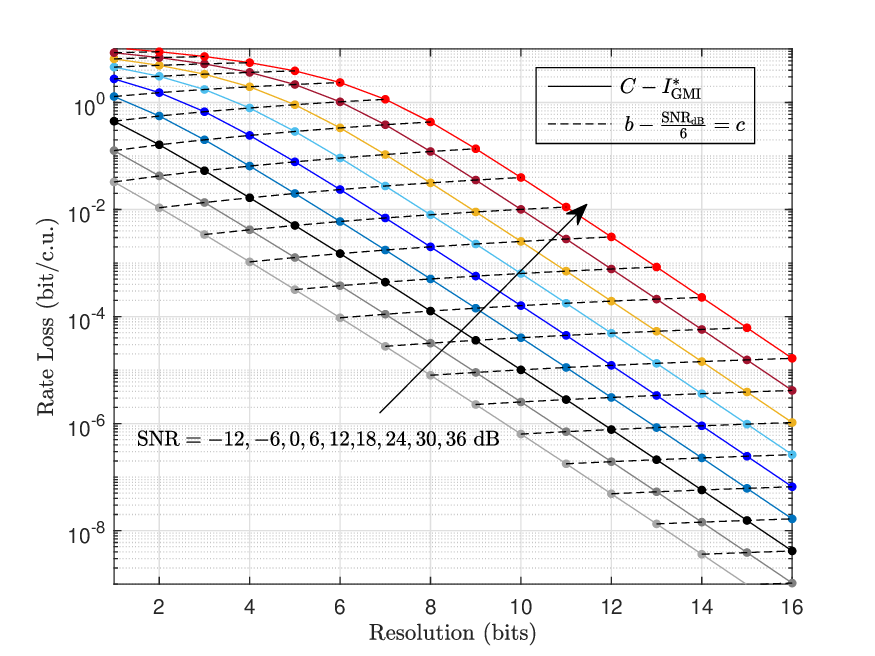}
\caption{For irreducible rate loss, a 6-dB-per-bit rule is inaccurate.}\label{6dB}
\end{figure}

The second per-bit rule can be observed from results in Table I, suggesting a roughly 5-dB-per-bit increase of $\mathsf{SNR}_q$:
\begin{align}\label{snrqrule}
10\log_{10}\mathsf{SNR}_q\approx 5b.
\end{align}
Since the rate loss in (\ref{GMIMMSE}) is determined by $\mathsf{SNR}/\mathsf{SNR}_q$, (\ref{snrqrule}) implies that, in our range of interest, a $5$ dB increase of SNR requires an extra bit of resolution to maintain the same rate loss.
More specifically, we have the following rule.

\begin{itemize}
\item
\emph{A 5-dB-per-bit rule for irreducible rate loss}:
We require a resolution of at least
\begin{align}\label{lossrule}
2\log_{10}\mathsf{SNR} + b_0
\end{align}
bits so that the irreducible loss $C-I^*_\textrm{GMI}$ can be as small as $10^{-b_0}$ bits/c.u., where $b_0\ge0$.
\end{itemize}
This rule can be confirmed by numerical results in  Fig. \ref{5dB}.
Although the 6-dB-per-bit rule is well-known in quantization theory and gives the correct asymptotics (as shown in (\ref{6db})), it is less accurate unless the rate loss is extremely small, see Fig. \ref{6dB}.

\section{Discussions on Further Quantization Rules at the Receiver}

Based on the analytical framework given in Theorem 2, a major part of this paper has focused on the simplest receiver quantization scheme, namely scalar uniform quantization with mid-rise levels.
This section briefly discusses nonuniform and other types of quantization rules in several aspects.

\subsection{Relationship Between MSE and Achievable Rate: Numerical Examples}

\begin{figure*}
\centering
\subfigure[Uniform quantizer: $\mspace{188mu}$ thresholds $0, \pm 0.25, \pm 0.5, \pm 0.75,\mspace{108mu}$ levels $\pm 0.125, \pm 0.375, \pm 0.625, \pm 0.875$]{\includegraphics[scale=0.4]{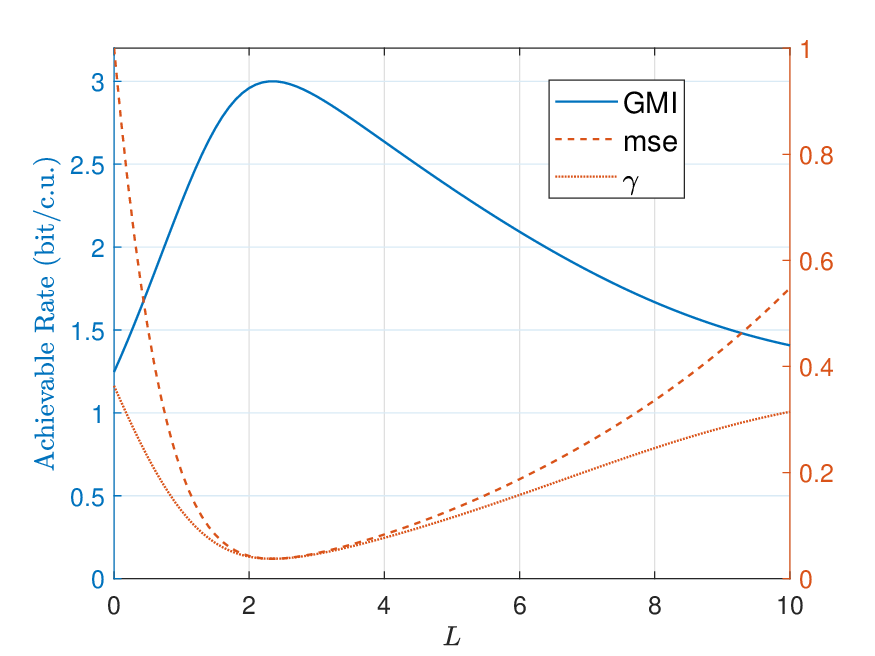}\label{gmimmse}}
\subfigure[Optimal nonuniform quantizer: $\mspace{108mu}$ thresholds $0, \pm 0.5006, \pm 1.0500, \pm 1.7480, \mspace{48mu}$ levels $\pm 0.2451, \pm 0.7560, \pm 1.3440, \pm 2.1520$]{\includegraphics[scale=0.4]{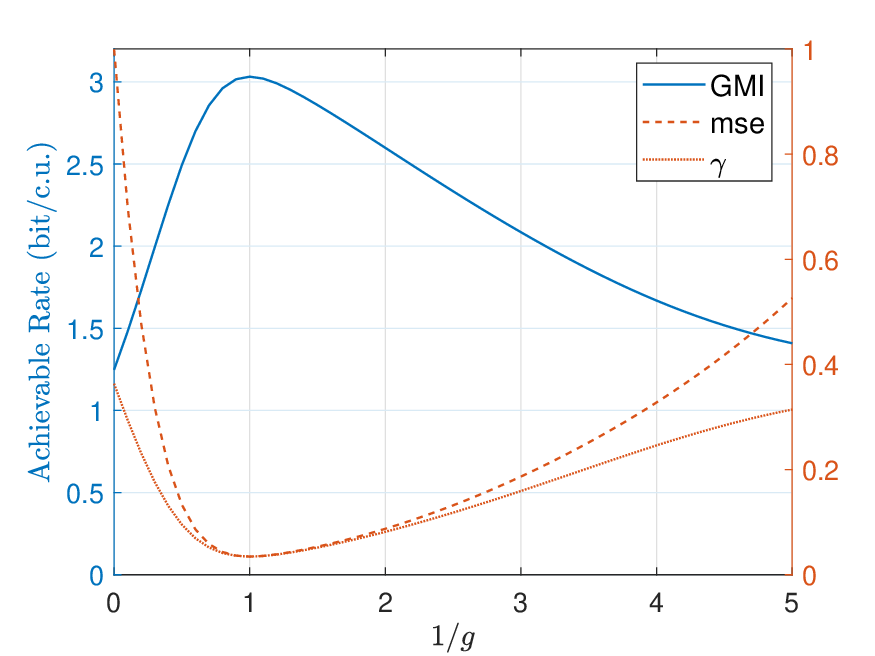}\label{gmimmseopt}}
\subfigure[Nonuniform quantizer: $\mspace{160mu}$ thresholds $0, \pm 0.2, \pm 0.6, \pm 0.7, \mspace{128mu}$ levels $\pm 0.1, \pm 0.5, \pm 0.7, \pm 0.9$]{\includegraphics[scale=0.4]{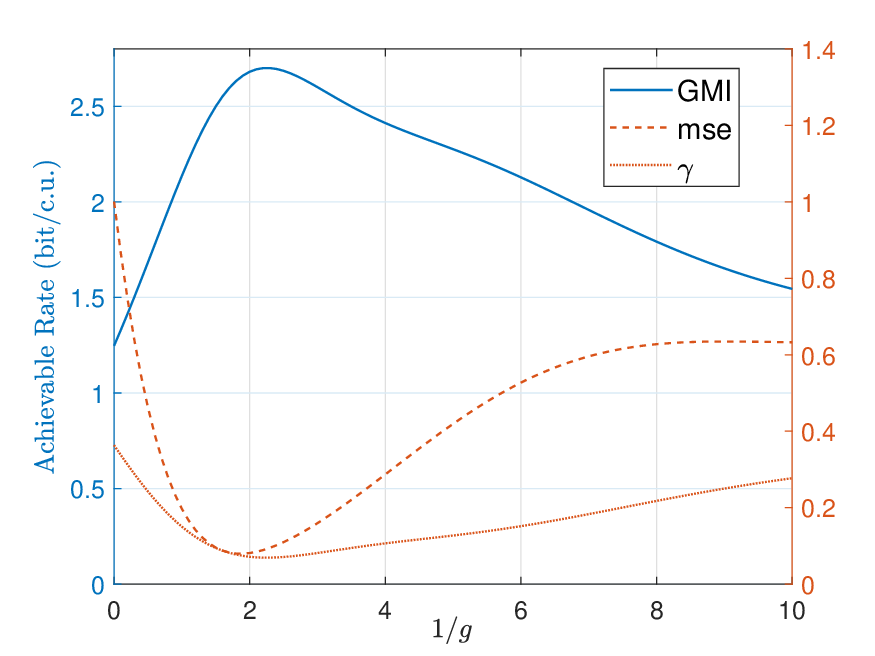}\label{gmimmsenu}}
\newline
\subfigure[Part of (a): $2.1\leq L\leq 2.6 \mspace{128mu}$]{\includegraphics[scale=0.4]{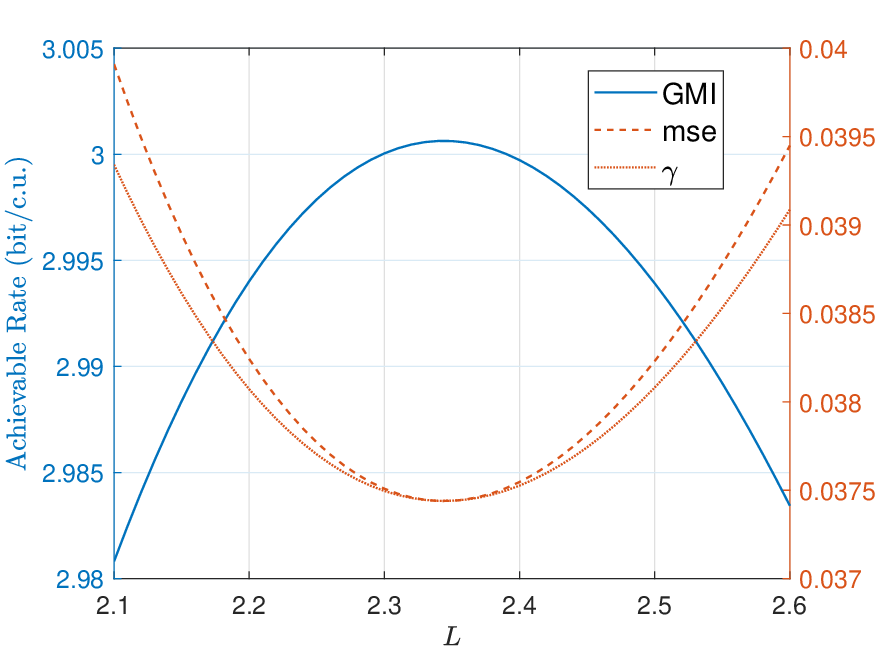}\label{gmimmseu1}}
\subfigure[Optimized quantizer with equispaced thresholds: $\mspace{256mu}$
thresholds $0, \pm 0.5646, \pm 1.1292, \pm 1.6937, \mspace{48mu}$ levels $\pm  0.2749, \pm 0.8247, \pm 1.3747, \pm 2.1049$]{\includegraphics[scale=0.4]{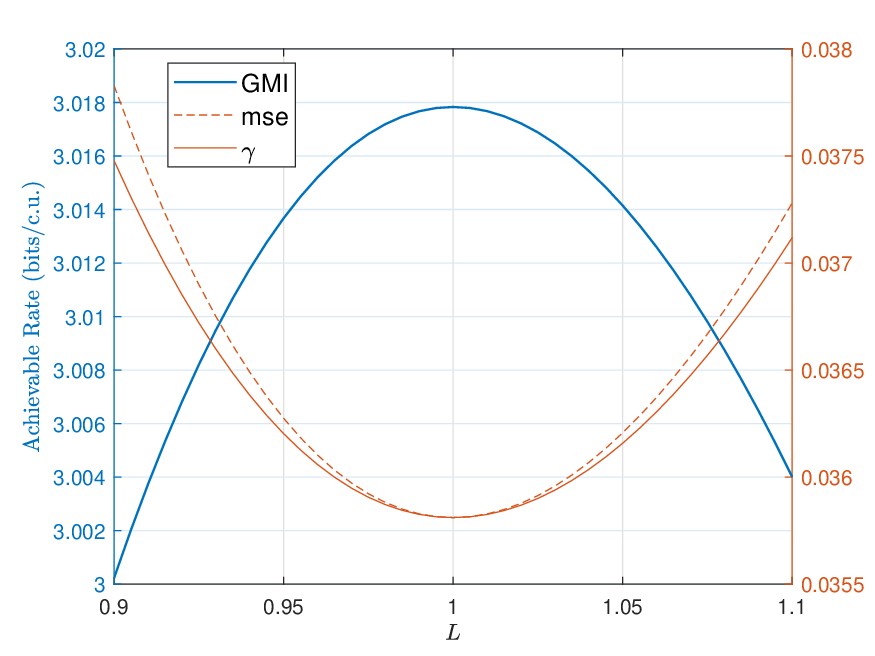}\label{gmimmseuo}}
\subfigure[Nonuniform quantizer: $\mspace{160mu}$ thresholds $0, \pm 0.2, \pm 0.6, \pm 0.7, \mspace{128mu}$ levels $\pm 0.1, \pm 0.9, \pm 0.8, \pm 0.2$]{\includegraphics[scale=0.4]{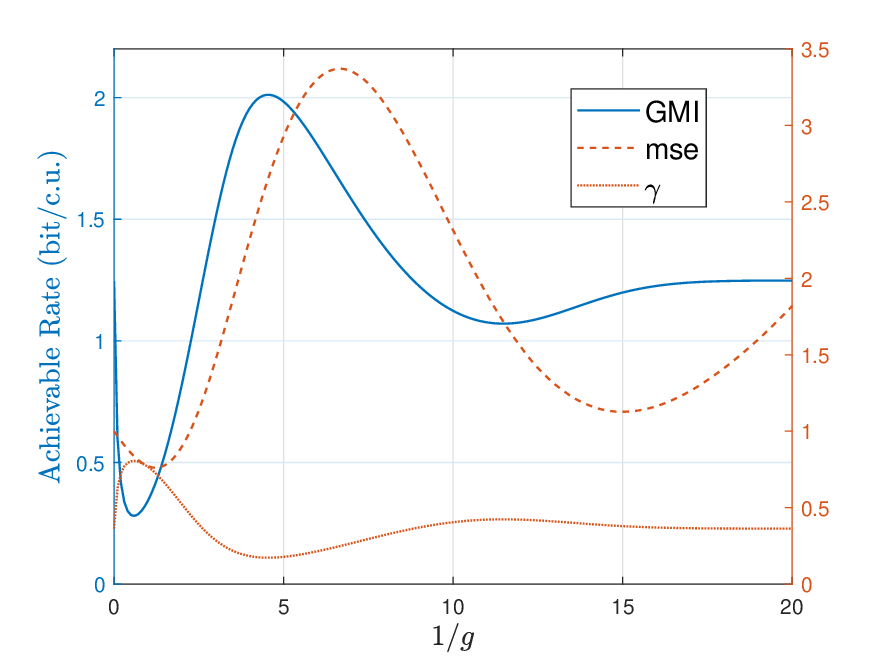}\label{gmimmsenus}}
\caption{GMI, MSE and $\gamma$ under gain control: $\mathsf{SNR}=10$ dB.}\label{gmimsegamma}
\end{figure*}

For the channel (\ref{YqXZ}) with a uniform quantizer $q(\cdot)$ whose resolution is at least two bits, we have shown that the unique gain control factor that minimizes the MSE also maximizes the GMI.
When non-uniform quantization rules are allowed, one may vary the thresholds $\{\ell_k\}$ and levels $\{y_k\}$ of $q(\cdot)$ (possibly under some constraints) to reduce $\gamma$ so that the GMI (\ref{main}) can be optimized, yielding thresholds $\{\ell_k^*\}$ and levels $\{y^*_k\}$.
However, when this optimized quantization rule is applied in a given channel, to achieve that optimized GMI, we need to set the gain control factor $g$ appropriately according to $\sigma_v$ (the standard deviation of the quantizer input) to guarantee that the thresholds satisfy $\ell_k=l_k/g=\ell^*_k$ for $1\leq k< K$.
Otherwise, the performance will be degraded.
We next explore the impact of gain control on the MSE and the GMI of the channel (\ref{YqXZ}) under nonuniform quantization rules by examples; see numerical results given in Fig. \ref{gmimsegamma}, where we set $\mathsf{SNR}=10$ dB.
The examples are described as follows, each of which satisfies $2K=4$, i.e., $b=2$.

\begin{itemize}
\item The uniform quantizer (with equispaced thresholds and mid-rise levels); see Fig. \ref{gmimmse} and Fig. \ref{gmimmseu1}.
\item The optimal nonuniform quantizer: the thresholds and levels are optimized to maximize the GMI; see Fig. \ref{gmimmseopt}.
\item Two nonuniform quantizers without optimization: the first one has monotonically increasing levels, and the second one is highly nonlinear since its levels are no longer monotonically increasing; see Fig. \ref{gmimmsenu} and Fig. \ref{gmimmsenus}, respectively.
\item An optimized quantizer with equispaced thresholds: its thresholds and levels are optimized to maximize the GMI under the constraint that the thresholds must be equispaced; see Fig. \ref{gmimmseuo}.
\end{itemize}

From Fig. \ref{gmimmse}, we confirm that, under uniform quantization, the parameter $\gamma$ converges to $\mathsf{mmse}$ when the loading factor approaches its optimal value $L^*$ (in this example it is $2.3441$); see also Fig. \ref{gmimmseu1}.
Interestingly, the consistency also occurs in two examples: Fig. \ref{gmimmseopt} for the optimal $8$-level quantizer obtained numerically in \cite{Zhang12} by Lloyd's algorithm \cite{Lloyd}, and Fig. \ref{gmimmseuo} for the optimized quantizer with equispaced thresholds (also obtained in \cite{Zhang12}).
But for the nonuniform quantizer given in Fig. \ref{gmimmsenu} and Fig. \ref{gmimmsenus}, the gain control factor that maximizes the GMI and the one that minimizes the MSE are different.
The highly nonlinear quantizer given in Fig. \ref{gmimmsenus} provides an example of unusual behavior of $I_\textrm{GMI}(g)$, which achieves its maximum when the MSE is relatively large.

\emph{Remark}:
In the achievable rate evaluation, if we use the additive noise model of the quantizer, i.e., treating the MSE as additive Gaussian noise, then we obtain an estimate as
\begin{subequations}
\begin{align}
\hat R &= \log\left( 1 + \frac {|h|^2\sigma_x^2} {\sigma^2+2\cdot\mathsf{mse}\cdot\sigma_v^2} \right)\\
&= \log(1+\mathsf{SNR}) - \log\left( 1 + \frac{\mathsf{mse}}{1+\mathsf{mse}} \mathsf{SNR} \right),
\end{align}
\end{subequations}
which may cause underestimate or overestimate.
In particular, it overestimates the achievable rate when the optimal uniform quantization is used, while it may significantly underestimate the achievable rate when $\mathsf {mse}$ is much larger than $\gamma$, which is possible; see the numerical results in Fig. \ref{gmimsegamma} when the gain control factor (or loading factor) is far away from its optimal value.

\subsection{On Improving Uniform Quantization by Post-Processing}

In \cite{ZWSL19,GNND}, it has been shown that the GMI (\ref{ZhangGMI}) in Proposition 1 can be improved by introducing a post-processing of the channel output as $Y\to\tilde{Y}$, thereby extending the channel $X\to Y$ to $X\to\tilde{Y}$;
The GMI is maximized by letting
\begin{align}
\tilde{Y}=\mathrm E\left[X|Y\right],
\end{align}
i.e., using a conditional expectation operator as the post-processor, which gives an MMSE estimation of $X$ under the observation $Y$.
Utilizing this result, we may improve a uniform quantizer by changing its levels.
In fact, the changed levels, denoted by $\{y^{\textrm{c}}_k\}$, can be obtained analytically by its step size $\ell$ as \cite{Zhang12}
\begin{align}\label{olvs}
y^{\textrm{c}}_k = \frac { \phi((k-1)\ell)-\phi(k\ell) } { Q((k-1)\ell)-Q(k\ell) }, \;k=1,...,K-1
\end{align}
and
\begin{align}\label{olvsK}
y^{\textrm{c}}_K = \frac {\phi((K-1)\ell)} {Q((K-1)\ell)}.
\end{align}
Such levels satisfy $y^{\textrm{c}}_k=\mathrm E\left[\frac{V}{\sigma_v}|\frac{V}{\sigma_v}\in[(k-1)\ell,k\ell)\right],\;k=1,...,K-1$ and $y^{\textrm{c}}_K=\mathrm E\left[\frac{V}{\sigma_v}|\frac{V}{\sigma_v}\ge(K-1)\ell\right]$, where $V\sim\mathcal N(0,\sigma_v^2)$ is the input of the quantizer;
i.e., the optimal level is the \emph{centroid} of the corresponding interval.
This is consistent with the aforementioned result that a conditional expectation operator maximizes the GMI.
In particular, the optimized quantizer with equispaced thresholds considered in Sec. VII-A (see Fig. \ref{gmimmseuo}) is equivalent to a uniform quantizer with the same step size $\ell^\circ=0.5646$ (found by numerical optimization) combined with an MMSE post-processor which changes mid-rise levels to centroids.
Comparing Fig. \ref{gmimmseuo} and Fig. \ref{gmimmseu1}, we see that such a post-processing indeed improves the maximum GMI slightly.
Interestingly, the optimal step size $\ell^\circ$ that maximizes the GMI in Fig. \ref{gmimmseuo} is smaller than the optimal step size $\ell^*= 0.5860$ in Fig. \ref{gmimmseu1} (see also Table I).
Thus, when post-processing is available, the step size that maximizes the GMI is reduced.
Consequently, a concatenation of an $\ell^*$-spaced uniform quantizer and an MMSE post-processor is strictly worse than a single $\ell^\circ$-spaced optimized quantizer.

Since the example Fig. \ref{gmimmseuo} shows that the gain of the MMSE post-processing is limited for low-resolution uniform quantization, we may infer that the gain is also limited under high resolution.
In fact, as the step size decreases, the centroid of the interval $[(k-1)\ell,k\ell)$ given in (\ref{olvs}) tends to be its midpoint.\footnote{In \cite{Macro} the asymptotic convergence of centroids to midpoints was proved under uniform quantization with infinitely many levels.}
To see this, note that for an arbitrary $c>\ell/2$, we have
\begin{subequations}
\begin{align}
\lim\limits_{\ell\to 0} \frac { \phi\left(c-\frac{\ell}{2}\right) - \phi\left(c+\frac{\ell}{2}\right) }
{ Q\left(c-\frac{\ell}{2}\right) - Q\left(c+\frac{\ell}{2}\right) } 
=\;& \lim\limits_{\ell\to 0} \frac{ \phi\left(c-\frac{\ell}{2}\right) - \phi\left(c+\frac{\ell}{2}\right) }
{ \int_{c-\frac{\ell}{2}}^{c+\frac{\ell}{2}}\phi(t)\mathrm d t }\\
=\;& \lim\limits_{\ell\to 0}
\frac{ \frac {c-\frac{\ell}{2}} {2} \phi\left(c-\frac{\ell}{2}\right) + \frac {c+\frac{\ell}{2}} {2} \phi\left(c+\frac{\ell}{2}\right)}
{ \frac{1}{2} \left( \phi\left(c+\frac{\ell}{2}\right) + \phi\left(c-\frac{\ell}{2}\right) \right) } \\
=\;&c,
\end{align}
\end{subequations}
which follows from L'H\^{o}pital's rule.
Moreover, according to (\ref{ULBQ}), the largest centroid $y^{\textrm{c}}_K$ given in (\ref{olvsK}) satisfies
\begin{align}
(K-1)\ell > y^{\textrm{c}}_K > (K-1)\ell + \frac{1}{(K-1)\ell},
\end{align}
implying that $y^{\textrm{c}}_K/\ell$ converges to $K-1$ as $\ell\to 0$.
From these facts we can conclude that the gain of MMSE post-processing asymptotically vanishes as the resolution increases.

\subsection{On Possible Gain of Further Quantization Rules}
For communication receivers, nonuniform quantization rules can be realized directly, or indirectly by combining a uniform quantizer with a pre-processing of its input (e.g., companding in pulse-code modulation (PCM)), depending on the cost of implementation.
There have been some works on numerical optimization of nonuniform quantization rules with respect to different performance measures, e.g., cutoff rate \cite{Massey74}, mutual information \cite{SDM09,Medard}, GMI \cite{Zhang12}, and MSE \cite{Fan15,Jacobsson17}.
But no general characterization of the performance gain has been given.

As shown in Fig. \ref{gmimmseopt} and Fig. \ref{gmimmseuo}, for a nonuniform quantizer, if all the thresholds and levels, rather than only a single factor ($g$ or $L$), are optimized (possibly under a constraint that the thresholds are equispaced), then there is also a consistency between GMI maximization and MSE minimization.
In [\ref{Zhang12}, Appendix E] a proof of the consistency has been given (the proof goes through under a equispaced-thresholds constraint).
Thus, the maximum achievable rate can be expressed as
\begin{align}\label{mmsenu}
I^*_\textrm{GMI}=\log(1+\mathsf{SNR})-\log(1+\mathsf{mmse}^\textrm{nu}\cdot\mathsf{SNR}),
\end{align}
where $\mathsf{mmse}^\textrm{nu}$ is the normalized minimum MSE of all possible nonuniform quantizers.
This result can also be understood via a similar geometrical interpretation as that given in Sec. IV-C.
From (\ref{mmsenu}) we can infer that, under our setting of transceiver architecture, the maximum gain in achievable rate from replacing the optimal uniform quantizer by the optimal nonuniform quantizer is determined by the reduction of the normalized MMSE from $\mathsf{mmse}$ to $\mathsf{mmse}^{\textrm{nu}}$.
The numerical results in Fig. \ref{gmimmse} and Fig. \ref{gmimmseopt} show that the rate gain is marginal for $b=2$.
For higher resolutions the rate gain grows slowly.
When $b=4$, according to the numerical results in \cite{Zhang12}, the optimal non-uniform quantization improves the saturation rate $\log(1/\gamma)$ by no more than $4.4 \%$, and optimized quantization with equispaced thresholds improves the saturation rate by $1.5 \%$.
In fact, for nonuniform quantization with a Gaussian input, the asymptotic quantization theory, in particular the Panter-Dite formula \cite{PD}, yields a ``6-dB-per-bit'' rule\footnote{There is another version of the $6$-dB-per-bit rule as $B_\mathrm{eff}=(\mathsf {SNR}_q-1.76)/6.02$, which has been widely used in practice to calculate the effective number of bits (ENoB) of a quantizer \cite{ADC1,ADC2}.
It is derived under assumptions that the granular distortion is uniformly distributed over $[-\ell/2,\ell/2]$ and there is no overload distortion; see \cite{ADC1} for more details.} for the maximum SNR as \cite{Gray-Neuhoff98}
\begin{align}\label{6dbnu}
10\log_{10}\mathsf{SNR}_q^\textrm{nu} = 6.02 b - 4.35 + o(1),
\end{align}
where $\mathsf{SNR}_q^\textrm{nu}=1/\mathsf{mmse}^\textrm{nu}$.
Comparing with (\ref{6db}), we can infer that under our setting the rate gain is limited in the range of resolution of interest.
In particular, the gain in the maximum saturation rate (\ref{satrate}) is $\log\left(\mathsf{SNR}_q^\textrm{nu}/\mathsf{SNR}_q\right)$, which increases logarithmically with $b$.

According to (\ref{6dbnu}), even if the best nonuniform scalar quantization is used,
the MSE achieved in the high-resolution regime is still $2.72$ times larger than the theoretical limit implied by the rate-distortion function of a Gaussian source \cite{Berger}.\footnote{The comparison is made under the assumption that the signal at the receiver front-end is represented by $b$ bits per channel use before feeding into the decoder. The MSE of the representation can be reduced by lossy source coding (treating the received signal as a source). A scalar quantizer is a very simple lossy source coding scheme.}
If we allow \emph{coded} uniform scalar quantization (i.e., representing the quantizer output by a variable-rate lossless code), the gap can be reduced to only $1.53$ dB ($\pi e/6$) worse than the theoretical limit (see \cite{Gray-Neuhoff98} and references therein).
Unfortunately, coded quantization does not help in communication receivers since the bottleneck therein is the limited resolution of the quantizer rather than the cost of representing its output.
 Vector quantization may further reduce the gap, and it is possible to extend our analysis to vector quantizers.\footnote
{Note that the two quantizers we considered in the basic model (\ref{YqXZ}) are equivalent to a 2-dimensional vector quantizer with a rectangular grid of quantization regions. Replacing it by a polar quantizer \cite{Gray-Neuhoff98} may achieve better performance.}
But applying vector quantization to communication receivers is a more challenging topic.
Nevertheless, the gap to the fundamental limit reminds us that it is possible to alleviate the ADC resolution bottleneck in communication receivers by exploring new quantization mechanisms.

A final remark is about dithering, an intentional randomization technique in source quantization \cite{Gray-Neuhoff98}, which may also be beneficial in receiver quantization; see, e.g., \cite{Liang-Zhang16}, which showed that \emph{nonsubtractive} Gaussian dithering\footnote{In nonsubtractive dithering, one adds a dither signal $W_\mathrm{d}$ independent of the quantizer input $V$ before quantization, yielding an output $q(V+W_\mathrm{d})$. Subtractive dithering has an additional step that subtracts the dither signal after quantization and finally obtains $q(V+W_\mathrm{d})-W_\mathrm{d}$. See \cite{Gray-Neuhoff98} for more discussion.} improves the GMI under certain circumstances.
In our setting, although nonsubstractive dithering reduces the SNR, it also changes $\gamma$ by enlarging the standard deviation of the input so that all \emph{normalized} thresholds and levels are scaled by a factor smaller than one.
The overall impact on GMI is thus unclear because it depends on the change of the product $\gamma \cdot \mathsf{SNR}$.
Nevertheless, for uniform quantization, if we employ perfect gain control which guarantees $\gamma \equiv \mathsf{mmse}$, then nonsubstractive Gaussian dithering is always harmful since the GMI (\ref{GMIMMSE}) satisfies
\begin{align}
\frac {\mathrm d I^*_\textrm{GMI}} {\mathrm d\mathsf{SNR}} = \frac {1-\mathsf{mmse}} { (1+\mathsf{SNR}) (1+\mathsf{mmse}\mathsf{SNR}) } > 0.
\end{align}

\section{Concluding Remarks}

The goal of this study is to evaluate the impact of resolution reduction on information-theoretic limits of communications with receiver quantization.
Leveraging the GMI as a basic tool, which enables us to take the decoding rule into consideration, we establish an array of exact and asymptotic results under a standard transceiver architecture.
Our results indicate a critical issue in system design that arises as the resolution decreases, namely, optimizing the loading factor by gain control, which minimizes the loss in achievable rate by eliminating the loading loss.
The remaining irreducible loss is an appropriate evaluation of the inherent robustness of the considered transceiver architecture.
Our results also establish explicit connections between the MSE (affected by the gain control) and the rate loss.
Although for general receiver quantization rules, a smaller MSE does not necessarily imply a smaller rate loss, for the commonly used uniform quantizer we prove that the unique loading factor that minimizes the MSE also maximizes the GMI (i.e., achieves zero loading loss).
For perfect gain control, we show that the irreducible loss is determined by the product of the normalized MMSE and the SNR, and provide a geometrical interpretation for this result.
Performance approximations and per-bit rules in this case are also given.
For imperfect gain control, to understand its impact, we characterize the decay of small rate loss in the high-resolution regime, and propose approximations of the achievable rate as a function of the loading factor which are fairly accurate for moderate resolutions.
These results provide insight into transceiver design with nonnegligible quantization effect, especially regarding the choice of quantizer resolution and the design of the AGC.

A limitation of this work is that the obtained analytical results are derived under the Gaussian codebook.
Like many classical information-theoretic results given by the mutual information, for the GMI, it is also not easy to get closed-form or analytical results without assuming a Gaussian input.
For practical systems with finite alphabets, we can expect that our result applies well for Gaussian-like constellations that come from constellation shaping.
But it is natural to ask whether our results are still useful in the cases of QAM, phase-shift keying (PSK), and other commonly used inputs, which are all bounded (unlike the unbounded Gaussian input).
Since in these cases Proposition 1 does not apply, the GMI evaluation may rely heavily on numerical computation.
We left this problem to future study.
Nevertheless, the insight gained in this work will be very helpful for the finite-alphabet case.

Finally, we list some topics that can be addressed following our information-theoretic framework of receiver quantization.

\begin{itemize}
\item
The orthogonal frequency division multiplexing (OFDM), which typically generate Gaussian-like signal.
Therein, a new effect is that the nonlinear distortion due to quantization leads to intercarrier interference.
The problem becomes more complicated when the channel introduces time-dispersion.

\item
Multiuser channels, especially the Gaussian multiple-access channel.
In this case a new phenomenon due to receiver quantization is that, when successive interference cancellation (SIC) is used to decode a user, there exists residual interference from other users which may significantly reduces the achievable rate of that user.

\item
Multiantenna channels.
We need an extended version of the proposed analytical framework which applies for different receiver architectures (e.g., MMSE receiver, maximal ratio combining receiver).
\end{itemize}

\begin{appendices}

\section{Proof of Proposition 4}

We first note that all the adjustable levels in $\{\ell_k=l_k/g,\; k=1,...,K-1\}$ tend to zero as $g\to \infty$, and they tend to infinity as $g\to 0$.
For the case of $g\to\infty$, from (\ref{As}) and (\ref{Bs})
we have $\lim_{g\to \infty}\mathcal A/y_K=\sqrt{2\pi}\phi(0)$ and $\lim_{g\to \infty}\mathcal B/y_K^2=\pi Q(0)$.
For the case of $g\to 0$, noting that $\mathcal A$ and $\mathcal B$ can be expressed by
\begin{align}
\mathcal A = y_1 + \sqrt{2\pi}\sum\limits_{k=1}^{K-1} \phi(\ell_k)(y_{k+1}-y_{k})
\end{align}
and
\begin{align}
\mathcal B = \frac{\pi}{2} y_1^2 + \pi\sum\limits_{k=1}^{K-1} Q(\ell_k)(y^2_{k+1}-y^2_{k}),
\end{align}
respectively, it is direct to check that $\lim_{g\to 0}\mathcal A/y_1=1$ and $\lim_{g\to 0}\mathcal B/y_1^2=\pi/2$.
Therefore, in both cases $\gamma$ tends to $1-2/\pi$ and the limits of the GMI are the same, $I_\textrm{GMI}^\textrm{1-bit}$.

\section{Proof of Proposition 5}

We begin from the definition (\ref{MSE}) in which the quantization rule is given by (\ref{qs}) and is equivalent to
\begin{align}
q(gV) = y_k \cdot \mathrm{sgn}(V), \; \textrm{if} \; \ell_{k-1} \leq v <\ell_k,
\end{align}
where $v=|V|/\sigma_v\sim\mathcal N(0,1)$.
Utilizing the symmetries of the input distribution and the quantization rule with respect to the origin, we have
\begin{subequations}
\begin{align}
\mathsf{mse}=\;& 2\sum\limits_{k=1}^{K} \int_{\ell_{k-1}}^{\ell_k} y_k^2\phi(v)\mathrm d v
 -2\sum\limits_{k=1}^{K} \int_{\ell_{k-1}}^{\ell_k}2y_k v\phi(v) \mathrm d v + 1 \\
=\;& 1 - 4\sum\limits_{k=1}^{K} y_k \left( \phi(\ell_{k-1})-\phi(\ell_k) \right)
 + 2\sum\limits_{k=1}^{K} y_k^2 \left( Q(\ell_{k-1})-Q(\ell_k) \right) \\
=\;& 1 - \frac{2}{\pi} \left(\sqrt{2\pi}\mathcal A-\mathcal B\right).
\end{align}
\end{subequations}
The lower bound and condition of equality follow from
\begin{align}
\mathsf{mse} - \gamma = \frac{1}{\mathcal B} \left(\mathcal A - \sqrt\frac{2}{\pi}\mathcal B\right)^2 \ge 0.
\end{align}

\section{Proof of Corollary 6}

According to (\ref{As}), we can derive (\ref{Au}) straightforwardly as
\begin{subequations}\label{alphaU}
\begin{align}
\mspace{-12mu}\mathcal A =
\;& \sum\limits_{k=1}^{K-1} \left(k\ell-\frac{\ell}{2}\right) \left(\exp\frac{-(k-1)^2\ell^2}{2} - \exp\frac{-k^2\ell^2}{2}\right)
 + \left(K\ell-\frac{\ell}{2}\right) \exp\frac{-(K-1)^2\ell^2}{2} \\
= \;& \sum\limits_{k=0}^{K-2} (k+1) \ell\exp\frac{-k^2\ell^2}{2} - \sum\limits_{k=1}^{K-1} k\ell \exp\frac{-k^2\ell^2}{2}
 -\sum\limits_{k=0}^{K-2} \frac{\ell}{2}\exp\frac{-k^2\ell^2}{2} + \sum\limits_{k=1}^{K-1} \frac{\ell}{2}\exp\frac{-k^2\ell^2}{2} \notag\\
& + K\ell\exp\frac{-(K-1)^2\ell^2}{2} - \frac{\ell}{2}\exp\frac{-(K-1)^2\ell^2}{2} \\
=\;&  \sum\limits_{k=0}^{K-1} \ell\cdot\exp\frac{-k^2\ell^2}{2} - \frac{\ell}{2},
\end{align}
\end{subequations}
and according to (\ref{Bs}), we obtain (\ref{Bu}) as
\begin{subequations}\label{betaU}
\begin{align}
\mathcal B = \;& \pi\sum\limits_{k=1}^{K-1} \left(k-\frac{1}{2}\right)^2 \ell^2 \left(Q((k-1)\ell)-Q(k\ell)\right)
 + \pi\left(K-\frac{1}{2}\right)^2 \ell^2 Q((K-1)\ell) \\
=\;& \pi\ell^2 \Bigg(\sum\limits_{k=1}^{K-1}k^2\left(Q((k-1)\ell) - Q(k\ell)\right)
 - \sum\limits_{k=1}^{K-1} k\left(Q((k-1)\ell) - Q(k\ell)\right)
 + \sum\limits_{k=1}^{K-1} \frac{1}{4}\left(Q((k-1)\ell) - Q(k\ell)\right) \notag\\
& +\left(K^2-K+\frac{1}{4}\right)Q((K-1)\ell)\Bigg) \\
=\;& \pi\ell^2 \Bigg( \sum\limits_{k=0}^{K-2} (k+1)^2Q(k\ell) - \sum\limits_{k=0}^{K-1} k^2Q(k\ell)
 - \sum\limits_{k=0}^{K-2} (k+1)Q(k\ell) + \sum\limits_{k=0}^{K-1}k Q(k\ell)
 + \frac{1}{4}\sum\limits_{k=0}^{K-2}Q(k\ell) - \frac{1}{4}\sum\limits_{k=0}^{K-1}Q(k\ell)+\frac{Q(0)}{4} \notag\\
& +\left(K^2-K+\frac{1}{4}\right)Q((K-1)\ell)\Bigg) \\
=\;& \pi\ell^2 \Bigg( \sum\limits_{k=0}^{K-1}(2k+1) Q(k\ell) -K^2Q((K-1)\ell) - \sum\limits_{k=0}^{K-1} Q(k\ell) +KQ((K-1)\ell) -\frac{1}{4}Q((K-1)\ell)+ \frac{1}{8}\notag\\
& +\left(K^2-K+\frac{1}{4}\right)Q((K-1)\ell)\Bigg) \\
=\;& \pi\sum\limits_{k=0}^{K-1} 2k\ell^2 Q(k\ell) + \frac{1}{8}\pi\ell^2.
\end{align}
\end{subequations}

\section{Proof of Lemma 7}

Let $K>1$.
According to Theorem 2, the LHS of (\ref{condition}) is equivalent to $\frac{\mathrm d \gamma}{\mathrm d \ell}=0$, or
\begin{align}\label{condition1}
\frac {\mathrm d \mathcal A} {\mathrm d \ell} = \frac {\mathcal A} {2\mathcal B} \frac {\mathrm d \mathcal B} {\mathrm d \ell}.
\end{align}
According to Proposition 5, the RHS of (\ref{condition}) is equivalent to
\begin{align}\label{condition2}
\frac {\mathrm d\mathcal A} {\mathrm d \ell} = \frac {1} {\sqrt{2\pi}} \frac {\mathrm d\mathcal B} {\mathrm d \ell}.
\end{align}
Let
\begin{align}
\mathcal C = \sum\limits_{k=0}^{K-1} \ell \cdot \left(k^2\ell^2 \cdot \exp\frac{-k^2\ell^2}{2}\right),
\end{align}
which is strictly positive when $K>1$.
Noting that
\begin{subequations}
\begin{align}
\ell \cdot \frac {\mathrm d \mathcal A} {\mathrm d \ell}
=\; & \sum\limits_{k=0}^{K-1} \ell\cdot\exp\frac{-k^2\ell^2}{2} - \frac{\ell}{2}
     -\sum\limits_{k=0}^{K-1} \ell\cdot\left(k^2\ell^2 \cdot \exp\frac{-k^2\ell^2}{2}\right) \\
=\; & \mathcal A - \sum\limits_{k=0}^{K-1} \ell \cdot \left(k^2\ell^2 \cdot \exp\frac{-k^2\ell^2}{2}\right) \\
=\; & \mathcal A - \mathcal C
\end{align}
\end{subequations}
and
\begin{subequations}
\begin{align}
\frac{1}{\sqrt{2\pi}}\ell \cdot \frac {\mathrm d \mathcal B} {\mathrm d \ell}
=\;& \sqrt{2\pi}\sum\limits_{k=0}^{K-1} 2k\ell^2 Q(k\ell) + \frac{\sqrt{2\pi}}{8}\ell^2
    -\sum\limits_{k=0}^{K-1} \ell \cdot \left(k^2\ell^2 \exp \frac{-k^2\ell^2}{2}\right) \\
=\;& \sqrt{\frac{2}{\pi}}\mathcal B - \sum\limits_{k=0}^{K-1} \ell \cdot \left(k^2\ell^2 \exp\frac{-k^2\ell^2}{2}\right) \\
=\;& \sqrt{\frac{2}{\pi}}\mathcal B - \mathcal C,
\end{align}
\end{subequations}
it is direct to check that both (\ref{condition1}) and (\ref{condition2}) are equivalent to ({\ref{AB}}),
thereby completing the proof.

\section{Proof of Proposition 11}

First, (\ref{rhoxv}) and (\ref{rhovy}) can be obtained by the Pythagorean relations (\ref{VXZ}) and (\ref{VYe}), respectively.
According to the definitions of $\mathsf X$, $\mathsf Y$, and $\mathsf V$, we can rewrite (\ref{XYA}) and (\ref{YB}) as
\begin{align}
\mathrm E\left[\mathsf X \overline{\mathsf Y}\right]
= \mathrm E\left[|\mathsf X|^2\right] \sqrt{\frac{2}{\pi}} \mathcal A
= \mathrm E\left[\mathsf X\overline{\mathsf V}\right] \sqrt{\frac{2}{\pi}} \mathcal A
\end{align}
and
\begin{align}
\mathrm E\left[|\mathsf Y|^2\right]=\frac{|h|^2\sigma_x^2+\sigma^2}{2}\mathrm E\left[|Y|^2\right]= \mathsf E\left[|\mathsf V|^2\right] \frac{2}{\pi}\mathcal B,
\end{align}
respectively.
Combining them with (\ref{rhoxv}) we obtain
\begin{align}\label{wAB}
\frac {\mathrm E\left[\mathsf{X}\overline{\mathsf{Y}}\right]} {\mathrm E\left[|\mathsf{Y}|^2\right]}
=\frac {\mathrm E\left[\mathsf{X}\overline{\mathsf V}\right]} {\mathrm E\left[|\mathsf V|^2\right]} \sqrt{\frac{\pi}{2}}\frac{\mathcal A}{\mathcal B},
\end{align}
thereby implying that (\ref{AB}) is equivalent to (\ref{ww}).
According to (\ref{rhovy}), we have
\begin{align}\label{YVY}
\mathrm E\left[\mathsf V\overline{\mathsf{Y}}\right] = (\mathcal E_x+\sigma^2)(1-\mathsf {mmse}) = \mathrm E\left[|\mathsf Y|^2\right],
\end{align}
where the second equality follows from the length of $\mathbf Y$; see the Pythagorean relation (\ref{VYe}).
Substituting (\ref{YVY}) into (\ref{wAB}) and utilizing the fact $\mathrm E[|\mathsf X|^2]=\mathrm E[\mathsf X \overline{\mathsf V}]$ yield
\begin{align}\label{betaAB}
\frac {\mathrm E\left[\mathsf{X}\overline{\mathsf{Y}}\right]} {\mathrm E\left[|\mathsf X|^2\right]}
= \frac {\mathrm E\left[\mathsf{V}\overline{\mathsf Y}\right]} {\mathrm E\left[|\mathsf V|^2\right]}
\sqrt{\frac{\pi}{2}} \frac{\mathcal A}{\mathcal B},
\end{align}
thereby implying that (\ref{AB}) is equivalent to (\ref{betabeta}).
Combining (\ref{wAB}), its equivalent form (\ref{betaAB}), and the definition of the Pearson correlation coefficient (\ref{Pearson}), we obtain
\begin{align}
\rho^2_{XY}=\rho^2_{XV}\rho^2_{VY}\frac{\pi\mathcal A^2}{2\mathcal B^2},
\end{align}
thereby implying that (\ref{AB}) is equivalent to (\ref{rhorhorho}).

\end{appendices}

\end{document}